\documentclass[12pt,a4paper]{article}

\usepackage{amsmath,amssymb}
\usepackage[nosort]{cite}
\usepackage[hyperref,bulletsep]{collect}
\def\gfxon{\usepackage[final]{graphicx}}

\gfxon

\makeatletter
\let\old@startsection=\@startsection
\renewcommand{\@startsection}[6]{\old@startsection{#1}{#2}{#3}{#4}{#5}{#6\mathversion{bold}}}
\makeatother

\makeatletter \@addtoreset{equation}{section} \makeatother

\makeatletter
\let\old@makecaption=\@makecaption
\def\@makecaption{\small\old@makecaption}
\makeatother

\let\oldPhi=\Phi
\let\oldPsi=\Psi
\let\oldGamma=\Gamma
\let\oldDelta=\Delta
\let\oldSigma=\Sigma
\let\oldLambda=\Lambda
\let\oldTheta=\Theta
\let\oldPi=\Pi
\renewcommand{\Phi}{\mathnormal{\oldPhi}}
\renewcommand{\Psi}{\mathnormal{\oldPsi}}
\renewcommand{\Gamma}{\mathnormal{\oldGamma}}
\renewcommand{\Sigma}{\mathnormal{\oldSigma}}
\renewcommand{\Delta}{\mathnormal{\oldDelta}}
\renewcommand{\Theta}{\mathnormal{\oldTheta}}
\renewcommand{\Lambda}{\mathnormal{\oldLambda}}
\renewcommand{\Pi}{\mathnormal{\oldPi}}

\newcommand{\ham}{\mathcal{H}}

\newcommand{\gen}[1]{\mathfrak{#1}}

\newcommand{\smat}{\mathcal{S}}
\newcommand{\ident}{\mathcal{I}}
\newcommand{\perm}{\mathcal{P}}
\newcommand{\tmat}{\mathcal{T}}

\newcommand{\transfer}{\mathcal{T}}
\newcommand{\mono}{\mathcal{M}}
\newcommand{\qdet}{\mathcal{D}}
\newcommand{\qdetev}{\mathrm{D}}
\newcommand{\ushift}{\mathrm{U}}

\newcommand{\superN}{\mathcal{N}}
\newcommand{\fldZ}{\mathcal{Z}}
\newcommand{\fldY}{\mathcal{Y}}
\newcommand{\fldX}{\mathcal{X}}
\newcommand{\copro}{\oldDelta}

\newcommand{\gym}{g\indups{YM}}

\newcommand{\str}{\mathop{\mathrm{str}}}
\newcommand{\sdet}{\mathop{\mathrm{sdet}}}

\newcommand{\order}[1]{\mathcal{O}(#1)}

\newcommand{\Integers}{\mathbb{Z}}
\newcommand{\Reals}{\mathbb{R}}

\makeatletter
\newcommand{\ellSN}{\mathop{\operator@font sn}\nolimits}
\newcommand{\ellCN}{\mathop{\operator@font cn}\nolimits}
\newcommand{\ellDN}{\mathop{\operator@font dn}\nolimits}
\newcommand{\ellAM}{\mathop{\operator@font am}\nolimits}
\newcommand{\ellK}{\mathop{\smash{\operator@font K}\vphantom{a}}\nolimits}
\newcommand{\ellE}{\mathop{\smash{\operator@font E}\vphantom{a}}\nolimits}
\makeatother

\ifx\genfrac\sdflkaj

\else

\fi
\newcommand{\sfrac}[2]{{\textstyle\frac{#1}{#2}}}
\newcommand{\half}{\sfrac{1}{2}}
\newcommand{\ihalf}{\sfrac{i}{2}}
\newcommand{\quarter}{\sfrac{1}{4}}

\newcommand{\auxrep}{\mathrm{a}}

\newcommand{\indup}[1]{_{\mathrm{#1}}}
\newcommand{\indups}[1]{_{\mathrm{\scriptscriptstyle #1}}}
\newcommand{\supup}[1]{^{\mathrm{#1}}}
\newcommand{\rep}[1]{{\mathbf{#1}}}
\newcommand{\matr}[2]{\left(\begin{array}{#1}#2\end{array}\right)}
\newcommand{\lvl}[1]{^{\mathrm{#1}}}

\newcommand{\lrbrk}[1]{\left(#1\right)}
\newcommand{\bigbrk}[1]{\bigl(#1\bigr)}

\newcommand{\comm}[2]{[#1,#2]}

\newcommand{\acomm}[2]{\{#1,#2\}}

\newcommand{\eval}[1]{#1|}

\newcommand{\set}[1]{\{#1\}}
\newcommand{\state}[1]{\mathopen{|}#1\mathclose{\rangle}}

\newcommand{\srep}[1]{\langle #1\rangle}
\newcommand{\lrep}[1]{\{#1\}}

\newcommand{\xexp}[2]{x^{(#1)}_{#2}}
\newcommand{\xp}[1]{x^{+}_{#1}}
\newcommand{\xm}[1]{x^{-}_{#1}}
\newcommand{\xpm}[1]{x^{\pm}_{#1}}
\newcommand{\xmp}[1]{x^{\mp}_{#1}}

\newcommand{\alg}[1]{\mathfrak{#1}}
\newcommand{\grp}[1]{\mathrm{#1}}

\newcommand{\nn}{\nonumber}
\newcommand{\nln}{\nonumber\\}
\newcommand{\nl}[1][0pt]{\nonumber\\[#1]&\hspace{-4\arraycolsep}&\mathord{}}
\newcommand{\nlnum}{\\&\hspace{-4\arraycolsep}&\mathord{}}
\newcommand{\earel}[1]{\mathrel{}&\hspace{-2\arraycolsep}#1\hspace{-2\arraycolsep}&\mathrel{}}
\newcommand{\eq}{\earel{=}}

\def\[{\begin{equation}}
\def\]{\end{equation}}
\def\<{\begin{eqnarray}}
\def\>{\end{eqnarray}}

\makeatletter
\def\mr@ignsp#1 {\ifx\:#1\@empty\else #1\expandafter\mr@ignsp\fi}%
\newcommand{\multiref}[1]{\begingroup
\xdef\mr@no@sparg{\expandafter\mr@ignsp#1 \: }%
\def\mr@comma{}%
\@for\mr@refs:=\mr@no@sparg\do{\mr@comma\def\mr@comma{,}\ref{\mr@refs}}%
\endgroup}
\makeatother

\newcommand{\hypref}[2]{\ifx\href\asklfhas #2\else\href{#1}{#2}\fi}
\newcommand{\Secref}[1]{Section~\multiref{#1}}
\newcommand{\secref}[1]{Sec.~\multiref{#1}}

\newcommand{\appref}[1]{App.~\multiref{#1}}

\newcommand{\tabref}[1]{Tab.~\multiref{#1}}

\newcommand{\figref}[1]{Fig.~\multiref{#1}}
\renewcommand{\eqref}[1]{(\multiref{#1})}

\ifx\href\asklfhas\newcommand{\href}[2]{#2}\fi
\newcommand{\arxivno}[1]{\href{http://arxiv.org/abs/#1}{#1}}

\begin{document}

\thispagestyle{empty}
\begin{flushright}\footnotesize
\texttt{\arxivno{nlin.SI/0610017}}\\
\texttt{AEI-2006-074}\\
\texttt{PUTP-2211}%
\vspace{0.5cm}
\end{flushright}
\vspace{0.5cm}

\renewcommand{\thefootnote}{\fnsymbol{footnote}}
\setcounter{footnote}{0}

\begin{center}%
{\Large\textbf{\mathversion{bold}%
The Analytic Bethe Ansatz for a Chain\\%
with Centrally Extended $\alg{su}(2|2)$ Symmetry}\par} \vspace{1cm}%

\textsc{Niklas Beisert}\vspace{5mm}%

\textit{Max-Planck-Institut f\"ur Gravitationsphysik\\%
Albert-Einstein-Institut\\%
Am M\"uhlenberg 1, 14476 Potsdam, Germany}\vspace{3mm}%

and\vspace{3mm}

\textit{Joseph Henry Laboratories\\%
Princeton University\\%
Princeton, NJ 08544, USA}\vspace{3mm}

\texttt{nbeisert@aei.mpg.de}%
\par\vspace{1cm}

\vfill

\textbf{Abstract}\vspace{5mm}

\begin{minipage}{12.7cm}
We investigate the integrable structure of spin chain models
with centrally extended $\alg{su}(2|2)$ and $\alg{psu}(2,2|4)$ symmetry.
These chains have their origin in the planar AdS/CFT correspondence,
but they also contain the one-dimensional Hubbard model as a special case.
We begin with an overview of the representation 
theory of centrally extended $\alg{su}(2|2)$. 
These results are applied in the construction and investigation of
an interesting S-matrix with $\alg{su}(2|2)$ symmetry. 
In particular, they enable a remarkably simple proof of the Yang-Baxter relation.
We also show the equivalence of the S-matrix to Shastry{}'s R-matrix
and thus uncover a hidden supersymmetry in the integrable structure
of the Hubbard model.
We then construct eigenvalues of the corresponding transfer matrix 
in order to formulate an analytic Bethe ansatz.
Finally, the form of transfer matrix eigenvalues 
for models with $\alg{psu}(2,2|4)$ symmetry is sketched.
\end{minipage}

\vspace*{\fill}

\end{center}

\newpage
\setcounter{page}{1}
\renewcommand{\thefootnote}{\arabic{footnote}}
\setcounter{footnote}{0}

\section{Introduction and Overview}
\label{sec:Intro}


Gauge/string dualities give promise to explain 
stringy aspects of quantum chromodynamics and
to deepen our understanding of quantum gravity.
They relate two seemingly different quantum field theory models:
gauge theories in various spacetime dimensions 
and string theories based on a two-dimensional world sheet QFT.
The most elaborate such duality is Maldacena{}'s AdS/CFT 
correspondence \cite{Maldacena:1998re,Gubser:1998bc,Witten:1998qj}.
It identifies a string theory on an $AdS_{d+1}\times X$ background with a 
conformal field theory on the $d$-dimensional boundary of the $AdS_{d+1}$ space. 
The key example of AdS/CFT is the conjectured exact duality between 
IIB superstrings on $AdS_5\times S^5$ and 
$\superN=4$ extended supersymmetric gauge theory
in four spacetime dimensions.
We shall focus on this particular duality in the present work. 

AdS/CFT-dual models typically have at least two parameters:
a coupling constant $\lambda$ and a genus-counting parameter $g\indup{s}$.
While the genus-counting parameter is natural within string theory, 
it is given by $4\pi g\indup{s}/\lambda= 1/N\indup{c}$ 
in a $\grp{U}(N\indup{c})$ gauge theory.
The equivalence of the latter two parameters 
was shown a long time ago by 't Hooft \cite{'tHooft:1974jz}. 
A suitable coupling constant for gauge theory is the 't Hooft coupling
$\lambda=\gym^2N\indup{c}$ and for string theory it
is related to the string tension by $\lambda=1/\alpha'^2$.
The relationship between these parameters is less obvious
because the perturbative regimes of both models do not overlap:
String theory is strongly coupled where gauge theory is 
perturbative and vice versa. 
The distinctness of perturbative regimes is actually what
makes the AdS/CFT possible despite the fact that
the perturbative models do not resemble each other remotely.
The strong/weak nature of AdS/CFT can thus be viewed ambivalently:
On the one hand, it gives access to hitherto inaccessible regimes 
in both modes. 
However, these predictions would require us 
to put all our faith into the correspondence.
If we prefer not to, on the other hand, 
the strong/weak nature prevents almost all
tests of the conjectured duality as we cannot compute 
corresponding quantities in both participating models simultaneously.
Nevertheless some tests are possible and 
confirm the duality,
cf.~the reviews \cite{Aharony:1999ti,D'Hoker:2002aw}.
Most of these tests involve quantities
which are protected from receiving quantum corrections
and which can therefore be carried easily from 
one perturbative regime to the other.

At least in the planar limit, $N\indup{c}=\infty$,
some progress towards a comparison of quantities
which depend non-trivially on the coupling constant $\lambda$
has been made in recent years.
To absorb most factors of $\pi$ and $2$ we shall use a
normalised coupling constant 
\[g=\frac{\sqrt{\lambda}}{4\pi}\,.\]
The suspected exact integrability of
planar $\superN=4$ super Yang-Mills (SYM) theory 
\cite{Minahan:2002ve,Beisert:2003tq,Beisert:2003yb},
see also \cite{Lipatov:1997vu,Braun:1998id,Belitsky:1999qh}, 
and of non-interacting IIB superstring theory on
$AdS_5\times S^5$ \cite{Mandal:2002fs,Bena:2003wd,Arutyunov:2004vx}
provides hope that their spectrum can be computed exactly at finite coupling $g$
by means of Bethe equations \cite{Beisert:2005fw},
cf.~the reviews \cite{Beisert:2004ry,Zarembo:2004hp,Plefka:2005bk}
and \cite{Belitsky:2004cz}.

The underlying integrable model of AdS/CFT is best described as
a two-dimensional sigma model \cite{Metsaev:1998it} 
in the limit of perturbative string theory 
and as a spin chain in the limit of perturbative gauge theory. 
Many results and techniques have been developed 
for these two types of integrable models. 
For instance, a general framework exists 
for the solution of a large class of integrable spin chains. 
The Bethe equations for these chains can easily
be written down once the symmetry and representation content is given
\cite{Reshetikhin:1987bz}. 
They are in general founded strongly on
the symmetry algebra and representation theory of the model.
Unfortunately, the standard form of rational Bethe equations
does not apply to the spin chain of $\superN=4$ SYM,
a fact which is explained by its slightly unusual form: 
While almost all known integrable spin chain Hamiltonians 
induce interactions between two neighbouring spin sites, 
the $\superN=4$ SYM spin chain Hamiltonian  
consists of interactions with a longer range and between more than two sites. 
Moreover, in standard spin chains the Hamiltonian
alias the time translation generator
factors from the remaining symmetry group $G$ as $\Reals\times G$. 
Here, the Hamiltonian is merely
one generator of the irreducible symmetry group $\grp{PSU}(2,2|4)$ of AdS/CFT.
This has some important and puzzling implications for the 
representation theory of the model. 
Similar problems are encountered for
the stringy sigma model of AdS/CFT which is
not strictly Poincar\'e-invariant unlike many of 
the well-known integrable sigma models.
All this means that the standard solution for integrable models
does not apply.
Nevertheless the Bethe equations for AdS/CFT are somewhat similar
to standard rational Bethe equations and they display signs of 
the underlying $\alg{psu}(2,2|4)$ symmetry. 
It is therefore conceivable that some unified framework 
for the treatment of the AdS/CFT and standard spin chains can be found. 
Such a framework would provide more insight into the
foundations of the integrable structures of AdS/CFT, 
but may also contribute to 
the general understanding of integrable structures.
It is the aim of this paper to take some steps towards such a framework.

The main objects of investigation in this article
will be the residual algebra, the S-matrix and 
transfer matrices. 
We will assume that the $\superN=4$ SYM spin chain 
has already been transformed to a 
particle model by means of a coordinate space Bethe ansatz
\cite{Bethe:1931hc,Sutherland:1978aa,Staudacher:2004tk},
see also \cite{Berenstein:2002jq}.
In other words, spin flips about a ferromagnetic vacuum 
are considered as momentum-carrying particles.
For the string sigma model we will assume that a light cone gauge 
has reduced the spectrum to physical excitation modes
also yielding a particle model. 
In the particle model picture the full $\grp{PSU}(2,2|4)$ symmetry of AdS/CFT
is spontaneously broken by the vacuum to some residual symmetry. 
The latter consists of two copies of the supergroup
$\grp{PSU}(2|2)$ with central extensions \cite{Beisert:2005tm,Arutyunov:2006ak}.
\Secref{sec:Alg} deals with the representation theory of the 
centrally extended $\alg{psu}(2|2)$ algebra.

Symmetry is a central ingredient for the construction
and investigation of the S-matrix performed in the
subsequent \secref{sec:scat}.
The S-matrix \cite{Staudacher:2004tk,Beisert:2005fw,Beisert:2005tm} describes 
the asymptotic wave functions of multi-particle
states on a vacuum of infinite length.
In this section we shall derive and compare
different notations to deal with multi-particle states 
and then derive the S-matrix as well as its properties. 
Most importantly, we will find a simple proof
of the Yang-Baxter equation (YBE) making full use of
representation theory. The YBE then enables
us to diagonalise the S-matrix by means of a nested Bethe ansatz.
Finally we consider multi-particle states
on a compact vacuum by imposing
periodicity conditions on the wave function.
These are the (asymptotic) Bethe equations of the system.

In \secref{sec:transfer} we shall proceed towards 
an analytic Bethe ansatz \cite{Reshetikhin:1983vw} for a spin chain 
with centrally extended $\alg{psu}(2|2)$ symmetry. 
The central objects of the analytic Bethe ansatz are transfer matrices
whose eigenvalues we shall obtain by reverse-engineering.
In other words, we will assume that the analytic Bethe ansatz leads to 
the same Bethe equations that were derived before. 
This step fixes large parts of the structure of the transfer matrix eigenvalues.
Considering some explicit states, we can write down the spectrum
of transfer matrices in several representations.

Finally, we would like to sketch how to assemble
the transfer matrices of two $\alg{psu}(2|2)$ spin chains 
to a transfer matrix of the AdS/CFT model with $\alg{psu}(2,2|4)$ symmetry.
This \secref{sec:psu224} is of a very explorative character; 
it does not provide conclusive answers, but rather starting points 
and clues for further investigations. 
A rigorous treatment would require a full investigation of 
the abelian phase factor of the S-matrix
which is beyond the scope of the present paper.
This phase should obey a crossing relation derived by Janik \cite{Janik:2006dc} 
leading to a very intricate
analytic structure \cite{Beisert:2006zy,Beisert:2006ib}.
Hopefully, an analytic Bethe ansatz will be obtained
elsewhere by a rigorous treatment of the analytic structure 
of the transfer matrix eigenvalue. 

In the interlude of \secref{sec:Hubbard}
we investigate the connection between
the AdS/CFT correspondence and the one-dimensional Hubbard model \cite{Hubbard:1963aa}.
The latter is an exceptional spin chain model because 
it lies within the standard class of nearest-neighbour models described above, 
but its Bethe equations, the so-called Lieb-Wu equations \cite{Lieb:1968aa}, 
take a non-standard trigonometric form. 
Its integrable structures are known to a large extent, 
cf.~\cite{Essler:2005aa} for a review,
but they also take an unusual form. 
For example, Shastry{}'s R-matrix for the Hubbard chain \cite{Shastry:1986bb}
is not of a difference form;
it is rather a function which genuinely depends on 
two independent spectral parameters.
It is fair to say that the foundations of this integrable system
are not yet fully understood.
In particular the relation to representation theory,
which is a central aspect of standard integrable spin chains,
remains obscure.
In this paper we will show that the Lieb-Wu equations and Shastry{}'s R-matrix
appear within AdS/CFT as parts of the Bethe equations and the S-matrix.
Both models are therefore based on the same integrable structures.
This provides a novel way of looking at the Hubbard chain,
in particular at the underlying symmetry and representation theory.
It should be noted that this connection between the two models 
is complementary to the one discovered earlier in \cite{Rej:2005qt}:
There are some similarities between the two observations, but
they neither explain nor exclude each other.

\section{Centrally Extended $\alg{su}(2|2)$}
\label{sec:Alg}

In this section we introduce the algebra 
on which the spin chain model is based.
We shall denote it by $\gen{h}$.

\subsection{The Algebra}
\label{sec:Alg.Alg}

The centrally extended $\alg{su}(2|2)$ algebra $\gen{h}=\alg{psu}(2|2)\ltimes\Reals^3$,
see e.g.~\cite{Iohara:2001aa},
consists of the $\alg{su}(2)\times\alg{su}(2)$ rotation generators 
$\gen{R}^a{}_b$, $\gen{L}^\alpha{}_\beta$, the 
supersymmetry generators 
$\gen{Q}^\alpha{}_b$, $\gen{S}^a{}_\beta$ 
and the central charges $\gen{C},\gen{P},\gen{K}$.
The non-trivial commutators are 
\<\label{eq:Alg.CommSU22}
\comm{\gen{R}^a{}_b}{\gen{J}^c}\eq\delta^c_b\gen{J}^a
  -\half \delta^a_b\gen{J}^c,
\nln
\comm{\gen{L}^\alpha{}_\beta}{\gen{J}^\gamma}\eq
  \delta^\gamma_\beta\gen{J}^\alpha
  -\half \delta^\alpha_\beta\gen{J}^\gamma,
\nln
\acomm{\gen{Q}^\alpha{}_a}{\gen{S}^b{}_\beta}
\eq 
\delta^b_a\gen{L}^\alpha{}_\beta
+\delta^\alpha_\beta\gen{R}^b{}_a
+\delta^b_a\delta^\alpha_\beta\gen{C},
\nln
\acomm{\gen{Q}^\alpha{}_a}{\gen{Q}^\beta{}_b}
\eq 
\varepsilon^{\alpha\beta}\varepsilon_{ab}\gen{P},\nln
\acomm{\gen{S}^a{}_\alpha}{\gen{S}^b{}_\beta}
\eq 
\varepsilon^{ab}\varepsilon_{\alpha\beta}\gen{K}.
\>
The symbols $\gen{J}^a,\gen{J}^\alpha$ represent any generator with 
an upper index.

\subsection{Outer Automorphism}
\label{sec:Alg.Ext}

The algebra $\gen{h}$ has an $\alg{sl}(2)$ outer automorphism.
This can be seen most easily when we rearrange the generators
into multiplets of the automorphism:
\[
\gen{C}^{\mathfrak{a}}{}_{\mathfrak{b}}=
\matr{ll}{-\gen{C}&+\gen{P}\\-\gen{K}&+\gen{C}},
\qquad
\gen{J}^{a\beta \mathfrak{c}}=
\matr{c}{\varepsilon^{ad}\gen{Q}^\beta{}_{d}\\\varepsilon^{\beta\delta}\gen{S}^a{}_{\delta}}.
\]
The non-trivial commutators of $\gen{h}$ 
are now written in a manifestly $\alg{sl}(2)$-invariant way
\<
\comm{\gen{R}^a{}_b}{\gen{R}^c{}_d}
\eq \delta_b^c\gen{R}^a{}_d
   -\delta^a_d\gen{R}^c{}_b,
\nln
\comm{\gen{L}^\alpha{}_\beta}{\gen{L}^\gamma{}_\delta}
\eq \delta_\beta^\gamma\gen{L}^\alpha{}_\delta
   -\delta^\alpha_\delta\gen{L}^\gamma{}_\beta,
\nln
\comm{\gen{R}^a{}_b}{\gen{J}^{c\delta\mathfrak{e}}}
\eq \delta_b^c\gen{J}^{a\delta\mathfrak{e}}
   -\half\delta_b^a\gen{J}^{c\delta\mathfrak{e}},
\nln
\comm{\gen{L}^\alpha{}_\beta}{\gen{J}^{c\delta\mathfrak{e}}}
\eq \delta_\beta^\delta\gen{J}^{c\alpha\mathfrak{e}}
   -\half\delta_\beta^\alpha\gen{J}^{c\delta\mathfrak{e}},
\nln
\acomm{\gen{J}^{a\beta\mathfrak{c}}}{\gen{J}^{d\epsilon\mathfrak{f}}}
\eq 
\varepsilon^{ad}\varepsilon^{\epsilon\kappa}\varepsilon^{\mathfrak{cf}}
\gen{L}^\beta{}_\kappa
+\varepsilon^{ak}\varepsilon^{\epsilon\beta}\varepsilon^{\mathfrak{cf}}
\gen{R}^d{}_k
+\varepsilon^{ad}\varepsilon^{\epsilon\beta}\varepsilon^{\mathfrak{fk}}
\gen{C}^{\mathfrak{c}}{}_{\mathfrak{k}}.
\>
We can introduce the $\alg{sl}(2)$ generators $\gen{B}^{\mathfrak{a}}{}_{\mathfrak{b}}$,
and their non-trivial commutators are consequently
\<
\comm{\gen{B}^{\mathfrak{a}}{}_{\mathfrak{b}}}{\gen{B}^{\mathfrak{c}}{}_{\mathfrak{d}}}
\eq \delta_{\mathfrak{b}}^{\mathfrak{c}}\gen{B}^{\mathfrak{a}}{}_{\mathfrak{d}}
   -\delta^{\mathfrak{a}}_{\mathfrak{d}}\gen{B}^{\mathfrak{c}}{}_{\mathfrak{b}},
\nln
\comm{\gen{B}^{\mathfrak{a}}{}_{\mathfrak{b}}}{\gen{J}^{c\delta\mathfrak{e}}}
\eq \delta_{\mathfrak{b}}^{\mathfrak{e}}\gen{J}^{c\delta\mathfrak{a}}
   -\half\delta_{\mathfrak{b}}^{\mathfrak{a}}\gen{J}^{c\delta\mathfrak{e}},
\nln
\comm{\gen{B}^{\mathfrak{a}}{}_{\mathfrak{b}}}{\gen{C}^{\mathfrak{c}}{}_{\mathfrak{d}}}
\eq \delta_{\mathfrak{b}}^{\mathfrak{c}}\gen{C}^{\mathfrak{a}}{}_{\mathfrak{d}}
   -\delta^{\mathfrak{a}}_{\mathfrak{d}}\gen{C}^{\mathfrak{c}}{}_{\mathfrak{b}}.
\>
The enlarged algebra shall be called 
$\gen{h}\indup{out}=\alg{sl}(2)\ltimes \alg{h}$.
Note that we will mostly not consider the enlarged algebra
$\gen{h}\indup{out}$ because its representations
are substantially different from those of
$\gen{h}$ which are of interest to us. 
Nevertheless, we might keep the Cartan generator of $\alg{sl}(2)$; let
us denote it by $\gen{B}$ and the
extended algebra by $\alg{h}_+=\Reals\ltimes \alg{h}$. 
This generator is the same as the abelian automorphism of the algebra 
$\alg{pu}(2|2)=\Reals\times\alg{psu}(2|2)$.
One can embed it into the $\alg{sl}(2)$ matrix as 
\[
\gen{B}^{\mathfrak{a}}{}_{\mathfrak{b}}=
\matr{ll}{-\gen{B}&+\gen{B}^+\\-\gen{B}^-&+\gen{B}}.
\]
%


The appearance of the automorphism can also
be understood in terms of the contraction
of the exceptional superalgebra $\alg{d}(2,1;\varepsilon,\Reals)$ 
with $\varepsilon\to 0$ presented in \cite{Beisert:2005tm}.
When setting $\varepsilon=0$ without rescaling
some of the generators, however, the algebra would be
$\alg{d}(2,1;0,\Reals)=\alg{sl}(2)\ltimes\alg{psu}(2|2)$.
Therefore, the generators $\gen{B}^{\mathfrak{a}}{}_{\mathfrak{b}}$ and 
$\gen{C}^{\mathfrak{a}}{}_{\mathfrak{b}}$ 
originate from the same $\alg{sl}(2)$ factor in  
$\alg{d}(2,1;\varepsilon,\Reals)$ by
taking two different limits $\varepsilon\to 0$.
Curiously, both triplets of generators can coexist 
in $\alg{h}\indup{out}$. 

Note that the bi-linear combination 
\[
\label{eq:bili}
\gen{\vec C}^2
=
\half 
\gen{C}^{\mathfrak{a}}{}_{\mathfrak{b}}
\gen{C}^{\mathfrak{b}}{}_{\mathfrak{a}}
=\gen{C}^2-\gen{P}\gen{K}
\]
is invariant under $\alg{sl}(2)$ and 
therefore under the complete algebra $\alg{h}\indup{out}$.

\subsection{Representations}

Here we consider representations of $\alg{h}$ for which
all the central charges $\gen{C},\gen{P},\gen{K}$ 
have well-defined numerical eigenvalues $C,P,K$.
We furthermore demand that the $\alg{sl}(2)$-invariant combination
$C^2-PK$, cf. \eqref{eq:bili}, is positive.

In order to understand these representations 
we can make use of the
outer automorphism to 
relate them to representations of $\alg{su}(2|2)$, which have been studied in
\cite{Kamupingene:1989wj,Palev:1990wm,Zhang:2004qx,Gotz:2005ka,Gotz:2006qp}.
Under the automorphism, the charge eigenvalues $(C,P,K)$
transform as a space-like vector of $\alg{so}(2,1)=\alg{sl}(2)$.
We can thus transform $(C,P,K)$ to $(\pm\sqrt{C^2-PK},0,0)$.
The representation becomes a representation of $\alg{su}(2|2)$
with central charge $\pm\sqrt{C^2-PK}$.
The representations of $\alg{h}$ 
with well-defined eigenvalues of the central charges
are nothing but representations of $\alg{su}(2|2)$ 
modulo a $\alg{sl}(2)$-rotation of the algebra.

Let us now discuss two relevant types of finite-dimensional 
irreducible representations of $\alg{h}$. We shall call them long (typical) 
and short (atypical).

\begin{figure}\centering
\parbox{4cm}{\centering\includegraphics[scale=1]{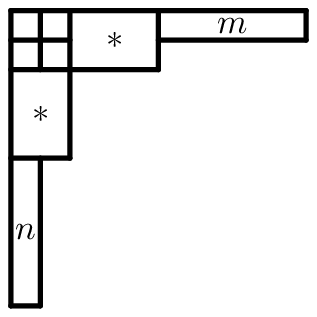}}
\parbox{3cm}{\centering\includegraphics[scale=1]{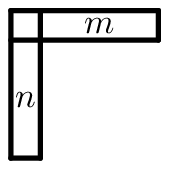}}
\caption{Young tableaux for long and short representations $\lrep{m,n}$ and $\srep{m,n}$.
The label $\ast$ represents a stack of arbitrarily many boxes. 
A single box represents the fundamental representation $\srep{0,0}$.}
\label{fig:tableaux}
\end{figure}

\paragraph{Long Multiplets.}

A long multiplet of $\alg{h}$ will be denoted by  
\[\{m,n;C,P,K\}=\{m,n;\vec C\}.\] 
Here $\vec C=(C,P,K)$ are the eigenvalues
of the central charges. 
The non-negative integers $m,n$ are Dynkin labels specifying 
multiplets of $\alg{su}(2)\times\alg{su}(2)$.
The overall dimension of this multiplet is $16(m+1)(n+1)$
distributed evenly among the gradings.

The corresponding representations 
of $\alg{su}(2|2)$ are 
specified by the Dynkin labels
${[m;r;n]}$
with $r=\pm\sqrt{C^2-PK}-\half n+\half m$.
Note that the middle Dynkin label $r$ 
is related to the $\alg{su}(2|2)$ central charge by
$C=\half n-\half m+r$. 
In fact, one class of representations with fixed $m,n,|\vec C|$ 
of $\alg{h}$ generally interpolates between two
representations of $\alg{su}(2|2)$ with $C=\pm|C|$
(unless $C=0$).
The Young tableaux for long representations 
are depicted in \figref{fig:tableaux},
see \cite{BahaBalantekin:1980qy} for the use of Young tableaux
in superalgebras.

In terms of the bosonic subalgebra $\alg{su}(2)\times\alg{su}(2)$
the multiplet decomposes into the components
\[
\left\{
\begin{array}{cccc}
[m+0,n+0]&[m+0,n+0]&[m+0,n+0]&[m+0,n+0]\\{}
[m+2,n+0]&[m+0,n+2]&[m-2,n+0]&[m+0,n-2]\\\hline
[m+1,n+1]&[m+1,n+1]&[m-1,n+1]&[m-1,n+1]\\{}
[m+1,n-1]&[m+1,n-1]&[m-1,n-1]&[m-1,n-1]
\end{array}
\right\}.
\]
Here $[m,n]$ specifies the $\alg{su}(2)\times\alg{su}(2)$ Dynkin labels, 
i.e.~$m,n$ are \emph{twice} the spins of the multiplets. 
The bar separates components with different grading. 

For small values of $m,n$ special care has to be taken 
in the decomposition. We should then treat all
components of the form $[-1,n]$ or $[m,-1]$ as absent. 
The components $[-2,n]$ and $[m,-2]$ should be treated
as $[0,n]$ and $[m,0]$, respectively, but with
multiplicity $-1$ (they will 
always cancel against some other component).

\paragraph{Short Multiplets.}

A short multiplet will be denoted by 
\[\label{eq:shortcond}
\srep{m,n;C,P,K}=\srep{m,n;\vec C}
\quad\mbox{with}\quad 
\vec{C}^2=C^2-PK=\quarter(n+m+1)^2.
\] 
Again $m,n$ are non-negative integers representing 
Dynkin labels of $\alg{su}(2)\times\alg{su}(2)$.
The overall dimension of this multiplet is $4(m+1)(n+1)+4mn$
distributed evenly among the gradings.

The Dynkin labels of the corresponding $\alg{su}(2|2)$ representations 
are $[m-1;m;n]$ (for $m=0$ we should pick $[0;0;n+1]$ instead)
or by $[m;-n;n-1]$ (for $n=0$ we should pick $[m+1;0;0]$ instead).
These representations are atypical.
The Young tableaux for short representations are presented in 
\figref{fig:tableaux}.

The $\alg{su}(2)\times\alg{su}(2)$ components of a short multiplet are 
\[
\left\{\begin{array}{cccc}
[m-1,n+0]&[m-1,n+0]&[m+1,n+0]&[m-1,n-2]\\\hline
[m+0,n-1]&[m+0,n-1]&[m+0,n+1]&[m-2,n-1]
\end{array}\right\}.
\]
Interesting special cases are the two series of multiplets 
\<\label{eq:symrep}
\srep{m,0;\vec C}\earel{\to}
\left\{\begin{array}{cc|c}
[m-1,0]&[m+1,0]&[m+0,1]
\end{array}\right\},
\nln
\srep{0,m;\vec C}\earel{\to}
\left\{\begin{array}{cc|c}
[0,m-1]&[0,m+1]&[1,m+0]
\end{array}\right\}.
\>
In particular, the single multiplet $\srep{0,0;\vec C}$
being part of both series
deserves further consideration:
It is the smallest non-trivial multiplet, it has
two bosonic and two fermionic components
\[
\left\{\begin{array}{c|c}
[1,0]&[0,1]
\end{array}\right\}.
\]
It can thus be viewed as the fundamental
multiplet of $\gen{h}$ in analogy to the one of $\alg{su}(2|2)$. 
It shall be denoted as
\[
\srep{C,P,K}:=
\srep{0,0;C,P,K}\quad\mbox{or}\quad
\srep{\vec C}:=\srep{0,0;\vec C},\]
and will be discussed in detail in \secref{sec:funda}.
The former multiplets \eqref{eq:symrep}
can be considered totally (anti)symmetric 
products of the fundamental multiplet.

\paragraph{Anomalous Multiplets.}

There are further types of finite-dimensional multiplets. 
These include at least the trivial multiplet $\srep{\cdot}$ 
and several types of adjoint multiplets. 
Both the singlet and the adjoint multiplets have 
all central charges equal zero, $C=P=K=0$.

The minimal adjoint multiplet $\srep{\mathrm{adj}_{\alg{psu}(2|2)}}$ has the components
\[
\left\{\begin{array}{cc|cc}
[2,0]&[0,2]&[1,1]&[1,1]
\end{array}\right\},
\]
corresponding to the algebra $\alg{psu}(2|2)$.
The bigger adjoints may have several of the additional components
of the extended algebra $\gen{h}\indup{out}$.
Note that the components corresponding to central charges may 
form submultiplets of $\gen{h}$.

\paragraph{Multiplet Splitting.}

\begin{figure}\centering
$\parbox{2cm}{\centering\includegraphics[scale=1]{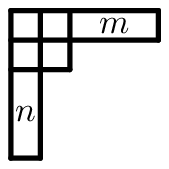}}
=
\parbox{2cm}{\centering\includegraphics[scale=1]{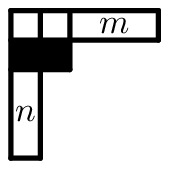}}
\oplus\parbox{2cm}{\centering\includegraphics[scale=1]{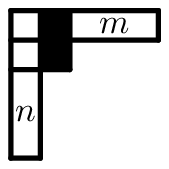}}
$
\caption{Diagrammatic representation of multiplet splitting \protect\eqref{eq:Splitting}
using Young tableaux. The filled boxes on the right hand side are to be removed
leaving short representations.}
\label{fig:splitting}
\end{figure}

A long multiplet $\{m,n;\vec C\}$
satisfying the condition $\vec{C}^2=C^2-PK=\quarter(m+n+2)^2$
is reducible. 
It splits into two short multiplets
as follows
\[\label{eq:Splitting}
\{m,n;\vec C\}=\srep{m+1,n;\vec C}\oplus\srep{m,n+1;\vec C}'.\]
The prime at the second multiplet indicates
that the grading of all components has been flipped.
Diagrammatically multiplet splitting can be understood 
as shown in \figref{fig:splitting}.

An anomalous decomposition involving the adjoint multiplet is the following
\[\label{eq:AdjSplitting}
\{0,0;\vec 0\}=\srep{\cdot}\oplus\srep{\mathrm{adj}_{\alg{psu}(2|2)}}\oplus\srep{\cdot}
=\srep{\mathrm{adj}_{\alg{u}(2|2)}}.
\]
Note that this includes a singlet $\srep{\cdot}$ as well as
the adjoint $\srep{\mathrm{adj}_{\alg{su}(2|2)}}$ of $\alg{su}(2|2)$
as closed submultiplets.

\paragraph{Tensor Products.}

For Lie algebras one is used to the fact that a product of 
irreducible representations yields a non-trivial sum of irreducible representations.
Furthermore, for superalgebras one is used to the fact that
the tensor product of atypical representations yields
a sum of atypical and typical representations. 
In the algebra $\alg{h}$ we find 
remarkable exceptions to these rules.

Firstly, a tensor product of two short representations
will generically yield \emph{no short} representations, 
but \emph{only long} ones. 
This is easily understood because the central charge
eigenvalues will add up in tensor products.
The characteristic quantity \eqref{eq:shortcond} for the determination
of short representations is however a quadratic form in the
charge eigenvalues. 
For example, let the two representations have central charges
$\vec{C}=(C,P,K)$ and $\vec{C}'=(C',P',K')$. 
Then the tensor product has central charges 
$\vec{C}+\vec{C}'$ whose quadratic form is
\[
\bigbrk{\vec{C}+\vec{C}'}^2=\vec{C}^2+\vec{C}^{\prime\,2}+2\vec{C}\cdot\vec{C}'
=\quarter(m+n+1)^2+\quarter(m'+n'+1)^2+2\vec{C}\cdot\vec{C}'.
\]
Generically, this will not be the square of a half-integer number
because $\vec{C}\cdot\vec{C}'$ is continuous.
Therefore, none of the irreducible representations in the tensor
product satisfy the shortening condition (in generic cases),
and they all have to be long. 

This statement has very remarkable consequences
for the fundamental multiplet $\srep{\vec C}$.
Its dimension is $4$ whereas the minimal dimension
of a long multiplet is $16$. For the tensor product it
follows immediately that
\[\label{eq:TensorSquare}
\srep{\vec C}
\otimes
\srep{\vec C'}
=
\{0,0;\vec C+\vec C'\}.
\]
In other words, we have found a rather unique
example of two irreducible representations
whose tensor product is again \emph{irreducible}!
This feature is reminiscent of tensor products in 
quantum algebras%
\footnote{I would like to thank a referee of
this manuscript for pointing this out in the report.}
which are used to describe integrable systems.
In fact, the structure of the algebra $\gen{h}_+$,
which has an outer automorphism and a central charge, 
is similar to that of affine Kac-Moody algebras. 
A key difference to affine Kac-Moody and quantum algebras
is that our algebra $\gen{h}_+$ is finite-dimensional.%
\footnote{It would be desirable to investigate
further this connection and a quantum version of $\gen{h}_+$,
see also the end of \protect\secref{sec:notations}.}

A generalisation of this formula is
\[
\srep{m,n;\vec{C}}\otimes\srep{\vec{C}'}
=\lrep{m,n;\vec C+\vec C'}\oplus\lrep{m-1,n-1;\vec C+\vec C'},
\]
where for $m=0$ or $n=0$ the second long multiplet has a label $-1$ and should be dropped.
It can be used to derive 
the product of three fundamentals
\[\label{eq:TensorCube}
\srep{\vec C_1}
\otimes
\srep{\vec C_2}
\otimes
\srep{\vec C_3}
=\{1,0;\vec C_1+\vec C_2+\vec C_3\}\oplus\{0,1;\vec C_1+\vec C_2+\vec C_3\}
\]
which plays an important role for the Yang-Baxter equation.

Another useful generalisation is the tensor product
\[
\srep{m,0;\vec{C}}\otimes\srep{n,0;\vec{C}'}
=\bigoplus_{k=0}^{\min(m,n)}\lrep{m+n-2k,0;\vec C+\vec C'}
\]
which has applications to the scattering of bound states
and which is strikingly similar to the tensor product 
of $\alg{su}(2)$ representations.

\subsection{Fundamental Representation}
\label{sec:funda}

We label the $2|2$ states of the fundamental multiplet 
$\srep{C,P,K}$ by $\state{\phi^a}$ and $\state{\psi^\alpha}$.
Each $\alg{su}(2)$ factor should act canonically on either of the 
two-dimensional subspaces%
\<\label{eq:Rep.ActionRL}
\gen{R}^a{}_b\state{\phi^c}\eq\delta^c_b\state{\phi^a}
  -\half \delta^a_b\state{\phi^c},
\nln
\gen{L}^\alpha{}_\beta\state{\psi^\gamma}\eq\delta^\gamma_\beta\state{\psi^\alpha}
  -\half \delta^\alpha_\beta\state{\psi^\gamma}.
\>
The supersymmetry generators should also act in a manifestly
$\alg{su}(2)\times\alg{su}(2)$ covariant way.
The most general transformation rules are thus
\<\label{eq:Rep.ActionQS}
\gen{Q}^\alpha{}_a\state{\phi^b}\eq a\,\delta^b_a\state{\psi^\alpha},\nln
\gen{Q}^\alpha{}_a\state{\psi^\beta}\eq b\,\varepsilon^{\alpha\beta}\varepsilon_{ab}\state{\phi^b},\nln
\gen{S}^a{}_\alpha\state{\phi^b}\eq c\,\varepsilon^{ab}\varepsilon_{\alpha\beta}\state{\psi^\beta},\nln
\gen{S}^a{}_\alpha\state{\psi^\beta}\eq d\,\delta^\beta_\alpha\state{\phi^a}.
\>
The central charge eigenvalues are then determined
by the commutation relations \eqref{eq:Alg.CommSU22}
\[\label{eq:Rep.ActionCPK}
C=\half (ad+bc),\quad P=ab,\quad K=cd.
\]
Furthermore, closure of $\gen{h}$ in \eqref{eq:Alg.CommSU22} requires
the constraint $ad-bc=1$. The latter is equivalent
to the shortening condition 
\[
\vec{C}^2=C^2-PK=\quarter.
\]
%

\paragraph{$\xpm{}$ Parameters.}

It is convenient to replace the four parameters $a,b,c,d$
by a new set of parameters
$\xp{},\xm{},g,\eta,\zeta$ as follows
\[\label{eq:abcd}
a=\sqrt{g}\,\eta,\quad
b=\sqrt{g}\,\frac{\zeta}{\eta}\lrbrk{1-\frac{\xp{}}{\xm{}}},\quad
c=\sqrt{g}\,\frac{i\eta}{\zeta \xp{}}\,,\quad
d=\sqrt{g}\,\frac{\xp{}}{i\eta}\lrbrk{1-\frac{\xm{}}{\xp{}}}.
\]
The constraint $ad-bc=1$ in the new parameters takes the form 
\[\label{eq:xpmconstr}
\xp{}+\frac{1}{\xp{}}
-\xm{}-\frac{1}{\xm{}}=\frac{i}{g}\,.
\]
Note that the parameter $g$ appears unnecessary here; 
one could pick an arbitrary value and adjust $\eta,\zeta,\xpm{}$ 
accordingly to match any $a,b,c,d$. 
However, $g$ will represent a global parameter of the model later on 
while $\eta,\zeta,\xpm{},p,u$ vary between different representations.
For fixed $g$ the constraint \eqref{eq:xpmconstr} defines a complex torus 
\cite{Janik:2006dc}. It can be solved explicitly using elliptic
functions, see \appref{sec:torus}.
We shall introduce two further parameters $p$ and $u$ related to $\xpm{}$ by 
\[
e^{ip}=\frac{\xp{}}{\xm{}}\,,
\qquad
u=
\xp{}+\frac{1}{\xp{}}-\frac{i}{2g}
=
\xm{}+\frac{1}{\xm{}}+\frac{i}{2g}\,.
\]
For later purposes, it is useful to note the relationship between 
the differentials
\[\label{eq:diffs}
d\xpm{}=\frac{du}{1-1/\xpm{}\xpm{}}\,.
\]
The new parameters determine the central charge $C$ 
which can be written in various ways using the constraint 
\eqref{eq:xpmconstr}
\[
C=-ig\xp{}+ig\xm{}-\frac{1}{2}
=\frac{1}{2}+\frac{ig}{\xp{}}-\frac{ig}{\xm{}}
=\frac{1}{2}\,\frac{1+1/\xp{}\xm{}}{1-1/\xp{}\xm{}}\,.
\]
When we consider $C$ an energy and $p$ as a momentum,
it obeys a mixture of a relativistic and a lattice dispersion relation
\[\label{eq:latticeshell}
C^2-4g^2 \sin^2(\half p)=\quarter.
\]
On the one hand, 
when setting $P=2g\sin(\half p)$, 
it becomes a standard relativistic mass shell condition. 
On the other hand, the relation is periodic under shifts 
of $p$ by $2\pi$, typical of a discrete system. 
The other two central charges are given by
\[
P=g\zeta \lrbrk{1-\frac{\xp{}}{\xm{}}}=g\zeta \bigbrk{1-e^{+ip}},\qquad
K=\frac{g}{\zeta}\lrbrk{1-\frac{\xm{}}{\xp{}}}=\frac{g}{\zeta} \bigbrk{1-e^{-ip}}.
\]
The parameter $\eta$ is an
internal parameter of the representation, it
does not influence the central charges.
It corresponds to a rescaling of the states $\state{\psi^\alpha}$
w.r.t.~$\state{\phi^a}$, i.e.~changing $\eta$ has
the same effect as rescaling $\state{\psi^\alpha}$
w.r.t.~$\state{\phi^a}$.

\paragraph{Unitarity Conditions.}

While we will mostly deal with general complex representations,
it is sometimes useful to know when the representation becomes unitary. 
The conditions for a canonically unitary representation 
can be written as
\[
|g||\eta|^2=-ig\xp{}+ig\xm{},\qquad
\frac{|\zeta|^2}{|\xm{}|^2}
=\frac{1}{\xp{}\xm{}}\,.
\]
If we assume that $g$ is positive,
it follows that $\xp{}$ and $\xm{}$ must be complex conjugates. 
Furthermore, up to a complex phase, $\eta$ and $\zeta$ are given by 
\[\label{eq:repuni}
\eta=\sqrt{-i\xp{}+i\xm{}},\qquad \zeta=1.
\]
%

\section{Scattering Matrix}
\label{sec:scat}

In the following we will discuss a factorised scattering matrix 
which acts on two or more particles transforming under $\alg{h}$. 
Its structure is based on planar $\superN=4$ SYM,
$AdS_5\times S^5$ string theory or more explicitly
on the dynamic spin chain model described in \cite{Beisert:2003ys}.
We shall start with the representation 
structure when the S-matrix acts on particles and 
chains of particles.
We will then construct the S-matrix as in \cite{Beisert:2005tm} 
and investigate it.

\subsection{Pairwise Scattering}

\begin{figure}\centering
\includegraphics[scale=0.8]{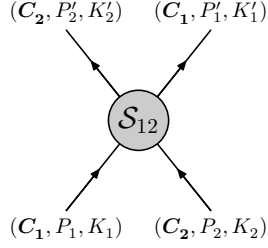}
\caption{Scattering process and transformation of central charges.}
\label{fig:scatter}
\end{figure}

The scattering matrix is an invariant operator $\smat_{12}$ acting on 
two multiplets $\srep{\vec C}$ 
(we shall refer to these as particles or sites), cf.~\figref{fig:scatter}
\[\label{eq:ScatTwoRep}
\smat_{12}:
\srep{\vec C_1}\otimes
\srep{\vec C_2}
\mapsto
\srep{\vec C'_2}\otimes
\srep{\vec C'_1}.
\]
In this section we will focus on fundamental multiplets
$\srep{\vec C}=\srep{C,P,K}$. 
Nevertheless, much of the following discussion 
will also be valid for arbitrary 
(short) multiplets.

\paragraph{Representation Structure.}

First of all, we demand that $\smat_{12}$ exchanges the central charges $C$
of the two involved particles
\[\label{eq:CPermute}
C'_1=C_1,\quad C'_2=C_2.
\]
The reason is that $C_k$ is interpreted as an energy of a particle 
and should therefore be tightly attached to it. 
The transformation of the other central charges
is then determined by the symmetry:
Firstly, the charges should add up to the same values
leading to the constraints
\[
P_1+P_2=P'_1+P'_2,\qquad
K_1+K_2=K'_1+K'_2.
\]
Secondly, the shortening conditions must remain true, i.e.
\[
P_jK_j=P'_jK'_j\quad\mbox{for }j=1,2.
\]
This system of four equations has the obvious solution
where also the other central charges are merely exchanged
\[\label{eq:RepPerm}
P'_j=P_j,\quad
K'_j=K_j.
\]
Due to the equations' quadratic nature, a second solution exists
where the other charges are transformed non-trivially
\[\label{eq:RepScatter}
P'_j=K_j\,\frac{P_1+P_2}{K_1+K_2}\,,\quad
K'_j=P_j\,\frac{K_1+K_2}{P_1+P_2}\,.
\]
%

\paragraph{Uniqueness.}

The degrees of freedom of an invariant operator acting
on a tensor product equal the number of irreducible
components in the tensor product. 
What is special about the fundamental representation
of $\alg{h}$ is that its tensor product
with another fundamental contains merely
\emph{one} irreducible component \eqref{eq:TensorSquare}
\[\label{eq:UniqueEq}
\srep{\vec C_1}\otimes
\srep{\vec C_2}=
\lrep{0,0;\vec C_1+\vec C_2}.
\]
Therefore the S-matrix is defined uniquely up to one overall factor.

The choice \eqref{eq:RepPerm} for the representations
would obviously lead to a graded permutation 
$\smat_{12}\sim\perm_{12}$ (up to an overall factor).
This yields a perfectly well-defined, albeit boring
S-matrix, so we shall discard this case.
The other choice \eqref{eq:RepScatter} leads to non-trivial
results and we will only consider this case in what follows.

\subsection{Chains and Factorised Scattering}

Now consider a sequence of $K$ fundamental multiplets
\[\label{eq:chain}
\srep{\vec C_1}\otimes
\srep{\vec C_2}\otimes\ldots\otimes
\srep{\vec C_K};
\]
this will be called a chain. 
Let $\smat_{kj}$ act on a pair of adjacent multiplets
while acting as the identity on the others.
Note that the labels of $\smat$ shall always
refer to the labels of the central charges $C_k$.
After some initial permutations these will not be
in proper order anymore, but in some
sequence $C_{\pi(k)}$ with the overall permutation specified by $\pi$. 
Acting on such a state, $\smat_{\pi(k),\pi(k+1)}$ would be a proper nearest-neighbour
S-matrix.

\paragraph{Permutation Group.}

\begin{figure}\centering
\parbox[c]{4cm}{\centering\includegraphics[scale=0.6]{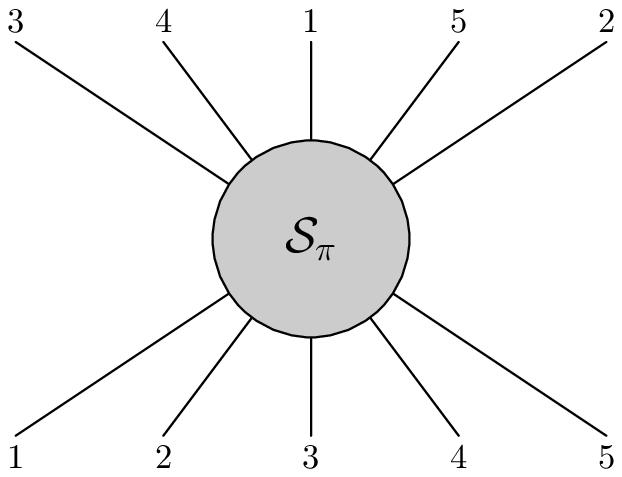}}
$=$
\parbox[c]{4cm}{\centering\includegraphics[scale=0.6]{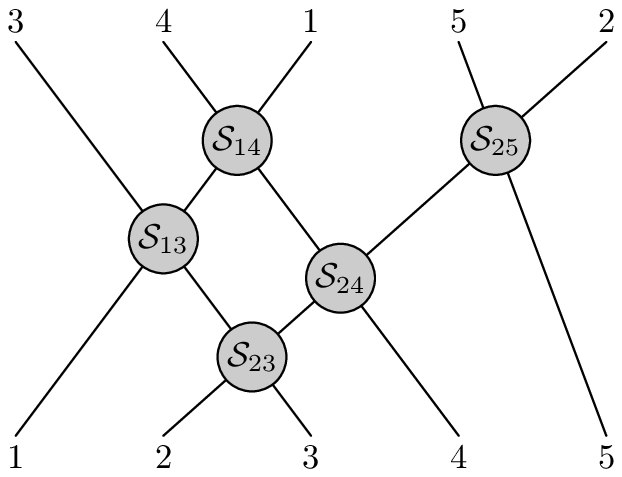}}
\caption{Factorised scattering.}
\label{fig:factorised}
\end{figure}

\begin{figure}\centering
\parbox[c]{1.5cm}{\centering\includegraphics[scale=0.6]{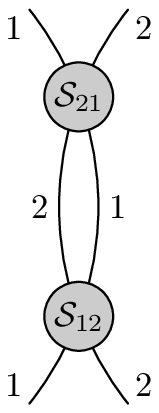}}
$=$
\parbox[c]{1.5cm}{\centering\includegraphics[scale=0.6]{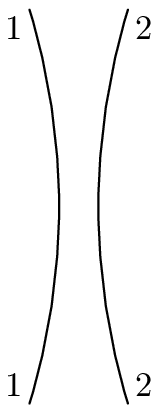}}
\qquad\qquad
\parbox[c]{3cm}{\centering\includegraphics[scale=0.6]{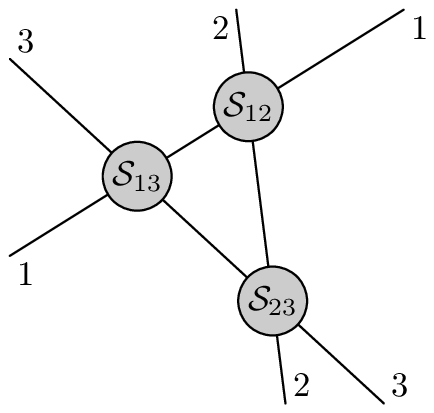}}
$=$
\parbox[c]{3cm}{\centering\includegraphics[scale=0.6]{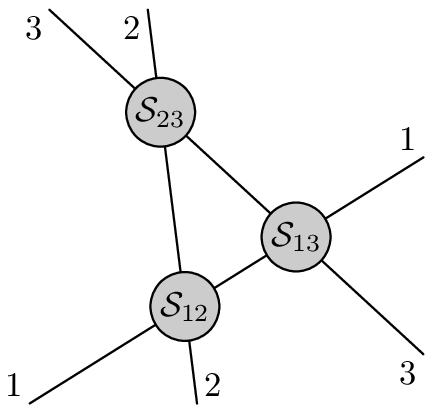}}
\caption{Unitarity condition and Yang-Baxter equation.}
\label{fig:ybe}
\end{figure}

A basic requirement for a consistent definition of
a factorised $K$-particle S-matrix $\smat_\pi$ 
for any permutation $\pi\in S_K$ is that it forms a
representation of the permutation group $S_K$,
i.e.~$\smat_{\pi'}\smat_{\pi}=\smat_{\pi'\pi}$.
As $S_K$ is generated by permutations of nearest neighbours,
we can always write $\smat_{\pi}$ as a product of 
nearest-neighbour S-matrices $\smat_{kj}$, 
i.e.~the S-matrix factorises.
See the example in \figref{fig:factorised}.
The necessary and sufficient conditions for 
the nearest-neighbour S-matrix $\smat_{kj}$
to generate a representation of $S_K$ 
are (see \figref{fig:ybe}) the unitarity condition
\[\label{eq:UniRep}
\smat_{21}\smat_{12}= \ident
\]
and the Yang-Baxter equation
\[\label{eq:YBErep}
\smat_{12}\smat_{13}\smat_{23}=
\smat_{23}\smat_{13}\smat_{12}.
\]
Together they allow to bring any product of $\smat_{kj}$ to
some standard form $\smat_\pi$ depending only
on the permutation $\pi\in S_K$.

\paragraph{Labels.}

Let us now focus on the representation labels: 
According to the definition \eqref{eq:ScatTwoRep},
the action of $\smat_\pi$ results in 
a new sequence of labels 
\[\label{eq:SGroupRep0}
\smat_{\pi}:
\srep{\vec C_1}\otimes\ldots\otimes\srep{\vec C_K}
\mapsto
\srep{\vec C^{\pi}_1}\otimes\ldots\otimes\srep{\vec C^{\pi}_K}
\]
with $C^\pi_k=C_{\pi(k)}$ due to \eqref{eq:CPermute}, and $P^\pi_k,K^\pi_k$ are 
defined through repeated application of \eqref{eq:RepScatter}.
Furthermore, the group multiplication rule $\smat_{\pi'}\smat_{\pi}=\smat_{\pi'\pi}$
requires
\[\label{eq:SGroupRep}
\smat_{\pi'}:
\srep{\vec C^\pi_1}\otimes\ldots\otimes\srep{\vec C^\pi_K}
\mapsto
\srep{\vec C^{\pi'\pi}_1}\otimes\ldots\otimes\srep{\vec C^{\pi'\pi}_K}.
\]
It is straightforward to verify that 
\eqref{eq:RepScatter} satisfies the relations
\eqref{eq:UniRep,eq:YBErep} in terms of the representation labels
and thus \eqref{eq:SGroupRep} holds.

\paragraph{Components.}

Let us now consider the implications of \eqref{eq:UniRep,eq:YBErep}
for the action of the S-matrix in components.
For that purpose we shall denote the state with the highest weight 
w.r.t.~one $\alg{su}(2)$ of the fundamental multiplet
as $\state{\phi}$.
The highest-weight state w.r.t.~the other $\alg{su}(2)$ 
shall be denoted as $\state{\psi}$.
Let us act with the S-matrix on a pair of such states.
Due to their highest-weight nature it is clear that
the resulting states will be of the same form,
\[
\smat_{12}\state{\phi_1\phi_2}
=A_{12}\state{\phi_2\phi_1},\qquad
\smat_{12}\state{\psi_1\psi_2}
=D_{12}\state{\psi_2\psi_1},
\]
where we defined the corresponding matrix elements as $A_{12}$ and $D_{12}$.
The coefficients $A_{12}$ and $D_{12}$ will be functions
of the central charges $\vec C_k$.
The above claim of uniqueness is that fixing $A_{12}$ also determines $D_{12}$ 
uniquely.

\paragraph{Unitarity.}

Let us consider \eqref{eq:UniRep} first. 
As mentioned below \eqref{eq:SGroupRep}
the unitarity condition holds in terms of representations, i.e.
\[\label{eq:SUniAct}
\smat_{21}\smat_{12}:
\srep{\vec C_1}\otimes
\srep{\vec C_2}\mapsto
\srep{\vec C_1}\otimes
\srep{\vec C_2}.
\]
The combination $\smat_{21}\smat_{12}$ is an $\alg{h}$-invariant 
operator and for the same reasons as for $\smat_{12}$ given around \eqref{eq:UniqueEq}, 
it is uniquely defined up to an overall factor.
As the identity acts like \eqref{eq:SUniAct}, the two maps
must be proportional 
\[
\smat_{21}\smat_{12}\sim\ident.
\]

To find the factor of proportionality, we shall act on the state $\state{\phi_1\phi_2}$
and obtain
\[
\smat_{21}\smat_{12}\state{\phi_1\phi_2}
=
A_{12}A_{21}\state{\phi_1\phi_2}.
\]
It implies that the relation 
\[\label{eq:UniA}
A_{12}A_{21}=1
\]
is sufficient to ensure by $\alg{h}$-symmetry
that the unitarity condition \eqref{eq:UniRep} holds
for the entire S-matrix.
In particular this fact together with
$\smat_{21}\smat_{12}\state{\psi_1\psi_2}=D_{12}D_{21}\state{\psi_1\psi_2}$
implies that the relation 
\[\label{eq:UniD}
D_{12}D_{21}=1
\]
is equivalent to \eqref{eq:UniA} by $\alg{h}$-symmetry.

\paragraph{Yang-Baxter Equation.}

Let us now consider the Yang-Baxter equation \eqref{eq:YBErep}. 
By moving all the terms to one side, we can write it
as
\[\label{eq:YBEoneside}
\smat_{32}\smat_{31}\smat_{21}\smat_{23}\smat_{13}\smat_{12}=\ident.
\]
Reminding ourselves of the tensor product \eqref{eq:TensorCube}
\[
\srep{\vec C_1}
\otimes
\srep{\vec C_2}
\otimes
\srep{\vec C_3}
=\lrep{1,0;\vec C_1+\vec C_2+\vec C_3}\oplus\lrep{0,1;\vec C_1+\vec C_2+\vec C_3}
\]
we find that it is sufficient to prove \eqref{eq:YBEoneside}
for one state in each of the two irreducible components.
The simplest such states are given by
$\state{\phi_1\phi_2\phi_3}$ and
$\state{\psi_1\psi_2\psi_3}$. 
For these it is straightforward
to evaluate \eqref{eq:YBErep}:
\[\label{eq:YBEscalar}
 A_{\underline{23}1}A_{2\underline{13}}A_{\underline{12}3}
=A_{3\underline{12}}A_{\underline{13}2}A_{1\underline{23}},\qquad
 D_{\underline{23}1}D_{2\underline{13}}D_{\underline{12}3}
=D_{3\underline{12}}D_{\underline{13}2}D_{1\underline{23}}.
\]
Here $A_{2\underline{13}}$ represents the element $A_{13}$ 
when the particles have been permuted to the sequence $213$ 
in the initial state before applying $\smat_{13}$. 
Note that it is important to keep track of all the labels 
because the S-matrix is based on the relation \eqref{eq:RepScatter}
which changes the central charges $P,K$ non-trivially.

We can ensure that one of the equations holds by
choosing a suitable function $A$ or $D$. 
However, the ratio of the equations
depends on $A/D$ only which is determined
by $\alg{h}$-symmetry as explained below \eqref{eq:UniD}.
We thus have to ensure
\[\label{eq:YBEscalar2}
 A_{\underline{23}1}A_{2\underline{13}}A_{\underline{12}3}
=A_{3\underline{12}}A_{\underline{13}2}A_{1\underline{23}},\qquad
\frac{A_{\underline{23}1}}{D_{\underline{23}1}}\,
\frac{A_{2\underline{13}}}{D_{2\underline{13}}}\,
\frac{A_{\underline{12}3}}{D_{\underline{12}3}}
=
\frac{A_{3\underline{12}}}{D_{3\underline{12}}}\,
\frac{A_{\underline{13}2}}{D_{\underline{13}2}}\,
\frac{A_{1\underline{23}}}{D_{1\underline{23}}}\,.
\]
To prove them, it would suffice to ensure
\[\label{eq:ADshift}
A_{\underline{12}3}=A_{3\underline{12}}
\quad\mbox{and}\quad
\frac{A_{\underline{12}3}}{D_{\underline{12}3}}=\frac{A_{3\underline{12}}}{D_{3\underline{12}}}\,,
\]
i.e.~that the scattering of two highest-weight states of 
one $\alg{su}(2)$ is independent of their position within the chain.
Note that the representation labels do depend on the position, 
see \eqref{eq:SGroupRep,eq:RepScatter}, and therefore \eqref{eq:ADshift} is far from evident.

\subsection{Constrained Labels}
\label{sec:labconstr}

So far the representations labels $P_j,K_j$ in
the natural ordering of particles
have been considered to be completely general. 
However, we will have to impose certain relations among the labels 
in order to satisfy the Yang-Baxter equation or \eqref{eq:ADshift}. 
To see this would require a substantial amount of work involving the explicit
expression for the S-matrix. 
Here we will present a plausible constraint
related to the S-matrix bootstrap, cf.~\cite{Zamolodchikov:1978xm}. 
This constraint leads to strong simplifications
and it will turn out to solve the YBE. 
We can however not determine rigorously whether the derived constraint 
is minimal or not.

\paragraph{Scattering with a Pair of Particles.}

Consider the permutation $\pi$ that interchanges three modules as follows
\[
\smat_\pi:
\srep{\vec C_1}
\otimes
\srep{\vec C_2}\otimes\srep{\vec C_3}
\mapsto 
\srep{\vec C'_3}\otimes\srep{\vec C'_1}
\otimes
\srep{\vec C'_2}.
\]
Applying \eqref{eq:RepScatter} two times we find
\[
P'_3=P_3\,\frac{P_1K_2+(P_1+P_2+P_3)K_3}{K_1P_2+(K_1+K_2+K_3)P_3}\,.
\]
Alternatively, in the S-matrix bootstrap we could interpret the pair
$\srep{\vec C_1}\otimes\srep{\vec C_2}$
as a composite multiplet $\lrep{0,0;\vec C_1+\vec C_2}$ 
which we scatter with $\srep{\vec C_3}$. 
While in general the long multiplet is irreducible, for particular
values of $\vec C_1+\vec C_2$ it splits into two
short multiplets, cf.~\eqref{eq:Splitting}.
In that case, the analog of formula \eqref{eq:RepScatter} must apply 
to preserve the shortness of the composite multiplet
\[
P'_3=K_3\,\frac{(P_1+P_2)+P_3}{(K_1+K_2)+K_3}\,.
\]
A natural way to ensure consistency in general
is to let the relation hold in general.
Equating the two expressions for $P'_3$,
we obtain the constraint
\[\label{eq:RepConstraint}
\frac{P_2(P_1+P_2+P_3)}{P_1P_3}=
\frac{K_2(K_1+K_2+K_3)}{K_1K_3}\,.
\]
There are further ways of expressing the consequences of this constraint:
First of all, it is certainly not a bad idea to have
\eqref{eq:RepScatter} hold also for scattering of composite objects.
Secondly, the invariant $(\vec C_1+\vec C_2)^2$ is preserved by the scattering.
Thirdly, we have $P'_1/P_1=P'_2/P_2=K_1/K'_1=K_2/K'_2$; in other words 
$P_1$ and $P_2$ are multiplied by a common factor and $K_1,K_2$ 
are multiplied by its inverse. 
And finally, it appears that the Yang-Baxter equation \eqref{eq:YBErep}
can only be satisfied if \eqref{eq:RepConstraint} holds.
To confirm this we would need the explicit form of 
the S-matrix which is derived only later in this section.
In conclusion, the above constraint ensures that the scattering of composite objects
modifies the labels of the chain in the least disruptive way.

\paragraph{Representation Structure.}

The constraint \eqref{eq:RepConstraint} 
should hold for all neighbouring triplets
which reduces the $2K$ quantities $P_k,K_k$ to
merely $K+2$ independent ones.
The unique solution to this constraint is
\<\label{eq:GoodPK0}
P_k\eq g\alpha\lrbrk{1-\exp(+ip_k)}\prod_{j=1}^{k-1}\exp(+ip_j)\,,\nln
K_k\eq \frac{g}{\alpha}\lrbrk{1-\exp(-ip_k)}\prod_{j=1}^{k-1}\exp(-ip_j)\,.
\>
Here $g$ and $\alpha$ are global constants while the momenta $p_k$ 
are defined individually for each particle.
Altogether they make up the $K+2$ independent parameters of the chain.
Using the shortening condition $C_k^2-P_kK_k$ 
we have also obtained a dispersion relation,
cf.~\eqref{eq:latticeshell}
\[\label{eq:GoodC0}
C_k=\pm\sqrt{\quarter+4g^2 \sin^2 (\half p_k)}\,,
\]
which relates the momenta $p_k$ to central charges $C_k$.
Note that the charges $P_k,K_k$ do not only depend on the momenta $p_k$ at site $k$, 
but also on all momenta $p_j$ of sites $j<k$ to the left of $k$.
Conversely, the charges $C_k$ are defined 
only through the momentum $p_k$ at the same site.

\paragraph{Scattering of Chains.}

\begin{figure}\centering
\includegraphics[scale=0.8]{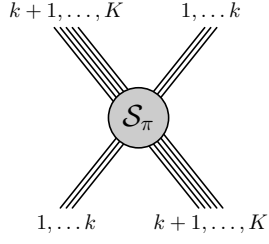}
\caption{Scattering of two chains.}
\label{fig:scatchain}
\end{figure}

Now let us consider the scattering of two composites of length $k$ and $K-k$,
see \figref{fig:scatchain} 
\[
\smat_\pi:
\srep{\vec C_1}\otimes\ldots\otimes\srep{\vec C_k}
\otimes
\srep{\vec C_{k+1}}\otimes\ldots\otimes\srep{\vec C_K}
\mapsto 
\srep{\vec C'_{k+1}}\otimes\ldots\otimes\srep{\vec C'_K}
\otimes
\srep{\vec C'_1}\otimes\ldots\otimes\srep{\vec C'_k}.
\]
In other words, all the particles labelled $j=1,\ldots,k$
are moved past the particles $j=k+1,\ldots,K$. 
Initially the charges $\vec{C}_k$ are given by
\eqref{eq:GoodPK0,eq:GoodC0}. 
The resulting central charges after the scattering process are determined
by repeated application of \eqref{eq:RepScatter} and read
\[\label{eq:chainscattercharge}
\begin{array}[b]{rclcrclcl}
P'_l\eq \displaystyle P_l \prod_{j=k+1}^K\exp(+ip_j),&\quad&
K'_l\eq \displaystyle K_l \prod_{j=k+1}^K\exp(-ip_j)&\quad&\mbox{for }l\leq k,\\[3pt]
P'_l\eq \displaystyle P_l \prod_{j=1}^k \exp(-ip_j),&&
K'_l\eq \displaystyle K_l \prod_{j=1}^k \exp(+ip_j) &\quad&\mbox{for }l>k.
\end{array}
\]
We see that the central charges are multiplied 
by the net $\exp(ip)$ of the composite they scatter with. 
As a consequence, the invariants $(\vec C_1+\ldots+\vec C_k)^2$ 
and $(\vec C_{k+1}+\ldots+\vec C_K)^2$ are not changed by the scattering.
In particular it means that whenever a shortening condition applies
to a subchain $1\ldots k$, it will also apply after scattering
with anything else. This feature is required for consistency 
with the S-matrix bootstrap and fusion of particles. 
This shows that the constraint \eqref{eq:RepConstraint} between
the central charges of three adjacent sites is sufficient
for a consistent factorised scattering. 

\subsection{Notations}
\label{sec:notations}

Due to \eqref{eq:SGroupRep} the representation labels $\vec{C}^\pi_k$
depend on the particular ordering of particles which one starts with.
Furthermore, in \secref{sec:labconstr} it became clear that 
$\vec{C}^\pi_k$ depends on the parameters of the individual particles
in a non-local way. Here we present four different notations 
to deal with this complication and compare them.

\paragraph{Non-Local Notation.}

After we apply a general permutation $\pi\in S_K$ to any chain of length
$K$, the momenta will be redefined as follows, cf.~\eqref{eq:SGroupRep0}
\<\label{eq:GoodPK}
P^\pi_k\eq g\alpha\lrbrk{1-\exp(+ip_{\pi(k)})}\prod_{j=1}^{k-1}\exp(+ip_{\pi(j)})\,,\nln
K^\pi_k\eq \frac{g}{\alpha}\lrbrk{1-\exp(-ip_{\pi(k)})}\prod_{j=1}^{k-1}\exp(-ip_{\pi(j)})\,.
\>
In other words, they still depend on the momenta $p_{\pi(k)}$ at site $\pi(k)$, 
as well as on all momenta of sites to the left.
Conversely, the resulting central charges $C^\pi_k$
are just permutations of the original $C_k$'s
\<
C^\pi_k=\pm\sqrt{\quarter+P^\pi_k K^\pi_k}
\eq\pm\sqrt{\quarter+g^2 \lrbrk{1-\exp(+ip_{\pi(k)})}\lrbrk{1-\exp(-ip_{\pi(k)})}}
\nln\eq 
\pm\sqrt{\quarter+4g^2 \sin^2 (\half p_{\pi(k)})}=C_{\pi(k)}\,.
\>

Dealing with quantities like \eqref{eq:GoodPK} 
which do not only depend on
the parameters of one site $k$, but also on the other sites $j$
and their particular ordering $\pi$
inevitably leads to a cluttering of notation. 
There are at least three ways to simplify the notation
which we shall now present.

\paragraph{Cumulative Notation.}

\begin{figure}\centering
\parbox[c]{6cm}{\centering\includegraphics[scale=0.6]{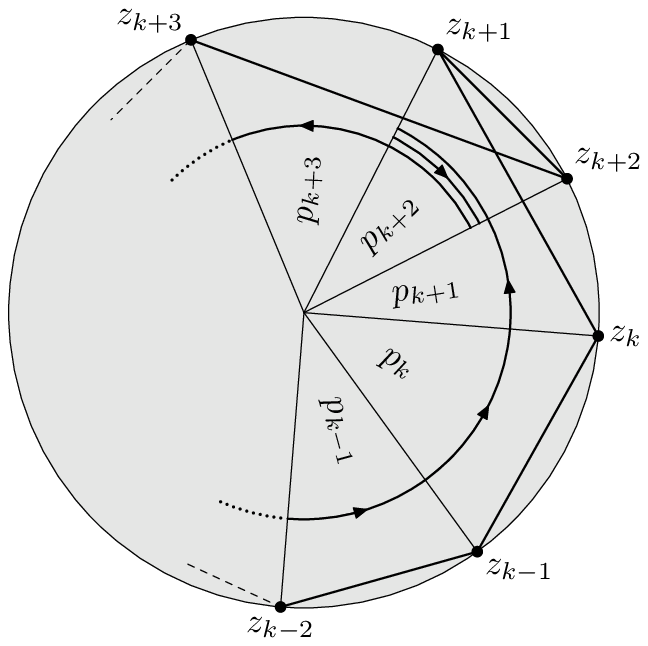}}
\qquad
\parbox[c]{3cm}{\centering\includegraphics[scale=0.8]{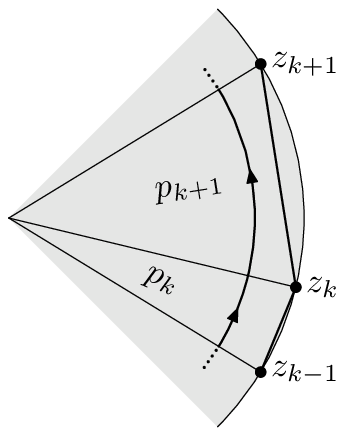}}
$\stackrel{\smat_{k,k+1}}{\longrightarrow}$
\parbox[c]{3cm}{\centering\includegraphics[scale=0.8]{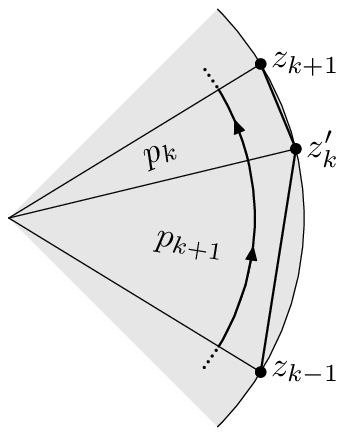}}

\caption{Cumulative picture of a chain and 
of a scattering process.}
\label{fig:cumulative}
\end{figure}

For the first notation based on \cite{Hofman:2006xt}
we introduce cumulative momenta $\varphi^\pi_k$ 
and their exponentials $z^\pi_k$ via
\[\label{eq:cumuphiz}
\varphi^\pi_k=-i\log\alpha+\sum_{j=1}^k p_{\pi(j)},
\qquad
z^\pi_k=\exp(i\varphi^\pi_k)=\alpha\prod_{j=1}^k \exp(ip_{\pi(j)}).
\]
The relation between the momenta $p$ and the $z$ parameters
is best displayed pictorially in \figref{fig:cumulative}.
These cumulative parameters allow us to write the central charges $P,K$ 
simply as differences
\[\label{eq:CumuPK}
P^\pi_k=gz^\pi_{k-1}-gz^\pi_{k},
\qquad
K^\pi_k=g/z^\pi_{k-1}-g/z^\pi_{k}.
\]
This notation is closest in nature to string theory on $AdS_5\times S^5$
using the coordinates introduced in \cite{Lin:2004nb}.
Here the parameters $\varphi_k$ represent angles on a great circle of $S^5$. 
In the figure, a string excitation with momentum $p_k$ corresponds
to the line segments joining the points $z_{k-1}$ and $z_k$ \cite{Hofman:2006xt}. 
Conversely, the latter determine the position of the string between two excitations. 
Scattering of two particles $k,k+1$ is illustrated on the r.h.s.~of \figref{fig:cumulative}.
It interchanges two neighbouring line segments and consequently modifies
only the intermediate point $z_k\mapsto z'_k$.
The overall central charges $P,K$ for a chain of
particles are immediately determined by \eqref{eq:CumuPK} 
as $P=gz^\pi_0-gz^\pi_K$ and $K=g/z^\pi_0-g/z^\pi_K$.
As the endpoints of a chain remain fixed, $P,K$ are obviously preserved
in any scattering process. Moreover, a closed chain with
$z^\pi_K=z^\pi_0$ has vanishing charges $P=K=0$.

All in all, this notation gets rid of the non-local contributions but still requires
to carry along the dependence on the order of particles in $\pi$.
It is therefore only partially suitable to tidy up the expressions.
Before proceeding to a different notation, let us make some comments:

It is curious to see that in this picture both sides of 
the constraint \eqref{eq:RepConstraint}
take the form of a conformal cross ratio,
cf.~\eqref{eq:CumuPK}
\[
\frac{(z_1-z_2)(z_0-z_3)}{(z_0-z_1)(z_2-z_3)}
=\frac{(1/z_1-1/z_2)(1/z_0-1/z_3)}{(1/z_0-1/z_1)(1/z_2-1/z_3)}\,.
\]
The conformal cross ratio is invariant under inversion $z_k\mapsto 1/z_k$
showing that \eqref{eq:RepConstraint} follows from
\eqref{eq:CumuPK}.
It remains a question if there is a meaning to 
conformal transformations of this $z$-plane which map the 
unit circle to itself.

Finally, this notation sheds some light on the nature 
of the constant $\alpha$. The definition \eqref{eq:cumuphiz} 
implies $z^\pi_0=\alpha$, therefore 
$\alpha$ determines the origin for a chain of particles in $z$-space.

\paragraph{Twisted Notation.}

In a different notation due to \cite{Beisert:2005tm} 
we allow for certain markers $\fldZ^n$ 
to be inserted between the particles of a chain, e.g.
\[
\state{\ldots\fldX_4\,\fldZ^{+1}\,\fldX_5\fldX_6\,\fldZ^{-2}\,\fldX_7\ldots},
\]
where $\fldX_k$ represents some state in $\srep{\vec C_k}$.
These can be shifted around the chain by picking up phases $\exp(ip_k)$ as
follows
\[
\state{\ldots\fldX_k\fldZ^n\ldots}=
\exp(ip_k)^n
\state{\ldots\fldZ^n\fldX\ldots}.
\]
They combine by adding their exponents $\fldZ^n\fldZ^m=\fldZ^{n+m}$.
In \cite{Beisert:2005tm} the markers represent insertion or deletion
of background sites $\fldZ$ within 
the coordinate Bethe ansatz for planar $\superN=4$ gauge theory.
The action of $\gen{P},\gen{K}$ on a single multiplet can then be defined as
\[\label{eq:ActPK1}
\gen{P}_k\state{\ldots\fldX_k\ldots}=P_k \state{\ldots\fldZ^+\fldX_k\ldots}\,,\qquad
\gen{K}_k\state{\ldots\fldX_k\ldots}=K_k \state{\ldots\fldZ^-\fldX_k\ldots},
\]
where $P_k,K_k$ now only depend on the momentum $p_k$ of the respective site 
\[\label{eq:GoodPK1}
P_k=g\alpha\lrbrk{1-\exp(+ip_k)},\qquad
K_k=\frac{g}{\alpha}\lrbrk{1-\exp(-ip_k)}.
\]
Within a chain, $\gen{P}$ acts as
\<\label{eq:ActPexps}
\gen{P}\state{\fldX_1\fldX_2\fldX_3\ldots}
\eq
P_1\state{\fldZ^+\fldX_1\fldX_2\fldX_3\ldots}
+P_2\state{\fldX_1\fldZ^+\fldX_2\fldX_3\ldots}
+P_3\state{\fldX_1\fldX_2\fldZ^+\fldX_3\ldots}
+\ldots
\nln\eq
\bigbrk{P_1+\exp(ip_1)P_2+\exp(ip_1+ip_2)P_3+\ldots}
\state{\fldZ^+\fldX_1\fldX_2\fldX_3\ldots},
\>
where we have shifted the marker $\fldZ^+$ 
always to the left end of the chain.
The additional factors of $\exp(ip_k)$ 
supply the terms of \eqref{eq:GoodPK0}
which are missing in \eqref{eq:GoodPK1}.
We thus obtain precisely the original picture when 
we ignore any marker that is at the very left of a chain. 
The benefit of this notation is that 
all the particles are completely independent of each other. 
The markers $\fldZ^\pm$ provide the missing non-local terms. 
By comparing \eqref{eq:GoodPK} to \eqref{eq:ActPK1,eq:GoodPK1} it should
become clear how to translate between the two:
For example, an insertion of $\fldZ^+$ leads to
multiplication with $\xp{j}/\xm{j}$ for all
particles which are to the left of particle $k$.
We will thus continue to work in the twisted notation.

\paragraph{Hopf Algebra Notation.}

Another suitable framework to deal with the above spin chain representations 
is given by Hopf algebras \cite{Gomez:2006va,Plefka:2006ze}.
This is also the standard framework for the investigation and description 
of integrable structures for spin chains. It is likely to play an 
important role for the understanding of the present integrable model,
let us therefore outline the relationship to the above notation. 

In the Hopf algebra 
one introduces a new generator $\gen{U}$ which acts by returning $e^{ip}$
\cite{Gomez:2006va,Plefka:2006ze}
\[
\gen{U}_k\state{\ldots\fldX_k\ldots}=\exp(ip_k)\state{\ldots\fldX_k\ldots}\,.
\]
To reproduce \eqref{eq:ActPexps} one defines the action 
of $\gen{P}$ on a chain of particles as
\[
\gen{P}
=\sum_{j=1}^K \gen{U}_1\ldots\gen{U}_{j-1}\gen{P}_j
=\sum_{j=1}^K \gen{U}^{\otimes j-1}\otimes 
\gen{P}\otimes
\gen{I}^{\otimes K-j-1}.
\]
Here $\gen{I}$ is the identity operator.
This representation of $\gen{P}$ on the chain is
easily obtained using the coproduct $\copro:\alg{h}\to\alg{h}\otimes\alg{h}$.
For example, it acts on $\gen{P}$ and $\gen{U}$ as
\[
\copro\gen{P}=\gen{P}\otimes\gen{I}+\gen{U}\otimes\gen{P},\qquad
\copro\gen{U}=\gen{U}\otimes\gen{U}\,.
\]
Then $\gen{P}$ acting on the chain is simply given by the multiple
coproduct $\copro^{K-1}\gen{P}$.
The eigenvalue $P_k$ in \eqref{eq:ActPK1} is 
obtained by setting $\gen{P}=g\alpha\gen{I}-g\alpha\gen{U}$. 
Consequently, the action of $\gen{P}$ on the chain 
yields $\copro^{K-1}\gen{P}=g\alpha \gen{I}^{\otimes K}-g\alpha \gen{U}^{\otimes K}$.
The construction of the Hopf algebra can be extended to all generators of 
$\alg{h}$, see \cite{Gomez:2006va,Plefka:2006ze} for details and further aspects.

It is interesting to see that for the present model 
the representation in terms of the Lie algebra $\alg{h}$ (non-local notation)
coexists to the Hopf algebra representation. This is apparently related
to the fact that the momentum parameters $p_k$ for the Hopf algebra 
are encoded into the representation labels $\vec{C}_k$ for the Lie algebra. 
For conventional integrable spin chains, the Lie algebra does not see the momentum parameters.

\subsection{Chain of Fundamentals}

We would now like to set up the action of the symmetry generators 
on a chain of fundamental multiplets. 
Here the twisted notation turns out to be very useful.

\paragraph{Twisted Notation.}

The action of the bosonic $\alg{su}(2)\times\alg{su}(2)$ generators is canonical.
The action of fermionic generators on each particle
is as in \eqref{eq:Rep.ActionQS}, but with markers $\fldY,\fldZ$ inserted
\<\label{eq:Rep.ActionQStwist}
\gen{Q}^\alpha{}_a{}_{,k}\state{\ldots\phi^b_k\ldots}\eq a_k\,\delta^b_a\state{\ldots\fldY^+\psi^\alpha_k\ldots},\nln
\gen{Q}^\alpha{}_a{}_{,k}\state{\ldots\psi^\beta_k\ldots}\eq b_k\,\varepsilon^{\alpha\beta}\varepsilon_{ab}\state{\ldots\fldZ^+\fldY^-\phi^b_k\ldots},\nln
\gen{S}^a{}_\alpha{}_{,k}\state{\ldots\phi^b_k\ldots}\eq c_k\,\varepsilon^{ab}\varepsilon_{\alpha\beta}\state{\ldots\fldZ^-\fldY^+\psi^\beta_k\ldots},\nln
\gen{S}^a{}_\alpha{}_{,k}\state{\ldots\psi^\beta_k\ldots}\eq d_k\,\delta^\beta_\alpha\state{\ldots\fldY^-\phi^a_k\ldots}.
\>
The coefficients $a_k$ are given in \eqref{eq:abcd}
\[\label{eq:abcdk}
a_k=\sqrt{g}\,\gamma_{k},\quad
b_k=\sqrt{g}\,\frac{\alpha}{\gamma_{k}}\lrbrk{1-\frac{\xp{k}}{\xm{k}}},\quad
c_k=\sqrt{g}\,\frac{i\gamma_{k}}{\alpha \xp{k}}\,,\quad
d_k=\sqrt{g}\,\frac{\xp{k}}{i\gamma_{k}}\lrbrk{1-\frac{\xm{k}}{\xp{k}}}
\]
with arbitrary $g,\alpha,\gamma_k,\xpm{k}$ subject to the constraint
\[\label{eq:xpmconstrk}
\xp{k}+\frac{1}{\xp{k}}
-\xm{k}-\frac{1}{\xm{k}}=\frac{i}{g}\,.
\]
The action of the central charges is as follows
\<
\gen{C}_{k}\state{\ldots\fldX_k\ldots}\eq C_k \state{\ldots\fldX_k\ldots},\nln
\gen{P}_{k}\state{\ldots\fldX_k\ldots}\eq P_k \state{\ldots\fldZ^+\fldX_k\ldots},\nln
\gen{K}_{k}\state{\ldots\fldX_k\ldots}\eq K_k \state{\ldots\fldZ^-\fldX_k\ldots},
\>
with 
\[
C_k=-ig\xp{k}+ig\xm{k}-\frac{1}{2}\,,\qquad
P_k=g\alpha\lrbrk{1-\frac{\xp{k}}{\xm{k}}},\qquad
K_k=\frac{g}{\alpha}\lrbrk{1-\frac{\xm{k}}{\xp{k}}}.
\]

The marker $\fldZ$ can be shifted around as explained above
\[\label{eq:MarkerZ}
\state{\ldots\fldX_k\fldZ^\pm\ldots}=
\frac{\xpm{k}}{\xmp{k}}\,
\state{\ldots\fldZ^\pm\fldX_k\ldots}
\]
and the new marker $\fldY$ behaves similarly
\[\label{eq:MarkerY}
\state{\ldots\fldX_k\fldY^\pm\ldots}=
(\xi_k)^{\pm 1}\,
\state{\ldots\fldY^\pm\fldX_k\ldots}
\]
with some new constants $\xi_k$.
In conclusion, the representation on the chain is specified by 
two global constants $g,\alpha$ as well as
the local parameters $\xpm{k},\gamma_k,\xi_k$ 
subject to \eqref{eq:xpmconstrk}.

\paragraph{Non-Local Notation.}

Let us rewrite the above representation in terms of the non-local notation
to illustrate the differences. 
For that we use the representation \eqref{eq:Rep.ActionQStwist} without insertion of the markers.
We define the constants $\zeta^\pi_k$ and $\eta^\pi_k$
\[
\zeta^\pi_k=\alpha \prod_{j=1}^{k-1}\frac{\xp{\pi(j)}}{\xm{\pi(j)}}\,,
\qquad
\eta^\pi_k= \gamma_{\pi(k)} \prod_{j=1}^{k-1}\xi_{\pi(j)}.
\]
The parameters $a,b,c,d$ in \eqref{eq:abcd} are consequently given by 
\<\label{eq:abcdnl}
a^\pi_k\eq\sqrt{g}\,\gamma_{\pi(k)} \prod_{j=1}^{k-1}\xi_{\pi(j)},\nln
b^\pi_k\eq\sqrt{g}\,\frac{\alpha}{\gamma_{\pi(k)}}\,\lrbrk{1-\frac{\xp{\pi(k)}}{\xm{\pi(k)}}}
\prod_{j=1}^{k-1}\lrbrk{\frac{\xp{\pi(j)}}{\xm{\pi(j)}}\,\frac{1}{\xi_{\pi(j)}}},\nln
c^\pi_k\eq\sqrt{g}\,\frac{i\gamma_{\pi(k)}}{\alpha \xp{\pi(k)}}
\prod_{j=1}^{k-1}\lrbrk{\frac{\xm{\pi(j)}}{\xp{\pi(j)}}\,\xi_{\pi(j)}},\nln
d^\pi_k\eq \sqrt{g}\,\frac{\xp{\pi(k)}}{i\gamma_{\pi(k)}}\,\lrbrk{1-\frac{\xm{\pi(k)}}{\xp{\pi(k)}}}
\prod_{j=1}^{k-1}\frac{1}{\xi_{\pi(j)}}\,.
\>
We see that this agrees with \eqref{eq:abcdk}
when we take the action of the markers in \eqref{eq:MarkerZ,eq:MarkerY} into account. 
This shows how much leaner than the twisted notation
is in comparison to the non-local one.

\subsection{Components of the S-Matrix}

\begin{table}
\<
\smat_{12}\state{\phi_1^a\phi_2^b}
\eq
A_{12}\state{\phi_2^{\{a}\phi_1^{b\}}}
+B_{12}\state{\phi_2^{[a}\phi_1^{b]}}
+\half C_{12}\varepsilon^{ab}\varepsilon_{\alpha\beta}\state{\fldZ^-\fldY^+\fldY^+\psi_2^\alpha\psi_1^\beta},
\nln
\smat_{12}\state{\psi_1^\alpha\psi_2^\beta}
\eq
D_{12}\state{\psi_2^{\{\alpha}\psi_1^{\beta\}}}
+E_{12}\state{\psi_2^{[\alpha}\psi_1^{\beta]}}
+\half F_{12}\varepsilon^{\alpha\beta}\varepsilon_{ab}\state{\fldZ^+\fldY^-\fldY^-\phi_2^a\phi_1^b},
\nln
\smat_{12}\state{\phi_1^a\psi_2^\beta}
\eq
G_{12}\state{\psi_2^\beta\phi_1^{a}}
+H_{12}\state{\phi_2^{a}\psi_1^\beta},
\nln
\smat_{12}\state{\psi_1^\alpha\phi_2^b}
\eq
K_{12}\state{\psi_2^\alpha\phi_1^{b}}
+L_{12}\state{\phi_2^{b}\psi_1^\alpha}\nonumber.
\>

\<
A_{12}\eq S^0_{12}\,\frac{\xp{2}-\xm{1}}{\xm{2}-\xp{1}}\,,\nln
B_{12}\eq S^0_{12}\,\frac{\xp{2}-\xm{1}}{\xm{2}-\xp{1}}\lrbrk{1
          -2\,\frac{1-1/\xm{2}\xp{1}}{1-1/\xp{2}\xp{1}}\,\frac{\xm{2}-\xm{1}}{\xp{2}-\xm{1}}},\nln
C_{12}\eq S^0_{12}\,\frac{2\gamma_1\gamma_2\xi_2}{\alpha \xp{1}\xp{2}}\,
          \frac{1}{1-1/\xp{1}\xp{2}}\,\frac{\xm{2}-\xm{1}}{\xm{2}-\xp{1}}\,,\nln
D_{12}\eq -S^0_{12}\,\frac{\xi_2}{\xi_1}\,,\nln
E_{12}\eq -S^0_{12}\,\frac{\xi_2}{\xi_1}\lrbrk{1
          -2\,\frac{1-1/\xp{2}\xm{1}}{1-1/\xm{2}\xm{1}}\,\frac{\xp{2}-\xp{1}}{\xm{2}-\xp{1}}},\nln
F_{12}\eq -S^0_{12}\,\frac{2\alpha(\xp{1}-\xm{1})(\xp{2}-\xm{2})}{\gamma_1\gamma_2\xi_1\xm{1}\xm{2}}\,
          \frac{1}{1-1/\xm{1}\xm{2}}\,\frac{\xp{2}-\xp{1}}{\xm{2}-\xp{1}}\,,\nln
G_{12}\eq S^0_{12}\,\frac{1}{\xi_1}\frac{\xp{2}-\xp{1}}{\xm{2}-\xp{1}}\,,\nln
H_{12}\eq S^0_{12}\,\frac{\gamma_1\xi_2}{\gamma_2\xi_1}\,\frac{\xp{2}-\xm{2}}{\xm{2}-\xp{1}}\,,\nln
K_{12}\eq S^0_{12}\,\frac{\gamma_2}{\gamma_1}\,\frac{\xp{1}-\xm{1}}{\xm{2}-\xp{1}}\,,\nln
L_{12}\eq S^0_{12}\,\xi_2\,\frac{\xm{2}-\xm{1}}{\xm{2}-\xp{1}}\,.\nonumber
\>
\caption{The fundamental S-matrix of $\alg{h}$.}
\label{tab:SCoeff}
\end{table}

We can now solve for the S-matrix
by demanding its invariance under $\alg{h}$
\[
\comm{\gen{J_1}+\gen{J_2}}{\smat_{12}}=0.
\]
As discussed above this will lead to a unique result
up to one overall phase factor $S^0_{12}$.
The results are summarised in \tabref{tab:SCoeff}.
The expressions are slightly more general than the ones found
in \cite{Beisert:2005tm} due to the introduction of the marker $\fldY^\pm$, but
otherwise they agree.

\paragraph{Factorised Scattering.}

We can now confirm unitarity and the Yang-Baxter equation.
First of all \eqref{eq:UniA} implies a constraint for the overall phase factor
\[
S^0_{12}S^0_{21}=1.
\]
Furthermore, we see that 
\[
A_{12}
=S^0_{12}\,\frac{\xp{2}-\xm{1}}{\xm{2}-\xp{1}}\,,\qquad
\frac{A_{12}}{D_{12}}
=-\frac{\xi_1}{\xi_2}\,\frac{\xp{2}-\xm{1}}{\xm{2}-\xp{1}}
\]
depend only on the parameters of sites $1,2$. Therefore
the YBE is proved via \eqref{eq:ADshift}.

Note that for two sites the choice made in 
\secref{sec:labconstr} is not a restriction
but fully general. 
There are four independent new parameters $\alpha,g,\xp{1},\xp{2}$
to replace the original independent charges $P_1,P_2,K_1,K_2$.
Thus we could use the expression
in \tabref{tab:SCoeff} to investigate the YBE in the
more general case. 
Nevertheless we should emphasise that all of $\alpha,g,\xpm{1},\xpm{2}$
will become dependent not only on the site, 
but also on the permutation of sites.
This will make the investigation of the YBE, even in the simplified
form \eqref{eq:YBEscalar2}, quite hard.

\paragraph{Non-Local Notation.}

Curiously, the markers $\fldZ^\pm$ and $\fldY^\pm$ appear
only for the contributions with coefficients $C_{12}$ and $F_{12}$
and even in the same combination.
Here it is worth emphasising that in the non-local notation
the insertions of $\fldZ^-\fldY^+\fldY^+$ would lead to an additional
factor of 
\[\label{eq:STwistFactor}
\prod_{j=1}^{k-1}\lrbrk{\frac{\xm{\pi(j)}}{\xp{\pi(j)}}\,(\xi_{\pi(j)})^2}
\]
in $C_{12}$ when the S-matrix is applied to position $k,k+1$ of
the chain. Likewise $\fldZ^+\fldY^-\fldY^-$ leads to the inverse of 
\eqref{eq:STwistFactor} as an additional factor in $F_{12}$.
To complete the non-local notation, 
all symbols $\xpm{j},\gamma_j,\xi_j$ with $j=1,2$ in \tabref{tab:SCoeff}
will have to be replaced
by the symbols
$\xpm{\pi(k+j-1)},\gamma_{\pi(k+j-1)},\xi_{\pi(k+j-1)}$, 
respectively.

\subsection{Crossing Symmetry}
\label{sec:crossing}

\begin{figure}\centering
\includegraphics[scale=0.8]{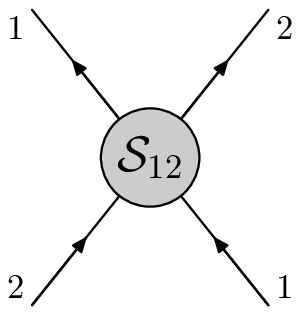}
\qquad
\includegraphics[scale=0.8]{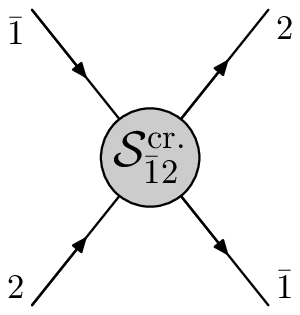}
\caption{S-matrix and crossed S-matrix.}
\label{fig:crossing}
\end{figure}

S-matrices of integrable models
are commonly expected to obey crossing relations.
Crossing of the S-matrix replaces one particle with its
conjugate particle propagating backwards in space and time,
see \figref{fig:crossing}.
The kinematic parameters of the conjugate particle are obtained 
by the antipode map
\[\label{eq:antipode}
\xpm{k}\mapsto \xpm{\bar k}=1/\xpm{k}.
\]
In \cite{Janik:2006dc} the condition for a crossing-symmetric
S-matrix was derived to be 
\[X_{12}=1,\]
where $X_{12}=X(\xpm{1},\xpm{2})$ 
is a function depending on the phase factor $S^0_{12}$
\[\label{eq:cross}
X_{12}=
\frac{S^0_{1\bar2}S^0_{12}}{\xi_1^2}\,
\frac{\xp{2}-\xm{1}}{\xm{2}-\xm{1}}\,
\frac{1/\xp{2}-\xp{1}}{1/\xm{2}-\xp{1}}\,.
\]
Here $S^0_{1\bar 2}$ denotes the crossed phase factor
$S^0_{1\bar 2}=S^0(\xpm{1},1/\xpm{2})$.
A naive attempt to solve the crossing relation 
by finding a suitable phase factor $S^0_{12}$ must fail:
If $X_{12}=1$ holds, then also 
$X_{1\bar 2}=X(\xpm{1},1/\xpm{2})=1$
must be true. 
However, both equations are incompatible because
$X_{12}/X_{1\bar 2}$ is independent of $S^0_{12}$
and does not equal unity.
The resolution to this problem is to allow branch 
cuts in the function $S^0_{12}$. Then the antipode map
$\xpm{}\mapsto 1/\xpm{}$ is not 
an involution when applied to the phase factor.
The doubled map 
$\xpm{}\mapsto 1/\xpm{}\mapsto \xpm{}$ 
should instead correspond to change of Riemann sheets in 
the function $S^{0}_{12}$.

Indeed, the leading orders of the function $S^0_{12}$ 
extracted from perturbative string theory
\cite{Arutyunov:2004vx,Beisert:2005cw,Hernandez:2006tk,Freyhult:2006vr}
do have branch cuts and 
are consistent with the crossing relation \cite{Arutyunov:2006iu,Beisert:2006zy}.
Therefore it is natural to expect crossing symmetry to hold exactly. 
A crossing-symmetric phase for perturbative $AdS_5\times S^5$ string theory 
was proposed recently in \cite{Beisert:2006ib}. 

In this article we shall not assume that the phase factor
satisfies the crossing equation $X_{kj}=1$. 
In this way we avoid having to deal with an intricate
analytic structure involving multiple Riemann sheets, 
see \cite{Arutyunov:2006iu,Beisert:2006zy,Beisert:2006ib}.
Nevertheless, it is worth keeping in mind that 
occurrences of $X_{kj}$ may be set to $1$ in a crossing-symmetric model.

\subsection{Special Points and Bootstrap}
\label{sec:Points}

Let us briefly study special points of the S-matrix
and relate them to the structure of representations.

\paragraph{Identity.}

When the particle momenta become equal,
\[\xpm{2}=\xpm{1},\]
the S-matrix degenerates into a permutation operator
with negative sign. 
It interchanges the particles without interchanging their flavours
\[\smat_{12}
\state{\fldX_1\fldX'_2}
=-\state{\fldX_2\fldX'_1}.
\]
It also inverts the sign.
Essentially, as the particle representations
are equal, it acts like the identity $\smat_{12}=-\ident_{12}$,
it merely interchanges the particle labels.

\paragraph{Symmetric Products.}

It is straightforward to see that for
\[\label{eq:sprojanti}
\xp{2}=\xm{1}
\]
the coefficient $A_{12}$ in \tabref{tab:SCoeff} is zero.
Also several other combinations of coefficients vanish.
Therefore, at this point, $\smat_{12}$ becomes a projector. 
This feature is related to the multiplet splitting rule
\eqref{eq:Splitting}. In general, two sites together form 
a long multiplet $\lrep{0,0;\vec{C}}$. 
However, if $\vec{C}^2=1$, this multiplet splits up into
two short multiplets
\[
\lrep{0,0;\vec{C}}
=\srep{1,0;\vec{C}}\oplus\srep{0,1;\vec{C}}.
\]
As noted in \cite{Dorey:2006dq}, the condition 
$\vec{C}^2=1$ is met when \eqref{eq:sprojanti} holds.
At this point, the scattering matrix projects out the first of these
multiplets (symmetric product) and leaves only the latter (antisymmetric product).
Likewise, if instead
\[\label{eq:sprojsym}
\xm{2}=\xp{1},
\]
the S-matrix becomes a projector, but onto the other irreducible multiplet 
(symmetric product).%
\footnote{It may seem that the S-matrix has a pole here.
However, the pole can be absorbed into the overall phase factor $S^0_{12}$. 
What matters is the ratio $A_{12}/D_{12}$
for which the values $0$ and $\infty$ mark projector points.}

\paragraph{Higher Representations.}

The projective character of the S-matrix can be used to construct the S-matrix 
for various tensor product representations. One merely has to prepare 
composite states which transform in the desired representation
and then scatter them as a whole with other objects, see \figref{fig:scatchain}.
In particular, by chaining up 
particles with $\xpm{k}=\xmp{k+1}$ one obtains the totally
symmetric representations $\srep{m,0}$ in \eqref{eq:symrep}.
These composite states are also called bound states, cf.~\cite{Dorey:2006dq}.

\paragraph{Adjoint and Singlet.}

Another interesting point is 
\[
\xpm{2}=1/\xpm{1}.
\]
Here, all central charges vanish, $\vec{C}=0$,
and the long multiplet $\lrep{0,0;\vec{C}}$
splits up into a singlet and an adjoint. 
Poles can be observed in $B_{12},C_{12},E_{12},F_{12}$
and it would be necessary to investigate the action of $\smat_{12}$ further. 
Here, some complications arise due to the semi-reducible nature of 
the adjoint representation in $\alg{u}(N|N)$ algebras.

\begin{figure}\centering
\parbox[c]{4cm}{\centering\includegraphics[scale=0.8]{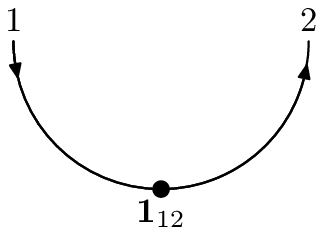}}
\qquad\qquad
\parbox[c]{4cm}{\centering\includegraphics[scale=0.8]{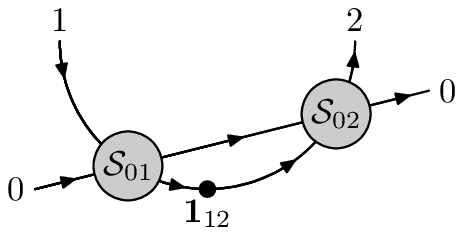}}
$=$
\parbox[c]{4cm}{\centering\includegraphics[scale=0.8]{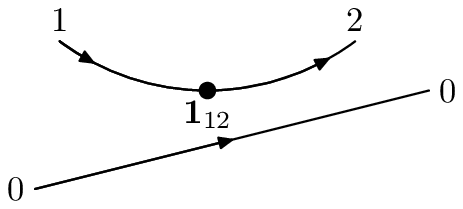}}
\caption{Singlet state and relation for trivial scattering.}
\label{fig:bootstrapcrossing}
\end{figure}

The singlet state however may be constructed easily, see \cite{Beisert:2005tm}.
Here it takes the form
\[
\state{\rep{1}_{12}}=
\frac{\alpha}{\gamma_1\gamma_2\xi_1}\lrbrk{\frac{\xp{1}}{\xm{1}}-1}\varepsilon_{ab}
\state{\fldY^-\fldY^-\fldZ^+\phi^a_1\phi^b_2}+
\varepsilon_{\alpha\beta}
\state{\psi^\alpha_1\psi^\beta_2}
\]
where $\xpm{1}=1/\xpm{2}$.
This composite has vanishing central charges, however the constituents
have non-zero central charges. When we flip space, time and $\alg{su}(2)$ charges for particle $1$,
then it becomes equivalent to particle $2$, i.e.~the two constituents are CPT-conjugates.
On l.h.s.~of \figref{fig:bootstrapcrossing}, 
particle $1$ can be viewed as the part of the 
worldline of particle $2$ which moves backwards in time.
We can thus view the singlet state as a curl of the worldline of
particle $2$ moving backwards in time for a while.
In other words, one might consider the singlet as a particle-hole fluctuation
of the vacuum. 
 
Let us now scatter the singlet state with an arbitrary site $\fldX_0$. 
We find \cite{Beisert:2005tm}%
\footnote{We must set $\xi_2=1/\xi_1$ as well
in order for \eqref{eq:scatsing} to hold
without rescaling fermions on the site $\fldX_0$.}
\[\label{eq:scatsing}
\smat_{02}\smat_{01}\,\state{\fldX_0\rep{1}_{12}}
=
X_{0,12}\,\state{\rep{1}_{12}\fldX_0}
\]
with the unique factor
\[\label{eq:Xdef0}
X_{0,12}=
\frac{S^0_{01}S^0_{02}}{\xi_0^2}\,
\frac{\xp{0}-\xp{1}}{\xp{0}-\xm{1}}\,
\frac{\xm{0}-\xp{2}}{\xm{0}-\xm{2}}\,.
\]
Recalling that $\xpm{1}=1/\xpm{2}$, we observe 
that the function $X_{0,12}=X_{02}$ 
is the same as the one encountered for crossing symmetry in
\cite{Janik:2006dc}, see \secref{sec:crossing}.

This observation can be understood diagrammatically using 
the r.h.s.~of \figref{fig:bootstrapcrossing}: 
Both arrows on $\smat_{01}$ come from the left 
as for the crossed S-matrix in \figref{fig:crossing}
and not from the bottom as usual. 
If we assume crossing symmetry to hold, then 
the equation in \figref{fig:bootstrapcrossing}
is equivalent to the unitarity condition in \figref{fig:ybe} rotated by $90^\circ$. 
Therefore, we should expect the equation in \figref{fig:bootstrapcrossing} to
hold if and only if the S-matrix is crossing-symmetric.

The crossing relation $X_{12}=1$ can now be interpreted alternatively 
as a bootstrap condition: The singlet state is a bound state of two particles
with zero total energy and momentum. 
As argued above, it represents an inessential vacuum fluctuation.
Therefore one may expect its scattering with any real particle to be trivial.

\subsection{Diagonalisation of the S-Matrix}

The diagonalisation of a slightly restricted version of the
S-matrix with $\xi_k=1$ was performed in \cite{Beisert:2005tm}. 
Let us merely highlight the differences due to the introduction 
of non-trivial $\xi_k$'s here as compared to App.~C of \cite{Beisert:2005tm}. 

\paragraph{Vacuum.}

The level-II vacuum is composed from only $\phi^1$'s.
\[\label{eq:lev2vac}
\state{0}\lvl{II}=\state{\phi^1_1\ldots\phi^1_K}.
\]
%

\paragraph{Excitations.}

We create an excitation $\psi^\alpha_k$ at site $k$ by acting with 
$(\gen{Q}^\alpha{}_1)_k$. Let us define
\[
(\gen{Q}^\alpha{}_1)^\pm_k=\frac{\xmp{k}}{\xmp{k}-\xpm{k}}\,
(\gen{Q}^\alpha{}_1)_k.
\]
Then a level-II excitation has the following form
\[
\state{\psi^\alpha}
=\sum_{k=0}^K\Psi_k(y)\bigbrk{
(\gen{Q}^\alpha{}_1)^-_k+
(\gen{Q}^\alpha{}_1)^+_{k+1}}
\state{0}\lvl{II}
\]
with the wave function
\[
\Psi_k(y)=\prod_{j=1}^k S\lvl{II,I}(y,x_j).
\]
The element $S\lvl{II,I}(y,x_k)$ of the diagonalised S-matrix in our case is
\[
S\lvl{II,I}(y,x_k)=\xi_k \frac{y-\xm{k}}{y-\xp{k}}\,.
\]
By comparing to (4.10) in \cite{Beisert:2005tm}, it is easy to see
that the additional factor of $1/\xi_1$ in $G_{12}$ requires
the compensating factor $\xi_1$ in $S\lvl{II,I}(y,x_1)$.

\paragraph{Scattering.}

A closer look at the first line in (4.17) of \cite{Beisert:2005tm} 
shows that there must be an overall factor of $\xi_1$ 
from $S\lvl{II,I}(y_2,x_1)$ in the state $\state{\psi^\alpha_1\psi^\beta_2}\lvl{II}$. 
Application of $\smat_{12}$ to the state
will then turn the factor $\xi_1$ into $\xi_2$ as desired. 
Consequently, the fourth and fifth lines must have the same overall factor
coming from $S\lvl{II,I}(y_1,x_1)$. 
The matrix elements $M,N$ of the level-II scattering matrix 
thus remain unchanged. The same holds for the diagonalised elements
$S\lvl{II,II}(y_1,y_2)$,
$S\lvl{III,II}(w_1,y_2)$ and
$S\lvl{III,III}(w_1,w_2)$.

\paragraph{Further Twisting.}

In fact, we could consider a slightly generalised setup, where
we introduce two more markers which rescale $\phi^2$ w.r.t.~$\phi^1$
and $\psi^2$ w.r.t.~$\psi^1$. These would thus break the two $\alg{su}(2)$ invariances.
The mechanism is the same as for $\fldY$ which 
rescales $\psi^\alpha$ w.r.t.~$\phi^a$ and which twists supersymmetry.
It is not difficult to convince oneself that this leads to 
the non-trivial element of the diagonalised S-matrix 
\[\label{eq:twistrho}
S\lvl{III,I}(w_1,x_2)=\rho_2
\]
with $\rho_k$ determining the shift rule for the marker as 
in \eqref{eq:MarkerY}. 
Although we had two independent rescalings, there is only one new coefficient. 
This is because the spin excitation $\phi^1\to\phi^2$ 
is a double excitation and therefore does not 
come along with an independent rescaling. 
It should be thought of as composed 
from two type-II, one type-III as well as a $\fldZ^-$ excitation.
The corresponding factor $\rho'$ for rescaling $\phi^2$ w.r.t.~$\phi^1$
must equal
\[\label{eq:rhoprime}
\rho'_j=\rho^{}_j\xi_j^2\,\frac{\xm{j}}{\xp{j}}\,,
\]
for self-consistency. Effectively, this means that 
we cannot have full manifest $\alg{h}$-symmetry at the local 
level. It may nevertheless still appear as a global symmetry.

Similarly, we can twist level-II w.r.t.~level-III
with a global parameter $\tau$, 
see e.g.~\cite{Beisert:2005if}.
This introduces a factor of $\tau$ in
the element $S\lvl{III,II}(w_1,y_2)$.

\paragraph{Elements.}

In conclusion, the elements of the diagonalised S-matrix are given by
\<\label{eq:sdiagelem}
S\lvl{I,0}(\xpm{1},\cdot)\eq \frac{\xp{1}}{\xm{1}}\,,
\nln
S\lvl{I,I}(\xpm{1},\xpm{2})\eq S^0(x_1,x_2)\,\frac{\xm{1}-\xp{2}}{\xp{1}-\xm{2}}\,,
\nln
S\lvl{II,I}(y_1,\xpm{2})\eq \xi_2\,\frac{y_1-\xm{2}}{y_1-\xp{2}}\,,
\nln
S\lvl{II,II}(y_1,y_2)\eq 1,
\nln
S\lvl{III,I}(w_1,\xpm{2})\eq\rho_2,
\nln
S\lvl{III,II}(w_1,y_2)\eq \tau\,\frac{w_1-y_2-1/y_2-\ihalf g^{-1}}{w_1-y_2-1/y_2+\ihalf g^{-1}}\,,
\nln
S\lvl{III,III}(w_1,w_2)\eq \frac{w_1-w_2+ig^{-1}}{w_1-w_2-ig^{-1}}\,.
\>
Note that $\rho_j$ and $\tau$ are the constants introduced above.
Furthermore, the function $X_{12}$ in \eqref{eq:cross} related to crossing
symmetry is modified to
\[\label{eq:crossrho}
X_{12}=
\frac{S^0_{1\bar2}S^0_{12}}{\rho_1\xi_1^2}\,
\frac{\xp{2}-\xm{1}}{\xm{2}-\xm{1}}\,
\frac{1/\xp{2}-\xp{1}}{1/\xm{2}-\xp{1}}\,.
\]

\subsection{Periodic States}

So far, we have been interested in chains with two ends and 
unspecified boundary conditions. Therefore all states made up 
of an arbitrary number of excitations with arbitrary momenta 
correspond to well-defined wave functions. 
If we however restrict to periodic boundary conditions, 
the wave function must be compatible with this periodicity.
Periodicity is ensured by the Bethe equations.

\paragraph{Bethe Equations.}

\begin{figure}\centering
\parbox[c]{4.5cm}{\centering\includegraphics[scale=0.5]{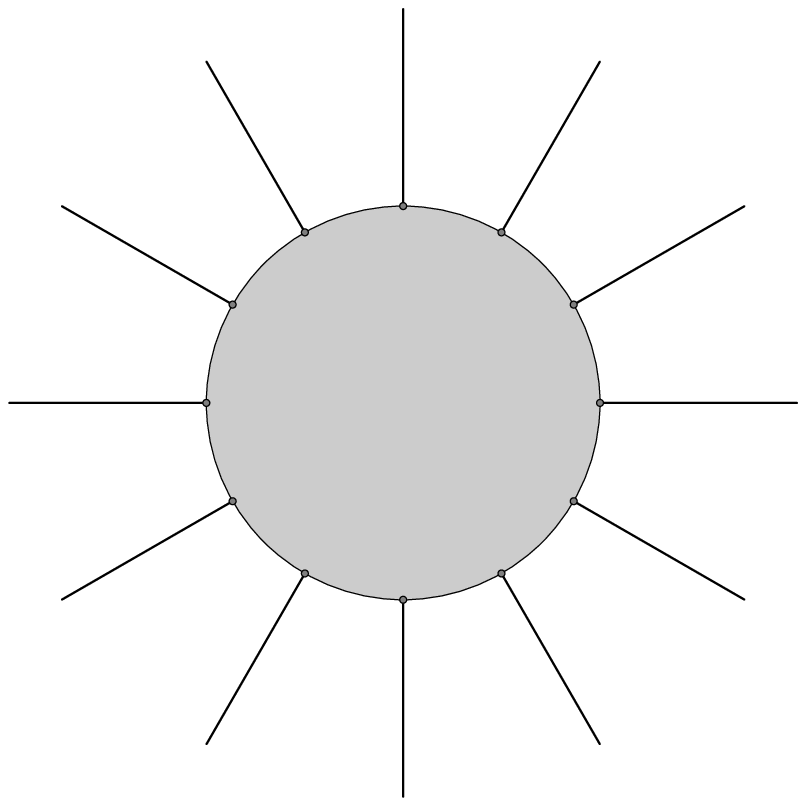}}
$=$
\parbox[c]{4.5cm}{\centering\includegraphics[scale=0.5]{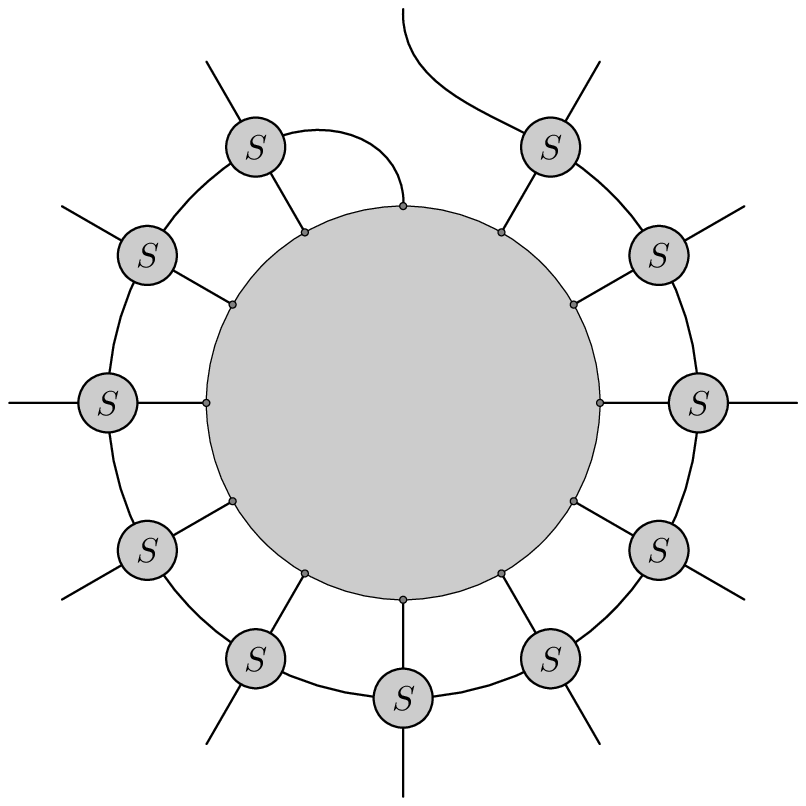}}
\caption{Bethe equations.}
\label{fig:bethe}
\end{figure}

Periodicity means that the total phase factor 
acquired by moving any excitation by one period 
must be trivial, cf.~\figref{fig:bethe}.
This condition introduces one equation 
per particle momentum and consequently leads to a discrete spectrum 
as expected for a compact space.

The scattering matrix is already in a diagonal form, 
so the Bethe equations for our model can be read off directly from
the diagonalised elements \eqref{eq:sdiagelem}.
The Bethe equations for levels II and III read as follows,
cf.~\cite{Beisert:2005fw}
\<\label{eq:Bethe23}
1\eq
\prod_{j=1}^{K}\frac{1}{\xi_j}
\prod_{j=1}^{K}\frac{y_k-\xp{j}}{y_k-\xm{j}}
\prod_{j=1}^{M}
\tau\,
\frac{y_k+1/y_k-w_j+\ihalf g^{-1}}{y_k+1/y_k-w_j-\ihalf g^{-1}}\,,
\nln
1\eq\prod_{j=1}^{K}\frac{1}{\rho_j}
\prod_{j=1}^{N}
\frac{1}{\tau}\,\frac{w_k-y_j-1/y_j+\ihalf g^{-1}}{w_k-y_j-1/y_j-\ihalf g^{-1}}
\mathop{\prod_{j=1}^{M}}_{j\neq k}\frac{w_k-w_j-ig^{-1}}{w_k-w_j+ig^{-1}}\,.
\>
Here $K$ is the number of sites
and $N,M$ are the number of level-II and level-III
excitations, respectively.
Note that the individual values of $\xi_j,\rho_j$ are completely irrelevant
as they should because they merely represent rescalings
of various types of spin orientations at different positions of the chain. 
However, for the Bethe equations it does matter 
how the spin orientations are periodically identified. 
This is determined by the product of all $\xi_j$'s and $\rho_j$'s,
respectively.

The Bethe equations \eqref{eq:Bethe23} are somewhat reminiscent of the
equations for a model with $\alg{su}(2|1)=\alg{osp}(2|2)$ symmetry
\cite{Maassarani:1994ac,Bracken:1995aa,Pfannmuller:1996vp}.
This is not surprising as there is a manifestly 
$\alg{su}(2|1)$-symmetric formulation of the S-matrix and the
Bethe ansatz \cite{Beisert:2005tm}.
Potentially the equations can even be matched precisely. This would require to relate the 
charge parameter of the four-dimensional spin representation
to the spectral parameter in a special way
along the lines of \cite{Beisert:2005wm}.

\paragraph{Dualisation.}

We can perform a dualisation or particle-hole transformation 
on the fermionic roots $y_j$ \cite{Woynarowich:1983aa,Bares:1992aa,Essler:1992nk,Beisert:2005di}:
The Bethe equation for $y_k$ in \eqref{eq:Bethe23} 
is in fact an algebraic equation in $y$ with coefficients
independent of the $y_j$'s. Therefore, the $N$ parameters $y_j$ are 
the roots of this equation and there exist further 
$\tilde N$ roots $\tilde y_j$ with 
\[\tilde N=K+2M-N.\]
We can reformulate this condition in terms of the function
\<\label{eq:dualityQ}
Q(y)\eq
\prod_{j=1}^{K}\xi_j\lrbrk{y-\xm{j}}
\prod_{j=1}^N\frac{1}{y-y_j}
\prod_{j=1}^{\tilde N}\frac{1}{y-\tilde y_j}
\prod_{j=1}^{M}\frac{y}{\tau}\lrbrk{y+1/y-w_j-\ihalf g^{-1}}
\nl-
\prod_{j=1}^{K}\lrbrk{y-\xp{j}}
\prod_{j=1}^N\frac{1}{y-y_j}
\prod_{j=1}^{\tilde N}\frac{1}{y-\tilde y_j}
\prod_{j=1}^{M}y\lrbrk{y+1/y-w_j+\ihalf g^{-1}}
\nln\eq
\prod_{j=1}^{K}\xi_j
\prod_{j=1}^{M}\frac{1}{\tau}-1.
\>
Demanding that $Q(y)$ is constant is equivalent to the Bethe equations
for the $y$'s and the $\tilde y$'s (which obey the same Bethe equation).

The property can be translated into a number of useful relations.
In particular we find
\[
\frac{Q(\xp{k})}{Q(\xm{k})}=
\xi_k\,
\mathop{\prod_{j=1}^{K}}_{j\neq k}
\xi_j\,
\frac{\xp{k}-\xm{j}}{\xm{k}-\xp{j}}
\prod_{j=1}^{N}\frac{\xm{k}-y_j}{\xp{k}-y_j}
\prod_{j=1}^{\tilde N}\frac{\xm{k}-\tilde y_j}{\xp{k}-\tilde y_j}
\prod_{j=1}^{M}\frac{\xp{k}}{\tau\xm{k}}
=1
\]
and, when setting $\xpm{w_k}+1/\xpm{w_k}=w_k\pm i/2g$, we further obtain
\<
\frac{Q(\xp{w_k})\,Q(1/\xp{w_k})}{Q(\xm{w_k})\,Q(1/\xm{w_k})}\eq
\prod_{j=1}^{K}\frac{1}{\xi_j^2\tau}\,\frac{\xp{j}}{\xm{j}}
\prod_{j=1}^{N}\tau\,\frac{w_k-y_j-1/y_j-\ihalf g^{-1}}{w_k-y_j-1/y_j+\ihalf g^{-1}}
\nlnum\nonumber
\times\prod_{j=1}^{\tilde N}\tau\,\frac{w_k-\tilde y_j-1/\tilde y_j-\ihalf g^{-1}}{w_k-\tilde y_j-1/\tilde y_j+\ihalf g^{-1}}
\mathop{\prod_{j=1}^{M}}_{j\neq k}\lrbrk{\frac{w_k-w_j+ig^{-1}}{w_k-w_j-ig^{-1}}}^2
=1.
\>

By using the second identity, the Bethe equations can now be written in a dual form
\cite{Beisert:2005fw},
\<\label{eq:BetheDual23}
1\eq\prod_{j=1}^{K}\xi_j
\prod_{j=1}^{K}\frac{\tilde y_k-\xm{j}}{\tilde y_k-\xp{j}}
\prod_{j=1}^{M}
\frac{1}{\tau}\,\frac{\tilde y_k+1/\tilde y_k-w_j-\ihalf g^{-1}}{\tilde y_k+1/\tilde y_k-w_j+\ihalf g^{-1}}\,,
\nln
1\eq
\prod_{j=1}^{K}\frac{1}{\rho_j\xi_j^2\tau}\,\frac{\xp{j}}{\xm{j}}
\prod_{j=1}^{\tilde N}
\tau\,
\frac{w_k-\tilde y_j-1/\tilde y_j-\ihalf g^{-1}}{w_k-\tilde y_j-1/\tilde y_j+\ihalf g^{-1}}
\mathop{\prod_{j=1}^{M}}_{j\neq k}\frac{w_k-w_j+ig^{-1}}{w_k-w_j-ig^{-1}}\,.
\>
%

\paragraph{Symmetries.}

Non-Abelian symmetries are realised in the Bethe equations by 
the possibility of adding Bethe roots at special points
without changing the equations.
These correspond to positive roots of the symmetry algebra $\alg{h}$.
Possible points are $y=\infty,y=0,w=\infty$
and certain combinations of these.
First of all the Bethe equations for the existing roots
will receive factors of $\tau$ for the introduction
of new roots. This requires $\tau=1$ in order for 
any generator of $\alg{h}$ to be preserved.
Let us therefore assume $\tau=1$ in the below.

Invariance under the two $\alg{su}(2)$ raising operators
represented by the sets of Bethe roots
$\set{w=\infty}$ and $\set{w=\infty, y=\infty, y=0}$
requires, respectively
\[\label{eq:symbos}
\mbox{1:}\quad 1=\prod_{j=1}^{K}\rho_j\qquad
\mbox{and}\qquad
\mbox{2:}\quad 1=\prod_{j=1}^{K}\rho_j\xi_j^2\,\frac{\xm{j}}{\xp{j}}\,.
\]
Invariance under the four fermionic raising generators requires
\[\label{eq:symferm}
\mbox{a:}\quad 1=\prod_{j=1}^{K}\xi_j,\qquad
\mbox{b:}\quad 1=\prod_{j=1}^{K}\rho_j\xi_j,\qquad
\mbox{c:}\quad 1=\prod_{j=1}^{K}\xi_j\,\frac{\xm{j}}{\xp{j}}\,,\qquad
\mbox{d:}\quad 1=\prod_{j=1}^{K}\rho_j\xi_j\,\frac{\xm{j}}{\xp{j}}\,.
\]

\begin{table}\centering
\begin{tabular}{|l|l|}\hline
symmetry&condition\\\hline
$\alg{su}(2|2)$&12abcd\\
$\alg{su}(2|1)$&1ab, 1cd, 2ad, 2bc\\
$\alg{su}(1|1)\times\alg{su}(1|1)$&ac, bd\\
$\alg{su}(1|1)$&abcd\\
$\alg{su}(2)\times\alg{su}(2)$&12\\
$\alg{su}(2)$&1, 2\\
$\cdot$&$\cdot$\\
\hline
\end{tabular}
\caption{Preserved non-abelian symmetries. The
second column lists all the combinations of 
conditions in \protect\eqref{eq:symbos,eq:symferm}}
\label{tab:symmetries}
\end{table}
The conserved generators can form one of the following non-abelian symmetry
algebras:
$\alg{su}(2|2)$, $\alg{su}(2)\times\alg{su}(2)$, $\alg{su}(2|1)$,
$\alg{su}(2)$, $\alg{su}(1|1)\times\alg{su}(1|1)$, $\alg{su}(1|1)$
or none at all.
The conditions for the various preserved symmetries 
are summarised in \tabref{tab:symmetries}.

\section{Hubbard Chain}
\label{sec:Hubbard}

In this section we will show how the Hubbard chain 
and Shastry{}'s R-matrix are related to our model.

\subsection{Qualitative Comparison}

The one-dimensional Hubbard model \cite{Hubbard:1963aa} is a spin chain 
of two bosonic and two fermionic spin degrees of freedom per site. 
It has a manifest $\alg{su}(2)$ symmetry and 
a so-called eta-pairing $\alg{su}(2)$ symmetry \cite{Lieb:1971aa,Yang:1989aa,Yang:1990aa}.
The latter is a symmetry which holds at the local level, but may be
destroyed by a global mismatch of phases. 
In the original formulation, the eta-pairing symmetry
holds exactly for even-length chains.
The integrability of the model was shown by Lieb and Wu
\cite{Lieb:1968aa} who also derived the corresponding Bethe equations. 
An R-matrix was constructed by Shastry \cite{Shastry:1986bb}.
The R-matrix has the remarkable property that it cannot
be written as a function of the difference of spectral parameters
of the two sites; it has full dependence on them.
There is a vast literature on this particular model,
see for instance \cite{Essler:2005aa}. 
One interesting recent development is the discovery of a connection 
to a sector of $\superN=4$ gauge theory \cite{Rej:2005qt}.

In fact, the above properties are reminiscent of the 
chain discussed in the previous section.%
\footnote{I thank Matthias Staudacher for suggesting this to me.}
First of all, the spin degrees of freedom clearly coincide 
with the fundamental multiplet of $\alg{h}$ introduced in \secref{sec:funda}.
Secondly, the above S-matrix%
\footnote{Our S-matrix serves the same purpose as the
R-matrix for the Hubbard chain. The difference in nomenclature
is related to the different applications of the models:
Our S-matrix arises at the first level of a nested Bethe ansatz
(for $\superN=4$ gauge theory). Conversely, Shastry{}'s R-matrix is
used to define the integrable structure of the Hubbard chain.}
is not of a difference form, just like Shastry{}'s R-matrix.
The symmetry algebra $\alg{h}$ of our chain is bigger, but at least it contains 
$\alg{su}(2)\times\alg{su}(2)$ as a subalgebra. 
Another difference is that markers apparently play no role in the Hubbard chain.

Here we will show that, despite the latter two points, 
our S-matrix is essentially equivalent to Shastry{}'s R-matrix.
To understand how this can be true, 
we note that the $\alg{h}$ symmetry acts similarly to 
the eta-pairing symmetry. It is present locally, but a mismatch of
phases generically prevents it from being a global symmetry, 
cf.~\eqref{eq:symbos,eq:symferm} and \tabref{tab:symmetries}.
In particular, for the Hubbard Hamiltonian, the supersymmetry is 
always absent. For Shastry{}'s R-matrix it however implies the existence
of a new supersymmetry in addition to the well-known 
$\alg{su}(2)\times\alg{su}(2)$ symmetry.
The effect of the markers will turn out to cancel out completely so that 
we can effectively work without them. 
The relationship to the results of \cite{Rej:2005qt} will remain unclear though:
The embedding of the $\alg{su}(2)$ sector of $\superN=4$ gauge theory 
into the Hubbard model is quite different from the embedding of
Shastry{}'s R-matrix into $\superN=4$ gauge theory and strings on $AdS_5\times S^5$.

\subsection{Comparison of Bethe Equations}

The Bethe equations for the Hubbard chain are the 
Lieb-Wu equations \cite{Lieb:1968aa}
\<
1
\eq
\exp(-ik_k K)
\prod_{j=1}^{M}\frac{2\sin k_k-2\Lambda_j+\ihalf U}{2\sin k_k-2\Lambda_j-\ihalf U}\,,
\nln
1
\eq
\prod_{j=1}^{N}\frac{2\Lambda_k-2\sin k_j+\ihalf U}{2\Lambda_k-2\sin k_j-\ihalf U}
\mathop{\prod_{j=1}^{M}}_{j\neq k}\frac{2\Lambda_k-2\Lambda_j-i U}{2\Lambda_k-2\Lambda_j+i U}\,.
\>
We can easily match them with most of the terms in \eqref{eq:Bethe23}
by making the replacements
\[\label{eq:relhub}
g=\frac{1}{U}\,,
\qquad
w_k=2\Lambda_k,
\qquad
y_k=\bigbrk{-i\exp(ik_k)}^{\pm 1}
\]
as well as setting $\rho_j=\tau=1$.
The matching of the remaining term in \eqref{eq:Bethe23}
\[
-i(y_k)^{\pm1}=\frac{1}{\xi_j}\,\frac{y_k-\xp{j}}{y_k-\xm{j}}\,.
\]
fixes $\xpm{j}$ and $\xi_j$.
Here we have a choice for $y_k$: If we set 
$y_k=-i\exp(ik_k)$ we need to take the limit
\[
\xp{j},\xi_j\to 0,\quad \xm{j}\to\infty, 
\qquad
\xp{j}=i\xi_j,\quad
\xm{j}=-i/\xi_j.
\]
Likewise, for
$y_k=i\exp(-ik_k)$ we should take the limit
\[
\xp{j},\xi_j\to\infty, \quad 
\xm{j}\to 0,
\qquad
\xp{j}=i\xi_j,\quad
\xm{j}=-i/\xi_j.
\]
Note that the points $(0,\infty)$ and $(\infty,0)$ are
perfectly valid solutions $(\xp{},\xm{})$ of the constraint \eqref{eq:xpmconstr}.
This shows that the Lieb-Wu equations
are a special case of the Bethe equations 
for the present model. 
They correspond to a homogeneous chain 
because all parameters $\xpm{j},\xi_j$ are independent of the site $j$.

The symmetries are easily understood with the help
of \eqref{eq:symbos,eq:symferm}.
All supersymmetry is broken because none of the relations in \eqref{eq:symferm} holds.
The equation 1 in \eqref{eq:symbos} always holds, while equation 2
\[
1=\prod_{j=1}^{K}\xi_j^2\rho_j\,\frac{\xm{j}}{\xp{j}}
=
(-1)^{K}
\]
requires the length of the chain to be even. 
Therefore the Lieb-Wu equations reproduce the well-known symmetry of the Hubbard model
\cite{Lieb:1971aa,Yang:1989aa,Yang:1990aa}:
either $\alg{su}(2)\times\alg{su}(2)$ for even length or
$\alg{su}(2)$ for odd length.

\subsection{Comparison of the S/R-Matrices}
\label{sec:compsr}

We will now compare the models by comparing directly their 
S/R-matrices. 
We will use a form of Shastry{}'s R-matrix \cite{Shastry:1986bb}
given by Ramos and Martins \cite{Ramos:1996us}.
This form has the benefit that both $\alg{su}(2)$ factors are realised 
manifestly. 
We match the parameters as follows
\[g=\frac{1}{U}\,,\qquad
\xp{k}=\frac{ib_k}{a_kU}\,\exp(2h_{k}),\qquad
\xm{k}=\frac{a_k}{ib_kU}\,\exp(2h_{k}).
\]
The constraints in \cite{Ramos:1996us}
\[\label{eq:constrhub}
a_k^2+b_k^2=1,\qquad \sinh(2h_k)=\half Ua_kb_k,
\]
turn out to be equivalent to \eqref{eq:xpmconstrk}.%
\footnote{The constraints \protect\eqref{eq:constrhub} for $a,b,\exp(h)$ 
define a genus-three surface \protect\cite{Goehmann:2006aa}.
Conversely, the constraint \protect\eqref{eq:xpmconstrk} for $\xpm{}$
defines merely a genus-one surface \protect\cite{Janik:2006dc}. 
This superficial mismatch is resolved in \protect\eqref{eq:hubaux}
which relates $\xi,\gamma$ to $\xpm{}$:
Together, the constraints \protect\eqref{eq:xpmconstrk,eq:hubaux} 
for $\xpm{},\xi,\gamma$ define a higher-genus surface.}
We also have to set the auxiliary parameters to
\[\label{eq:hubaux}
\xi_k=\frac{b_k}{a_k}\,,\qquad\gamma_k=\sqrt{\alpha}\,\frac{\exp(h_k)}{a_k}\,,\qquad \rho_k=1.
\]
This guarantees a more symmetry S-matrix, i.e.~$H_{12}=K_{12}$, $C_{12}=F_{12}$,
which holds by construction in Shastry{}'s R-matrix.
Note also that the relation $\xp{k}/\xm{k}=-\xi_k^2$ follows from the above.
The R-matrix in \cite{Ramos:1996us} is given in terms of ten coefficient functions $\alpha_{i}$.
We find the following relations to our coefficients in \tabref{tab:SCoeff}
\[\label{eq:compshastry}
\begin{array}{rclcrclcrcl}
\displaystyle\frac{\alpha_2}   {\alpha_1}\eq\displaystyle\frac{A_{12}}{D_{12}}\,,&&
\displaystyle\frac{\alpha_3}   {\alpha_1}\eq\displaystyle\frac{D_{12}+E_{12}}{2D_{12}}\,,&&
\displaystyle\frac{\alpha_4}   {\alpha_1}\eq\displaystyle\frac{A_{12}+B_{12}}{2D_{12}}\,,
\\[10pt]
\displaystyle\frac{\alpha_5}   {\alpha_1}\eq\displaystyle\frac{H_{12}}{D_{12}}=\frac{K_{12}}{D_{12}}\,,&&
\displaystyle\frac{\alpha_6}   {\alpha_1}\eq\displaystyle-\frac{D_{12}-E_{12}}{2D_{12}}\,,&&
\displaystyle\frac{\alpha_7}   {\alpha_1}\eq\displaystyle\frac{A_{12}-B_{12}}{2D_{12}}\,,
\\[10pt]
\displaystyle\frac{\alpha_8}   {\alpha_1}\eq\displaystyle\frac{G_{12}}{D_{12}}\,,&&
\displaystyle\frac{\alpha_9}   {\alpha_1}\eq\displaystyle-\frac{L_{12}}{D_{12}}\,,&&
\displaystyle\frac{\alpha_{10}}{\alpha_1}\eq \displaystyle-\frac{C_{12}}{2D_{12}}=-\frac{F_{12}}{2A_{12}}\,.
\end{array}
\]
Note that one of the functions on either side is undetermined and we may
only compare quotients.
Furthermore, the locations of the coefficients within the R-matrix agrees 
with the top of \tabref{tab:SCoeff} and \eqref{eq:compshastry}.
This shows that Shastry{}'s R-matrix is fact is invariant 
under $\alg{h}$, i.e.~it has a hidden $\alg{h}$-supersymmetry.

Crossing symmetry has also been considered in the context 
of Shastry{}'s R-matrix in \cite{Zhou:1996aa,Shiroishi:1997aa}.
We have not succeeded to match exactly this result 
to Janik{}'s crossing relation $X_{12}=1$,
however the functions $\rho$ and $\tilde\rho$ in \cite{Zhou:1996aa,Shiroishi:1997aa}
are at least similar to $X_{12}$. In this context it may be useful
to note that $\alpha_6/\alpha_1\sim -X_{1\bar 2}$ and
$\alpha_7/\alpha_2\sim 1/X_{12}$. 
Perhaps the crossing unitarity relation in \cite{Zhou:1996aa,Shiroishi:1997aa} 
is not literally the same as the one discussed in \secref{sec:crossing}.

\subsection{Relation to Rej-Serban-Staudacher}
\label{sec:RSS}

There is another relationship between the $\alg{su}(2)$ sector 
of $\superN=4$ SYM and the Hubbard chain 
which was recently discovered in \cite{Rej:2005qt}.
It is however of a different nature: 

Firstly, the connection in \cite{Rej:2005qt} is between the 
(slightly altered) Hamiltonian of the Hubbard chain
and the planar dilatation generator of $\superN=4$ SYM. 
Here, the connection is between the R-matrix of the Hubbard chain
and the S-matrix (one level up in the nested Bethe ansatz)
of the planar $\superN=4$ SYM chain.
Furthermore, the chain in \cite{Rej:2005qt} is homogeneous, here we
have different parameters for all the sites.

Secondly, the $\alg{su}(2)$ algebra of the sector does not (necessarily) 
correspond to one of the two $\alg{su}(2)$'s in $\alg{su}(2|2)$:
The $\alg{su}(2)$ sector namely consists of a vacuum state which is not
in our chain and an excitation which is one of the two bosonic states of our chain. 
By means of an $\alg{su}(4)$ rotation, it is however possible to 
make the two $\alg{su}(2)$ coincide. Perhaps this gives a formal 
explanation of why the Hubbard model appears in \cite{Rej:2005qt}. 
Nevertheless, it cannot really be made use of in terms
of sectors because our sites always correspond 
to excitations in $\superN=4$ SYM. 

It is encouraging to see though that precisely the same relationship 
between the coupling constants \eqref{eq:relhub} was also found in 
\cite{Rej:2005qt}. It would be remarkable if one could somehow join the two S-matrices 
for scattering of excitations with another S-matrix defining
the R-matrix for the first level of the nested Bethe ansatz as in \cite{Rej:2005qt}
and thus obtain an R-matrix with full $\alg{psu}(2,2|4)$ symmetry
suitable for $\superN=4$ SYM. On the other hand, this might be too much to
ask for.

\subsection{Hamiltonians}

Our spin chain model was constructed as the 
second level in the nested Bethe ansatz of $\superN=4$ SYM
and strings on $AdS_5\times S^5$.
We might however also consider spin chain models where our
S-matrix takes the role of an R-matrix at the first level 
of a nested Bethe ansatz. 
With the above results, these models represent 
generalisations of the Hubbard chain.
Numerous such models generalising the Hubbard model 
have appeared in the literature,
see e.g.~\cite{Essler:1992py,Maassarani:1994ac,Bracken:1995aa,Alcaraz:1999aa,Guan:2002aa}.

The present class of models has been outlined briefly at the end 
of section 5.1 in \cite{Martins:1997aa}. The generalised Bethe equations
in \cite{Gohmann:2001wh} will describe the model for a suitable
choice of parameter functions. Beyond this, there appears to be no further work 
on this particular class of models.

Here we shall derive a family of Hamiltonians from such models.
To limit the number of free parameters somewhat, 
we demand that the Hamiltonian is homogeneous, hermitian, 
and manifestly preserves $\alg{su}(2)\times\alg{su}(2)$ symmetry.

To ensure homogeneity, the parameters of all sites must be the same.
Hermiticity requires $\xp{}$ and $\xm{}$ to 
be complex conjugates and fixes $\gamma$, cf.~\eqref{eq:repuni}.
Finally, for manifest $\alg{su}(2)\times\alg{su}(2)$ symmetry
we set $\rho=\rho'=1$, cf.~\eqref{eq:rhoprime}.
We therefore set
\[\xi=\sqrt{\xp{}/\xm{}},\qquad
\gamma=\sqrt{-i\xp{}+i\xm{}}\,.\]
Interestingly, the choice of manifest $\alg{su}(2)\times\alg{su}(2)$
symmetry leads to trivial commutation of the combination of markers
$\fldY^\pm\fldY^\pm\fldZ^\mp$. As this is the only combination that appears 
in the S-matrix, we may as well drop them altogether. 

\paragraph{Hamiltonian.}

A nearest-neighbour Hamiltonian can be derived
from the S-matrix by expanding around 
coinciding spectral parameters $\xpm{1}=\xpm{2}$.
At this point the S-matrix becomes 
a permutation $\perm_{12}$
and the first order in the expansion yields
the Hamiltonian
\[
\ham_{12}(\xpm{})=-i\,\eval{\perm_{21}\frac{d}{du_1}\,\smat_{12}}_{\xpm{1,2}=\xpm{}}\,.
\]
For definiteness we have used $u$ as the single expansion parameter, 
$\xpm{}$ depend on it via \eqref{eq:diffs}.
The homogeneous Hamiltonian for the spin chain reads
\[
\ham = \sum_{k=1}^K \ham_{k,k+1}.
\]

The free parameters of the model are the coupling constant $g$ 
and the spectral parameter $\xpm{}$.
The fact that the spectral parameter will be a genuine parameter
of the spectrum is special to this model, cf.~\cite{Martins:1997aa}: 
In conventional integrable spin chains models
this is not the case, because the S-matrix is a function 
of the difference of spectral parameters. Therefore it does 
not matter around which point we expand. Here this is different.
With a non-trivial spectral parameter the model
turns out to be anisotropic or parity violating. 

The Hamiltonian can be shifted and rescaled without 
altering its spectrum qualitatively. We use this freedom
to bring the Hamiltonian to a simpler form $\ham'_{12}$
\[\label{eq:renormham}
\ham_{12}=\frac{(\xp{}+\xm{})^2(\xp{}\xm{}-1)\ham'_{12}+(-8\xp{}\xp{}\xm{}\xm{}+4\xp{}\xp{}+4\xm{}\xm{})\ident_{12}}
{4i(\xp{}-\xm{})(\xp{}\xp{}-1)(\xm{}\xm{}-1)}\,.
\]
The simplified pairwise Hamiltonian is given by the action
\<
\ham'_{12}\state{\phi^a\phi^b}
\eq
A'\state{\phi^{\{a}\phi^{b\}}}
+B'\state{\phi^{[a}\phi^{b]}}
+\half C'\varepsilon^{ab}\varepsilon_{\alpha\beta}\state{\psi^\alpha\psi^\beta},
\nln
\ham'_{12}\state{\psi^\alpha\psi^\beta}
\eq
D'\state{\psi^{\{\alpha}\psi^{\beta\}}}
+E'\state{\psi^{[\alpha}\psi^{\beta]}}
+\half F'\varepsilon^{\alpha\beta}\varepsilon_{ab}\state{\phi^a\phi^b},
\nln
\ham'_{12}\state{\phi^a\psi^\beta}
\eq
G'\state{\psi^\beta\phi^{a}}
+H'\state{\phi^{a}\psi^\beta},
\nln
\ham'_{12}\state{\psi^\alpha\phi^b}
\eq
K'\state{\psi^\alpha\phi^{b}}
+L'\state{\phi^{b}\psi^\alpha}
\>
with the coefficients
\<
A',D'\eq 1\mp 1\,,\nln
B',E'\eq 1\mp \lrbrk{1-\frac{8\xp{}\xm{}}{(\xp{}+\xm{})^2}},\nln
C',F'\eq \frac{8i\xp{}\xm{}(\xp{}-\xm{})}{(\xp{}+\xm{})^2(\xp{}\xm{}-1)}\,,\nln
G',L'\eq -\frac{4\xp{}\xm{}(\xmp{}\xmp{}-1)}{(\xp{}+\xm{})^2(\xp{}\xm{}-1)}\sqrt{\frac{\xpm{}}{\xmp{}}}\,,\nln
H',K'\eq 1\,.
\>
Note that typically for Hubbard-like models,
a fermionic spin notation is used. The map between states
and spin generators reads
\[
\state{\phi^1}\sim \state{0},\qquad
\state{\phi^2}\sim \mathrm{c}^\dagger_{1}\mathrm{c}^\dagger_{2}\state{0},
\qquad
\state{\psi^\alpha}\sim \mathrm{c}^\dagger_{\alpha}\state{0}.
\]
This dictionary can be used to cast the Hamiltonian 
\eqref{eq:renormham} in a spin form.


\paragraph{Bethe Ansatz.}

The vacuum state for this Hamiltonian consists of 
only bosons of one type $\phi^1$, its energy $E'$ is exactly zero. 
The two fermions $\psi^\alpha$ are the excitations. 
The dispersion relation for an excitation with momentum $p'$ is
\[
e'(p')=2-\frac{4\sqrt{\xp{}\xm{}}}{\xp{}+\xm{}}\,
\lrbrk{\cos p'-\frac{i(\xp{}-\xm{})(\xp{}\xm{}+1)}
{(\xp{}+\xm{})(\xp{}\xm{}-1)}\,\sin p'}.
\]
We can see that the Hamiltonian is not isotropic due to the $\sin p'$ term. 

The spectrum of the model is described by the above Bethe equations \eqref{eq:Bethe23}
\<\label{eq:BetheHubGen}
1\eq
\exp(-ip'_kK)
\prod_{j=1}^{M}
\frac{y_k+1/y_k-w_j+\ihalf g^{-1}}{y_k+1/y_k-w_j-\ihalf g^{-1}}\,,
\nln
1\eq
\prod_{j=1}^{N}
\frac{w_k-y_j-1/y_j+\ihalf g^{-1}}{w_k-y_j-1/y_j-\ihalf g^{-1}}
\mathop{\prod_{j=1}^{M}}_{j\neq k}\frac{w_k-w_j-ig^{-1}}{w_k-w_j+ig^{-1}}\,.
\>
Here the momentum of an excitation is related to the parameter $y$ as
\[
S\lvl{II,I}(y,\xpm{})=\sqrt{\xp{}/\xm{}}\,\frac{y-\xm{}}{y-\xp{}}=\exp(ip').
\]
As usual, the dispersion relation is obtained as the
derivative of the momentum taking into
account the prefactor in \eqref{eq:renormham}
\[
-i\frac{d}{du}\,\log S\lvl{II,I}(y,\xpm{})=
\frac{(\xp{}+\xm{})^2(\xp{}\xm{}-1)\,e'(p')}{4i(\xp{}-\xm{})(\xp{}\xp{}-1)(\xm{}\xm{}-1)}\,.
\]
We should note that these energies are not related to the central charges
in this model. The central charges are defined through the
spectral parameter $\xpm{}$ alone, while the energies are dynamical quantities.

\paragraph{Symmetries.}

Although the manifest symmetry is $\alg{su}(2)\times\alg{su}(2)$, 
it may be enhanced to $\alg{su}(2|2)$ at the global level.
This is the case if \eqref{eq:symferm} holds, i.e.~if
\[
\lrbrk{\frac{\xp{}}{\xm{}}}^{K/2}=1.
\]

In particular,
for the point $(\xp{},\xm{})=(\infty,\infty)$ one recovers
the $\alg{u}(2|2)$-invariant model in \cite{Essler:1992py}.
Then the simplified Hamiltonian becomes
$\ham'_{12}=\ident_{12}-\perm_{12}$ which is manifestly 
$\alg{u}(2|2)$-invariant. Somewhat disappointingly, the
coupling constant $g$ has dropped out from the system.
The way in which the coupling constant $g$ disappears
is however somewhat singular and it forces us to 
introduce another flavour of particle. The number
of different Bethe equations is thus three instead of two
for the more general model.

Another interesting model is obtained by setting $\xp{}=-\xm{}$. 
This model is supersymmetric on chains with a multiple of
four sites. It features a simplified Hamiltonian and a dispersion
law $e'(p')\sim \sin p'$ which is purely parity odd. Furthermore,
the coupling constant $g$ remains an essential parameter of the model.
Note that due to singularities in the simplified Hamiltonian $\ham'_{12}$
the latter will have to be renormalised before setting $\xp{}=-\xm{}$.
This model might deserve further investigation.

\section{Transfer Matrices}
\label{sec:transfer}

The transfer matrix is an element of central importance 
for the integrable structure of periodic chains. 
Here we will construct the transfer matrix for our model and
derive its eigenvalues.

\subsection{Monodromy Matrix}

First of all, let us introduce a chain with an 
auxiliary site at either the left or the right end.
The monodromy matrix shifts the auxiliary site
past the remaining chain
\[
\mono_{\auxrep}:
\srep{\vec C_{\auxrep}}\otimes
\srep{\vec C_1}\otimes\ldots\otimes\srep{\vec C_K}
\mapsto
\srep{\vec C'_1}\otimes\ldots\otimes\srep{\vec C'_K}
\otimes\srep{\vec C'_{\auxrep}}.
\]
It is therefore defined as the following product of S-matrices
\[
\mono_{\auxrep}=
\smat_{\auxrep K}\cdots\smat_{\auxrep 2}\smat_{\auxrep 1}.
\]
Note that according to \eqref{eq:chainscattercharge}
the central charges transform as follows
in the scattering process
\[\label{eq:MonoChargeMap}
P'_{\auxrep}=P_{\auxrep} \prod_{j=1}^K \frac{\xp{j}}{\xm{j}}\,,\qquad
K'_{\auxrep}=K_{\auxrep} \prod_{j=1}^K \frac{\xm{j}}{\xp{j}}\,,\qquad
P'_k=P_k\,\frac{\xm{\auxrep}}{\xp{\auxrep}}\,,\qquad
K'_k=K_k\,\frac{\xp{\auxrep}}{\xm{\auxrep}}\,,
\]
while the $C'_k=C_k$ are not modified.

\begin{figure}\centering
\parbox[c]{2cm}{\centering\includegraphics[scale=0.6]{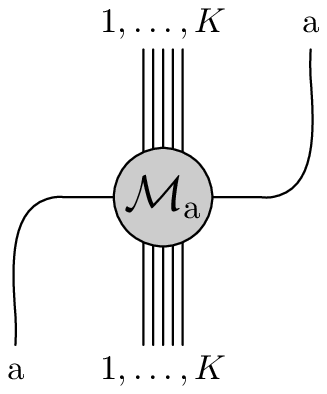}}
\qquad
\parbox[c]{2cm}{\centering\includegraphics[scale=0.6]{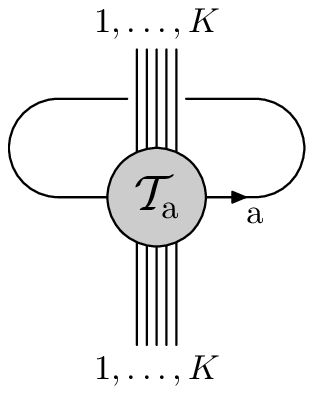}}
\qquad
\parbox[c]{2cm}{\centering\includegraphics[scale=0.6]{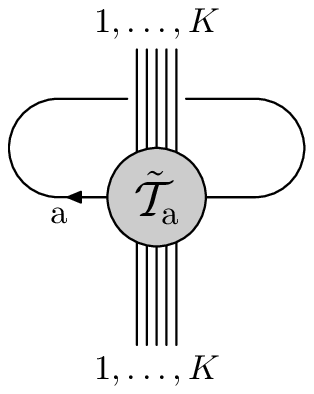}}
\qquad
\caption{Monodromy, transfer and reverse transfer matrices.}
\label{fig:transfer}
\end{figure}
%

\subsection{Transfer Matrix}

The transfer matrix is defined
as the trace of the monodromy matrix over the 
$2|2$-dimensional auxiliary space
\[
\transfer(\xpm{\auxrep})=\str\nolimits_{\auxrep}\mono_{\auxrep}.
\]
Note that the trace can only be invariant under $\alg{h}$ if
the representation acting on the auxiliary space is the
same before and after the scattering. This requires
$P'_{\auxrep}=P_{\auxrep}$ and $K'_{\auxrep}=K_{\auxrep}$ in \eqref{eq:MonoChargeMap}.
Furthermore, the other parameters of the representation, such as the $\xi$'s,
must not change. For full $\alg{h}$-invariance we are led
to the constraints
\[\label{eq:SU22constraint}
1=\prod_{j=1}^K \frac{\xp{j}}{\xm{j}}=\prod_{j=1}^K \xi_j=\prod_{j=1}^K \rho_j=\tau.
\]
This agrees with the conditions given above in \eqref{eq:symbos,eq:symferm}.
Note that for generic $\xpm{\auxrep}$ the individual central charges of the 
sites change according to \eqref{eq:MonoChargeMap}. 
Nevertheless, the tensor product of representations has 
\[
P=\sum_{j=1}^K P_j=0,\qquad
K=\sum_{j=1}^K K_j=0
\]
due to the momentum constraint \eqref{eq:SU22constraint}.
As the map in \eqref{eq:MonoChargeMap} is multiplicative, the
tensor product representation is invariant and so is the transfer matrix.

Let us mention though that we do not have to impose the above constraints for 
a consistent definition of the trace; it will simply fail to preserve the full $\alg{h}$ symmetry.
We will therefore continue to work with the most general set of parameters.

\subsection{Eigenvalues}
\label{sec:transeigen}

The procedure of finding eigenvalues of the transfer matrix is standard;
it can be applied to our model paying attention to the marker fields.

\paragraph{Vacuum Eigenvalue.}

To find the eigenvalues, let us first of all act with the 
transfer matrix on the vacuum state in \eqref{eq:lev2vac}.
This corresponds to the absence of level-II and level-III excitations.
We need the following elements of the scattering matrix 
\<
\smat_{\auxrep j}\state{\phi^1_{\auxrep}\phi^1_j}\eq A_{\auxrep j}\state{\phi^1_j\phi^1_{\auxrep}},
\\
\smat_{\auxrep j}\state{\psi^\alpha_{\auxrep}\phi^1_j}\eq 
L_{\auxrep j}\state{\phi^1_j\psi^\alpha_{\auxrep}}
+K_{\auxrep j}\state{\psi^\alpha_j\phi^1_{\auxrep}},
\nln\nonumber
\smat_{\auxrep j}\state{\phi^2_{\auxrep}\phi^1_j}\eq 
\half (A_{\auxrep j}-B_{\auxrep j})\state{\phi^1_j\phi^2_{\auxrep}}
+\half (A_{\auxrep j}+B_{\auxrep j})\state{\phi^2_j\phi^1_{\auxrep}}
-\half C_{\auxrep j}\varepsilon_{\alpha\beta}\state{\fldZ^-\fldY^+\fldY^+\psi^\alpha_j\psi^\beta_{\auxrep}}.
\>
We start by injecting a particle $\phi^1_{\auxrep}$ 
into the chain from the left. 
Repeated scattering leads to a product of $A_{\auxrep j}$'s. 
When we inject a particle $\psi^\alpha_{\auxrep}$ instead, 
it can either move right through the chain and we obtain 
a product of $L_{\auxrep j}$'s. Alternatively, 
it can be scattered into one of the sites. 
In the latter case we would extract a $\phi^1_{\auxrep}$ from the right
of the chain. This contribution drops out in taking the trace 
over the auxiliary space $\str_{\auxrep}$. Finally, scattering with
a $\phi^2_{\auxrep}$ leads to a product of 
$\half (A_{\auxrep j}-B_{\auxrep j})$'s. The overall eigenvalue of 
the level-II vacuum is
\[
T(\xpm{\auxrep})=
\prod_{j=1}^{K}A_{\auxrep j}
-
\prod_{j=1}^{K}L_{\auxrep j}
-
\prod_{j=1}^{K}\rho_j L_{\auxrep j}
+
\prod_{j=1}^{K}\half 
\rho_j\xi_j^2\, \frac{\xm{j}}{\xp{j}}
\bigbrk{A_{\auxrep j}-B_{\auxrep j}}.
\]
The prefactor $\rho$ in the third term stems from
the further twisting introduced in 
\eqref{eq:twistrho}. 
This twist is not reflected in \tabref{tab:SCoeff}
because it breaks manifest $\alg{su}(2)$ invariance
and would bloat the notation. 
The prefactor $\rho'$ in the fourth term,
cf.~\eqref{eq:rhoprime}, is related to a twist of the
other $\alg{su}(2)$. 
We have to introduce it because $\phi^2$ does not correspond to an
independent excitation but to a composite. In order for the wave function
to be periodic, the rescaling of $\phi^2$ by $\rho'$ must be consistent with 
the rescaling of the components by $\rho\xi^2\xm{}/\xp{}$. 

\paragraph{Analytic Bethe Ansatz.}

We can go on to directly derive the eigenvalues of
the transfer matrix for states with excitations. 
This is a rather tedious and not very illuminating procedure,
but there is a shortcut to obtain the correct expressions:
On the one hand, we may recycle the results of
\cite{Shastry:1988aa,Ramos:1996us,Yue:1996aa} for the Hubbard chain and modify them appropriately.
On the other hand, we can assume that the expression
for the eigenvalue $T(\xpm{\auxrep})$ will lead to 
the Bethe equations via an analytic Bethe ansatz \cite{Reshetikhin:1983vw}. 
Here we shall pursue the second method. 

To complete the analytic structure we note that
the quotient of two summand terms should constitute the r.h.s.~of 
some Bethe equation in \eqref{eq:Bethe23}. Indeed,
\[
\prod_{j=1}^{K}\frac{A_{\auxrep j}}{L_{\auxrep j}}
=
\prod_{j=1}^{K}
\frac{1}{\xi_j}\,
\frac{\xm{\auxrep}-\xp{j}}{\xm{\auxrep}-\xm{j}}\,,
\qquad
\prod_{j=1}^{K}
\frac{1}{\xi_j^2}\, \frac{\xp{j}}{\xm{j}}\,
\frac{2L_{\auxrep j}}{A_{\auxrep j}-B_{\auxrep j}}
=
\prod_{j=1}^{K}
\frac{1}{\xi_j}\,
\frac{1/\xp{\auxrep}-\xp{j}}{1/\xp{\auxrep}-\xm{j}}
\]
are both equivalent to level-II Bethe equations
when $\xm{\auxrep}=y_k$ and $1/\xp{\auxrep}=y_k$, respectively.
Therefore we shall introduce poles at these values of
the spectral parameter $\xpm{\auxrep}$
\[
\prod_{j=1}^N \frac{\ast}{\xm{\auxrep}-y_j}\,,\qquad
\prod_{j=1}^N \frac{\ast}{1-1/\xp{\auxrep}y_j}\,.
\]
The numerators will be determined by the level-III Bethe equation.

\paragraph{Full Eigenvalue.}

Taking a few steps at a time, the eigenvalue of the transfer matrix 
has to take the form
\[\label{eq:trans}
T(\xpm{\auxrep})
=T_1(\xpm{\auxrep})-T_2(\xpm{\auxrep})-T_3(\xpm{\auxrep})+T_4(\xpm{\auxrep})
\]
with
\<\label{eq:transcomp}
T_{1}(\xpm{\auxrep})\eq
\prod_{j=1}^{K}S^0_{\auxrep j}\,\frac{\xm{\auxrep}-\xp{j}}{\xp{\auxrep}-\xm{j}}
\prod_{j=1}^{N}\frac{1}{\xi_{\auxrep}}\,\frac{\xp{\auxrep}-y_j}{\xm{\auxrep}-y_j}
\prod_{j=1}^{M}\frac{1}{\rho_{\auxrep}}\,,
\nln
T_{2}(\xpm{\auxrep})\eq
\prod_{j=1}^{K}S^0_{\auxrep j}\,\xi_j\,\frac{\xm{\auxrep}-\xm{j}}{\xp{\auxrep}-\xm{j}}
\prod_{j=1}^{N}\frac{1}{\xi_{\auxrep}}\,\frac{\xp{\auxrep}-y_j}{\xm{\auxrep}-y_j}
\prod_{j=1}^{M}\frac{1}{\tau\rho_{\auxrep}}\,\frac{\xm{\auxrep}+1/\xm{\auxrep}-w_j-\ihalf g^{-1}}{\xm{\auxrep}+1/\xm{\auxrep}-w_j+\ihalf g^{-1}}\,,
\nln
T_{3}(\xpm{\auxrep})\eq
\prod_{j=1}^{K}S^0_{\auxrep j}\,\rho_j\xi_j\frac{\xm{\auxrep}-\xm{j}}{\xp{\auxrep}-\xm{j}}
\prod_{j=1}^{N}\frac{\tau}{\xi_{\auxrep}}\,\frac{1/\xm{\auxrep}-y_j}{1/\xp{\auxrep}-y_j}
\prod_{j=1}^{M}\frac{1}{\tau\rho_{\auxrep}}\,\frac{\xp{\auxrep}+1/\xp{\auxrep}-w_j+\ihalf g^{-1}}{\xp{\auxrep}+1/\xp{\auxrep}-w_j-\ihalf g^{-1}}\,,
\nln
T_{4}(\xpm{\auxrep})\eq
\prod_{j=1}^{K}S^0_{\auxrep j}\,\rho_j\xi_j^2\,\frac{\xm{j}}{\xp{j}}\,
               \frac{1-1/\xp{\auxrep}\xm{j}}{1-1/\xp{\auxrep}\xp{j}}\,
                \frac{\xm{\auxrep}-\xm{j}}{\xp{\auxrep}-\xm{j}}
\prod_{j=1}^{N}\frac{\tau}{\xi_{\auxrep}}\,\frac{1/\xm{\auxrep}-y_j}{1/\xp{\auxrep}-y_j}
\prod_{j=1}^{M}\frac{1}{\tau^2\rho_{\auxrep}}\,.
\>
The cancellation of poles at $\xm{\auxrep}=y_k$ and $1/\xp{\auxrep}=y_k$
between $T_{1,2}$ and between $T_{3,4}$, respectively,
is equivalent to the level-II Bethe equation. 
Furthermore, the poles at
\[u_{\auxrep}=\xp{\auxrep}+1/\xp{\auxrep}-\ihalf g^{-1}=\xm{\auxrep}+1/\xm{\auxrep}+\ihalf g^{-1}=w_k\]
cancel between $T_{3,4}$ provided that the level-III Bethe equation holds.
Thus, if the Bethe equations hold,
$T(\xpm{\auxrep})$ has poles at positions
determined through the $\xpm{j}$ alone. 

Note that as expected the form is consistent with the transfer matrix for the Hubbard
chain in \cite{Shastry:1988aa,Ramos:1996us,Yue:1996aa}
after the appropriate substitutions of parameters found in \secref{sec:compsr}.

\paragraph{Dualisation.}

In \eqref{eq:BetheDual23} above we displayed
an alternative form of the Bethe equations with
Bethe roots $\tilde y$'s dual to $y$'s. 
We can also write the eigenvalue of the transfer matrix in 
the dual picture. To derive it, it is easiest to use the 
constancy property of the function $Q(y)$ in \eqref{eq:dualityQ}
and demand
$Q(\xp{\auxrep})=Q(\xm{\auxrep})$ as well as 
$Q(1/\xp{\auxrep})=Q(1/\xm{\auxrep})$.
These two relations give alternative forms for the former and the latter two
lines in \eqref{eq:trans}, respectively. The
resulting expression is
\<
T(\xpm{\auxrep})
\eq
-
\prod_{j=1}^{K}S^0_{\auxrep j}\,\frac{\xi_j}{\xi_{\auxrep}}\,
\prod_{j=1}^{\tilde N}\xi_{\auxrep}\,\frac{\xm{\auxrep}-\tilde y_j}{\xp{\auxrep}-\tilde y_j}
\prod_{j=1}^{M}\frac{\xp{\auxrep}}{\tau\xi^2_{\auxrep}\rho_{\auxrep} \xm{\auxrep}}
\nln\earel{}
+
\prod_{j=1}^{K}S^0_{\auxrep j}\,\frac{1}{\xi_{\auxrep}}\,
\frac{\xp{\auxrep}-\xp{j}}{\xp{\auxrep}-\xm{j}}\,
\prod_{j=1}^{\tilde N}\xi_{\auxrep}\,\frac{\xm{\auxrep}-\tilde y_j}{\xp{\auxrep}-\tilde y_j}
\prod_{j=1}^{M}\frac{\xp{\auxrep}}{\xi^2_{\auxrep}\rho_{\auxrep}\xm{\auxrep}}\,\frac{\xp{\auxrep}+1/\xp{\auxrep}-w_j+\ihalf g^{-1}}{\xp{\auxrep}+1/\xp{\auxrep}-w_j-\ihalf g^{-1}}
\nln\earel{}
+
\prod_{j=1}^{K}S^0_{\auxrep j}\,
\frac{\rho_j\xi^2_j\tau \xm{j}}{\xi_{\auxrep}\xp{j}}
\frac{\xp{\auxrep}-\xp{j}}{\xp{\auxrep}-\xm{j}}\,
\prod_{j=1}^{\tilde N}
\frac{\xi_{\auxrep}}{\tau}\,\frac{1/\xp{\auxrep}-\tilde y_j}{1/\xm{\auxrep}-\tilde y_j}
\prod_{j=1}^{M}
\frac{\xp{\auxrep}}{\rho_{\auxrep}\xi^2_{\auxrep}\xm{\auxrep}}
\frac{\xm{\auxrep}+1/\xm{\auxrep}-w_j-\ihalf g^{-1}}{\xm{\auxrep}+1/\xm{\auxrep}-w_j+\ihalf g^{-1}}
\nln\earel{}
-
\prod_{j=1}^{K}S^0_{\auxrep j}\,
\frac{\rho_j\xi_j\tau}{\xi_{\auxrep}}\,
\frac{\xp{\auxrep}-\xp{j}}{\xp{\auxrep}-\xm{j}}\,
\frac{1-1/\xm{\auxrep}\xp{j}}{1-1/\xm{\auxrep}\xm{j}}\,
\prod_{j=1}^{\tilde N}
\frac{\xi_{\auxrep}}{\tau}\,\frac{1/\xp{\auxrep}-\tilde y_j}{1/\xm{\auxrep}-\tilde y_j}
\prod_{j=1}^M
\frac{\xp{\auxrep}\tau}{\xi^2_{\auxrep}\rho_{\auxrep}\xm{\auxrep}}
\,.
\>

\subsection{Reverse Transfer Matrix}
\label{sec:reverse}

We can define a reverse transfer matrix
$\tilde\tmat(\xpm{\auxrep})$ by scattering an auxiliary particle 
with the chain, but in the opposite direction,
cf.~\figref{fig:transfer}
\[\tilde\tmat(\xpm{\auxrep})=\str\nolimits_{\auxrep} 
\smat_{1\auxrep}\smat_{2\auxrep}\cdots
\smat_{K\auxrep}.
\]
By the same argument as above we obtain its eigenvalue 
\<\label{eq:transrev}
\tilde T(\xpm{\auxrep})
\eq
+
\prod_{j=1}^{K}S^0_{j\auxrep}\,\frac{\xp{\auxrep}-\xm{j}}{\xm{\auxrep}-\xp{j}}
\prod_{j=1}^{N}\xi_{\auxrep}\,\frac{\xm{\auxrep}-y_{j}}{\xp{\auxrep}-y_{j}}
\prod_{j=1}^{M}\rho_{\auxrep}
\nln\earel{}
-
\prod_{j=1}^{K}S^0_{j\auxrep}\,\frac{1}{\xi_j}\,\frac{\xp{\auxrep}-\xp{j}}{\xm{\auxrep}-\xp{j}}
\prod_{j=1}^{N}\xi_{\auxrep}\,\frac{\xm{\auxrep}-y_{j}}{\xp{\auxrep}-y_{j}}
\prod_{j=1}^{M}\tau\rho_{\auxrep}\,\frac{\xp{\auxrep}+1/\xp{\auxrep}-w_{j}+\ihalf g^{-1}}{\xp{\auxrep}+1/\xp{\auxrep}-w_{j}-\ihalf g^{-1}}
\nln\earel{}
-
\prod_{j=1}^{K}S^0_{j\auxrep}\,\frac{1}{\rho_j\xi_j}\,\frac{\xp{\auxrep}-\xp{j}}{\xm{\auxrep}-\xp{j}}
\prod_{j=1}^{N}\frac{\xi_{\auxrep}}{\tau}\,\frac{1-1/\xp{\auxrep}y_{j}}{1-1/\xm{\auxrep}y_{j}}
\prod_{j=1}^{M}\tau\rho_{\auxrep}\,\frac{\xm{\auxrep}+1/\xm{\auxrep}-w_{j}-\ihalf g^{-1}}{\xm{\auxrep}+1/\xm{\auxrep}-w_{j}+\ihalf g^{-1}}
\nln\earel{}
+
\prod_{j=1}^{K}S^0_{j\auxrep}\,\frac{1}{\rho_j\xi^2_j}\,\frac{\xp{j}}{\xm{j}}\,\frac{1-1/\xm{\auxrep}\xp{j}}{1-1/\xm{\auxrep}\xm{j}}\,
                                      \frac{\xp{\auxrep}-\xp{j}}{\xm{\auxrep}-\xp{j}}
\prod_{j=1}^{N}\frac{\xi_{\auxrep}}{\tau}\,\frac{1-1/\xp{\auxrep}y_{j}}{1-1/\xm{\auxrep}y_{j}}
\prod_{j=1}^{M}\tau^2\rho_{\auxrep}\,\,.
\>

The reverse transfer matrix is actually closely 
related to the forward transfer matrix after inverting the spectral parameter $\xpm{\auxrep}$.
It obeys the relation
\[\label{eq:revrel}
\tilde T(1/\xpm{\auxrep})=
T(\xpm{\auxrep})\,\xi_{\auxrep}^{2N}\rho_{\auxrep}^{2M}\tau^{2M-N}
\prod_{j=1}^{K}X_{j\auxrep},
\]
where curiously $X_{12}$ is precisely the function
\eqref{eq:crossrho} found in the context of crossing symmetry,
cf.~\cite{Janik:2006dc} and \secref{sec:crossing}.
Thus, in a crossing-symmetric model, 
the transfer matrix and its reverse have 
precisely the same analytic structure, up to inversion of
the spectral parameter. However, special care 
concerning the Riemann sheets of $S^0_{j\auxrep}$
may have to be taken in order to ensure $X_{j\auxrep}=1$.

Using the function $X_{12}$, we can also rewrite the transfer matrix \eqref{eq:trans}
in a very symmetric form as 
\<
T(\xpm{\auxrep})
\eq
+
\frac{\tau^{N/2-M}}{\xi_{\auxrep}^{N}\rho_{\auxrep}^{M}}
\prod_{j=1}^{K}S^0_{\auxrep j}\,\frac{\xm{\auxrep}-\xp{j}}{\xp{\auxrep}-\xm{j}}
\prod_{j=1}^{N}\frac{1}{\sqrt{\tau}}\,\frac{\xp{\auxrep}-y_j}{\xm{\auxrep}-y_j}
\prod_{j=1}^{M}\tau
\nln\earel{}
-
\frac{\tau^{N/2-M}}{\xi_{\auxrep}^{N}\rho_{\auxrep}^{M}}
\prod_{j=1}^{K}S^0_{\auxrep j}\,\xi_j\,\frac{\xm{\auxrep}-\xm{j}}{\xp{\auxrep}-\xm{j}}
\prod_{j=1}^{N}\frac{1}{\sqrt{\tau}}\,\frac{\xp{\auxrep}-y_j}{\xm{\auxrep}-y_j}
\prod_{j=1}^{M}\frac{\xm{\auxrep}+1/\xm{\auxrep}-w_j-\ihalf g^{-1}}{\xm{\auxrep}+1/\xm{\auxrep}-w_j+\ihalf g^{-1}}
\nln\earel{}
-
\frac{\tau^{N/2-M}}{\xi_{\auxrep}^{N}\rho_{\auxrep}^{M}}
\prod_{j=1}^{K}\frac{S^0_{j\bar\auxrep}}{\xi_jX_{j\auxrep}}\,\frac{1/\xp{\auxrep}-\xp{j}}{1/\xm{\auxrep}-\xp{j}}\,
\prod_{j=1}^{N}\sqrt{\tau}\,\frac{1/\xm{\auxrep}-y_j}{1/\xp{\auxrep}-y_j}
\prod_{j=1}^{M}\frac{\xp{\auxrep}+1/\xp{\auxrep}-w_j+\ihalf g^{-1}}{\xp{\auxrep}+1/\xp{\auxrep}-w_j-\ihalf g^{-1}}
\nln\earel{}
+
\frac{\tau^{N/2-M}}{\xi_{\auxrep}^{N}\rho_{\auxrep}^{M}}
\prod_{j=1}^{K}\frac{S^0_{j\bar\auxrep}}{X_{j\auxrep}}\frac{1/\xp{\auxrep}-\xm{j}}{1/\xm{\auxrep}-\xp{j}}\,
\prod_{j=1}^{N}\sqrt{\tau}\,\frac{1/\xm{\auxrep}-y_j}{1/\xp{\auxrep}-y_j}
\prod_{j=1}^{M}\frac{1}{\tau}
\,.
\>
Here the latter two terms equal,
up to the prefactor and factors of $X_{j\auxrep}$, 
the inverse of the former two when $\xpm{\auxrep}$ is replaced
by its inverse.

\subsection{Transfer Matrix from Diagonalised Scattering}

\begin{figure}\centering
\parbox[c]{2cm}{\centering\includegraphics[scale=0.6]{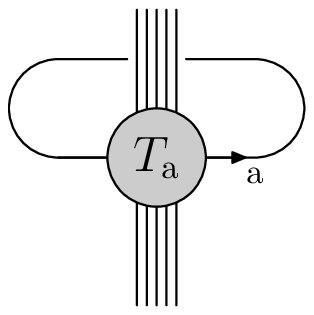}}
$=$
\parbox[c]{2cm}{\centering\includegraphics[scale=0.6]{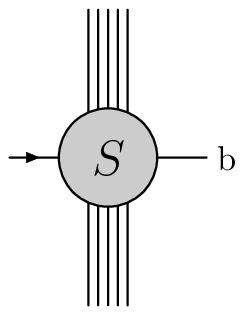}}
$-$
\parbox[c]{2cm}{\centering\includegraphics[scale=0.6]{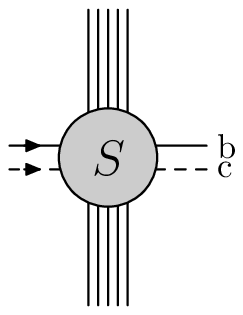}}
$-$
\parbox[c]{2cm}{\centering\includegraphics[scale=0.6]{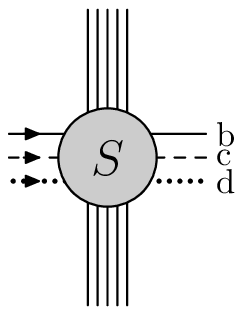}}
$+$ 
\parbox[c]{2cm}{\centering\includegraphics[scale=0.6]{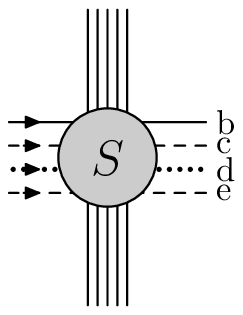}}
\caption{Eigenvalue of the fundamental transfer matrix $T(\xpm{\auxrep})$ from diagonalised excitation
scattering. 
The diagonalised excitations I, II, III constituting the fundamental multiplet are  
depicted as solid, dashed and dotted lines, respectively. Spectral parameters for particles
b, c, d, e are 
$\xpm{}=\xpm{\auxrep}$, $y=\xm{\auxrep}$, $w=u_{\auxrep}$, $y=1/\xp{\auxrep}$,
respectively. Compare to \protect\eqref{eq:transeigenscat}.}
\label{fig:fundtrans}
\end{figure}

The fundamental transfer matrix in \eqref{eq:trans} can be written in
terms of elements of the diagonalised S-matrix \eqref{eq:sdiagelem} as follows
\<\label{eq:transeigenscat}
T(\xpm{\auxrep})
\eq
+
\prod_{j=1}^{K}S\lvl{I,I}_{\auxrep j}(\xpm{\auxrep},\xpm{j})
\prod_{j=1}^{N}S\lvl{I,II}(\xpm{\auxrep},y_j)
\prod_{j=1}^{M}S\lvl{I,III}(\xpm{\auxrep},w_j)
\nln\earel{}
-
\prod_{j=1}^{K}
  S\lvl{I,I}_{\auxrep j}(\xpm{\auxrep},\xpm{j})
  S\lvl{II,I}_{\auxrep j}(\xm{\auxrep},\xpm{j})
\prod_{j=1}^{N}
  S\lvl{I,II}_{\auxrep j}(\xpm{\auxrep},y_{j})
  S\lvl{II,II}_{\auxrep j}(\xm{\auxrep},y_{j})
\prod_{j=1}^{M}\ldots
\nln\earel{}
-
\prod_{j=1}^{K}
  S\lvl{I,I}_{\auxrep j}(\xpm{\auxrep},\xpm{j})
  S\lvl{II,I}_{\auxrep j}(\xm{\auxrep},\xpm{j})
  S\lvl{III,I}_{\auxrep j}(u_{\auxrep},\xpm{j})
\prod_{j=1}^{N}\ldots
\prod_{j=1}^{M}\ldots
\\\nonumber\earel{}
+
\prod_{j=1}^{K}
  S\lvl{I,I}_{\auxrep j}(\xpm{\auxrep},\xpm{j})
  S\lvl{II,I}_{\auxrep j}(\xm{\auxrep},\xpm{j})
  S\lvl{III,I}_{\auxrep j}(u_{\auxrep},\xpm{j})
  S\lvl{II,I}_{\auxrep j}(1/\xp{\auxrep},\xpm{j})
\prod_{j=1}^{N}\ldots
\prod_{j=1}^{M}\ldots
\,.
\>
This gives us a way of expressing the four components of a fundamental 
multiplet in terms of elementary excitations of type I, II and III:
The transfer matrix can be viewed as scattering
a spin chain state with a fundamental multiplet 
and then summing over components. 
The first line corresponds to the first component (bosonic)
which is represented by a type-I excitation 
with spectral parameter $\xpm{}=\xpm{\auxrep}$. 
The second component (fermionic) has two excitations: the same
type-I excitation and a type-II excitation 
with spectral parameter $y=\xm{\auxrep}$.
The third component (fermionic) has in addition a type-III
excitation with spectral parameter
$w=u_{\auxrep}$ which is defined as 
\[u_{\auxrep}=
\xp{\auxrep}+1/\xp{\auxrep}-\ihalf g^{-1}
=\xm{\auxrep}+1/\xm{\auxrep}+\ihalf g^{-1}\,.
\]
The last component (bosonic) has another 
additional type-II excitation with 
parameter $y=1/\xp{\auxrep}$.
The transfer matrix can thus be represented graphically as
in \figref{fig:fundtrans}.

Likewise, the reverse transfer matrix is given by scattering 
(in the reverse direction) with the elementary excitations (in order of appearance)
$\xpm{}=\xpm{\auxrep}$, $y=\xp{\auxrep}$, $w=u_{\auxrep}$
and $y=1/\xm{\auxrep}$,
i.e.
\[
\tilde T(\xpm{\auxrep})
=+
\prod_{j=1}^{K}S\lvl{I,I}_{j\auxrep}(\xpm{j},\xpm{\auxrep})
\prod_{j=1}^{N}S\lvl{II,I}(y_j,\xpm{\auxrep})
\prod_{j=1}^{M}S\lvl{III,I}(w_j,\xpm{\auxrep})
\mp\ldots\,. 
\]
The relation between both transfer matrices is ensured by 
the following identities
(used in App.~D of \cite{Beisert:2005tm})
\[\arraycolsep0pt\begin{array}[b]{lrlrlrlrlrcl}
S\lvl{I,I}_{12}&(\xpm{j},\xpm{2})&\,
S\lvl{I,II}_{12}&(\xpm{1},\xm{2})&\,
S\lvl{I,III}_{12}&(\xpm{1},u_{2})&\,
S\lvl{I,II}_{12}&(\xpm{1},1/\xp{2})&\,
S\lvl{I,I}_{1\bar2}&(\xpm{1},1/\xpm{2})
\eq
X_{12},
\\[3pt]
S\lvl{II,I}_{12}&(y_1,\xpm{2})&\,
S\lvl{II,II}_{12}&(y_1,\xm{2})&\,
S\lvl{II,III}_{12}&(y_1,u_{2})&\,
S\lvl{II,II}_{12}&(y_1,1/\xp{2})&\,
S\lvl{II,I}_{1\bar 2}&(y_1,1/\xpm{2})
\eq \tau^{-1}\xi_2^2,
\\[3pt]
S\lvl{III,I}_{12}&(w_1,\xpm{2})&\,
S\lvl{III,II}_{12}&(w_1,\xm{2})&\,
S\lvl{III,III}_{12}&(w_1,u_{2})&\,
S\lvl{III,II}_{12}&(w_1,1/\xp{2})&\,
S\lvl{III,I}_{1\bar 2}&(w_1,1/\xpm{2})
\eq\tau^2\rho^2_2.
\end{array}\]
%

\subsection{Quantum Characteristic Function}
\label{sec:qchar}

Transfer matrices can be constructed for various
representations of the symmetry algebra. 
Of particular interest are the $m$-fold symmetric and antisymmetric products of the fundamental
representation
\[
\srep{m-1,0}\quad\mbox{and}\quad
\srep{0,m-1}.
\]
More explicitly, the former are the representations which appear for the bound
states discussed in \cite{Dorey:2006dq}. 
The latter appear in the decomposition of the non-compact spin representation 
of $\superN=4$ SYM, see \secref{sec:psu224} for further details.
Their central charges can be parametrised by $\xexp{+m}{}$ and $\xexp{-m}{}$
obeying the relation
\[
\xexp{+m}{}+\frac{1}{\xexp{+m}{}}-\xexp{-m}{}-\frac{1}{\xexp{-m}{}}=\frac{mi}{g}\,.
\]
Consequently, we expect the transfer matrices
to depend primarily on these parameters,
$\transfer_{\srep{m-1,0}}(\xexp{\pm m}{\auxrep})$
and 
$\transfer_{\srep{0,m-1}}(\xexp{\pm m}{\auxrep})$.

A useful object for the construction of transfer matrix eigenvalues 
in various symmetric representation is an operator
used in \cite{Krichever:1997qd}
in the context of the Baxter and Hirota equations. 
In the author's ignorance of an established name for this operator
we shall call it the quantum characteristic function $\qdet$.
Roughly speaking we may define it as
$\qdet_{\auxrep}=\sdet_{\auxrep}(\ushift^{-2}_{\auxrep}-\mono_{\auxrep})$,
i.e.~the characteristic function of the monodromy matrix $\mono$.
Note that $\ushift$ is in fact a shift operator for the spectral parameter
which is required for proper implementation of fusion.

Here we will make an educated guess on the eigenvalue $\qdetev_{\auxrep}$ 
of the quantum characteristic function 
of the present model. It generalises the proposal in \cite{Beisert:2005di} for 
supersymmetric spin chains and takes the form
\[\label{eq:qdet}
\qdetev_{\auxrep}
=
\bigbrk{1-\ushift_{\auxrep}T_{1,\auxrep}\ushift_{\auxrep}}
\bigbrk{1-\ushift_{\auxrep}T_{2,\auxrep}\ushift_{\auxrep}}^{-1}
\bigbrk{1-\ushift_{\auxrep}T_{3,\auxrep}\ushift_{\auxrep}}^{-1}
\bigbrk{1-\ushift_{\auxrep}T_{4,\auxrep}\ushift_{\auxrep}}.
\]
Here $T_{n,\auxrep}$ are the four terms which constitute the eigenvalue of the fundamental 
transfer matrix \eqref{eq:trans,eq:transcomp}.
The shift operator $\ushift_{\auxrep}$ acts 
on the spectral parameter $\xexp{m}{\auxrep}$
by shifting its index by one unit
\[\label{eq:shiftdef}
\ushift_{\auxrep}^{}\xexp{m}{\auxrep}\ushift_{\auxrep}^{-1}=\xexp{m+1}{\auxrep}.
\]
The relation between any two $\xexp{m}{\auxrep}$ and $\xexp{n}{\auxrep}$
and the parameter $u_{\auxrep}$ is defined as 
\[\label{eq:udef}
\xexp{m}{\auxrep}+\frac{1}{\xexp{m}{\auxrep}}-\frac{mi}{2g}
=\xexp{n}{\auxrep}+\frac{1}{\xexp{n}{\auxrep}}-\frac{ni}{2g}
=u_{\auxrep}\,.
\]
This shows that $\ushift_{\auxrep}$ essentially shifts $u_{\auxrep}$ by one unit of
$i/2g$
\[\ushift_{\auxrep}^{}u^{}_{\auxrep}\ushift_{\auxrep}^{-1}=u_{\auxrep}+i/2g.\]
Despite this simple action on $u_{\auxrep}$, the action on $\xexp{m}{\auxrep}$
is substantially more complex:
For a given $\xexp{m}{\auxrep}$ there are in general two
solutions for $\xexp{n}{\auxrep}$. Therefore, in order to define the shift operator 
in \eqref{eq:shiftdef} unambiguously, all the 
parameters $\xexp{m}{\auxrep}$ have to be fixed 
subject to the constraint \eqref{eq:udef}. As explained 
in \cite{Beisert:2006ib}, the set of solutions to \eqref{eq:udef}
forms an infinite-genus surface. In other words,
the operator $\qdetev$ is a function on the infinite-genus surface
defined by \eqref{eq:udef}.

\paragraph{Antisymmetric Representations.}

Eigenvalues of transfer matrices in totally antisymmetric representations
$T_{\srep{0,m-1}}(\xexp{\pm m}{\auxrep})$, cf.~\eqref{eq:symrep}, can be obtained 
by expanding the quantum characteristic function. We will assume that
the terms $T_k$ in \eqref{eq:qdet} are small compared to $1$. 
The expansion then takes the form 
\[\label{eq:genanti}
\qdetev_{\auxrep}=\sum_{m=0}^\infty (-1)^m \ushift_{\auxrep}^m T_{\srep{0,m-1}}(\xexp{\pm m}{\auxrep})\ushift_{\auxrep}^m
\]
from which the spectra of transfer matrices can be read off.
Explicitly, we find the following expression
\<\label{eq:transanti}
\earel{}
T_{\srep{0,m-1}}(\xexp{\pm m}{\auxrep})
\nln
\eq
(-1)^m
\prod_{j=1}^{K}\prod_{k=1}^m S^0(\xexp{+m-2k+2,+m-2k}{\auxrep},\xpm{j})\times
\nl
\Biggl(\mathord{}+
T_{[n]}(u_{\auxrep})
\prod_{j=1}^{K}\frac{\xexp{-m}{\auxrep}-\xm{j}}{\xexp{+m}{\auxrep}-\xm{j}}
\prod_{j=1}^{N}\frac{u_{\auxrep}-y_j-1/y_j}{(\xexp{-m}{\auxrep}-y_j)(1-1/\xexp{+m}{\auxrep}y_j)}
\nlnum\quad
-
T_{[n-1]}(u_{\auxrep}+\ihalf g^{-1})
\prod_{j=1}^{K}\frac{\xexp{-m}{\auxrep}-\xp{j}}{\xexp{+m}{\auxrep}-\xm{j}}
\prod_{j=1}^{N}\frac{u_{\auxrep}-y_j-1/y_j+\ihalf g^{-1}}{(\xexp{-m}{\auxrep}-y_j)(1-1/\xexp{+m}{\auxrep}y_j)}
\nl\quad
-
T_{[n-1]}(u_{\auxrep}-\ihalf g^{-1})
\prod_{j=1}^{K}\frac{\xexp{-m}{\auxrep}-\xm{j}}{\xexp{+m}{\auxrep}-\xm{j}}\,
\frac{1/\xexp{+m}{\auxrep}-\xm{j}}{1/\xexp{+m}{\auxrep}-\xp{j}}
\prod_{j=1}^{N}\frac{u_{\auxrep}-y_j-1/y_j-\ihalf g^{-1}}{(\xexp{-m}{\auxrep}-y_j)(1-1/\xexp{+m}{\auxrep}y_j)}
\nl\quad
+
T_{[n-2]}(u_{\auxrep})
\prod_{j=1}^{K}
\frac{\xexp{-m}{\auxrep}-\xp{j}}{\xexp{+m}{\auxrep}-\xm{j}}\,
\frac{1/\xexp{+m}{\auxrep}-\xm{j}}{1/\xexp{+m}{\auxrep}-\xp{j}}
\prod_{j=1}^{N}\frac{u_{\auxrep}-y_j-1/y_j}{(\xexp{-m}{\auxrep}-y_j)(1-1/\xexp{+m}{\auxrep}y_j)}
\Biggr).\nonumber
\>
The symbols $T_{[n]}$ represent standard $\alg{su}(2)$ transfer matrices
in spin-$(\half n)$ representations
\<\label{eq:transsu2}
T_{[n]}(u_{\auxrep})\eq
\sum_{k=0}^n
\prod_{j=1}^{N}\frac{u_{\auxrep}-y_j-1/y_j+(n-2k)\ihalf g^{-1}}{u_{\auxrep}-y_j-1/y_j}
\nlnum\nonumber\qquad\times
\prod_{j=1}^{M}
\frac{u_{\auxrep}-w_j-(n+1)\ihalf g^{-1}}{u_{\auxrep}-w_j+(n-1-2k)\ihalf g^{-1}}\,
\frac{u_{\auxrep}-w_j+(n+1)\ihalf g^{-1}}{u_{\auxrep}-w_j+(n+1-2k)\ihalf g^{-1}}\,.
\>
Note the the structure of the transfer matrix eigenvalue \eqref{eq:transanti} 
follows the $\alg{su}(2)\times\alg{su}(2)$ decomposition in \eqref{eq:symrep}.
We furthermore observe that $T_{\srep{0,m-1}}(\xexp{\pm m}{\auxrep})$
depends on $\xexp{\pm m}{\auxrep}$ only.
All the dependence on $\xexp{k}{\auxrep}$ with $|k|<m$ 
is in the form $\xexp{k}{\auxrep}+1/\xexp{k}{\auxrep}=u_{\auxrep}+\ihalf kg^{-1}$.
Thus the kinematic space of each of these transfer matrices is a torus 
(with modulus depending on $m$ and $g$).
Contributions from the undetermined factors $S^0_{12}$ may however spoil this rule.

\paragraph{Conjugate Representations.}

If we decide to consider the $T_k$ in \eqref{eq:qdet} to
be large compared to $1$, we obtain an expansion in terms
of the reverse transfer matrix eigenvalues
$\tilde T_{\srep{0,m-1}}(\xexp{\pm m}{\auxrep})$.
The first two terms in the expansion read
\[\label{eq:genconj}
\qdetev_{\auxrep}
=
\prod_{j=1}^{K}
\frac{\xexp{0}{\auxrep}-\xp{j}}{\xexp{0}{\auxrep}-\xm{j}}\,
\frac{1/\xexp{0}{\auxrep}-\xm{j}}{1/\xexp{0}{\auxrep}-\xp{j}}
-
\ushift_{\auxrep}^{-1}
\prod_{j=1}^{K}
\frac{\xm{\auxrep}-\xp{j}}{\xm{\auxrep}-\xm{j}}\,
\frac{1/\xm{\auxrep}-\xm{j}}{1/\xp{\auxrep}-\xm{j}}\,
\tilde T(\xpm{\auxrep})\,
\ushift_{\auxrep}^{-1}
+\ldots\,.
\]
The first term might be interpreted as the quantum determinant. 
For the higher representations there are some similar prefactors
which are yet to be interpreted.

\paragraph{Symmetric Representations.}

Eigenvalues of transfer matrices in totally symmetric representations
$T_{\srep{m-1,0}}(\xexp{\pm m}{\auxrep})$ can be obtained 
by expanding the inverse of the quantum characteristic function. 
Under the assumption that
the $T_k$ in \eqref{eq:qdet} are small compared to $1$,
the expansion takes the form 
\[\label{eq:gensym}
\qdetev^{-1}_{\auxrep}=\sum_{m=0}^\infty  \ushift_{\auxrep}^m T_{\srep{m-1,0}}(\xexp{\pm m}{\auxrep})\ushift_{\auxrep}^m.
\]
Likewise, by assuming that the $T_k$ are large compared to
one, we obtain the reverse transfer matrix eigenvalues
$\tilde T_{\srep{m-1,0}}(\xexp{\pm m}{\auxrep})$ 
as expansion coefficients.

\paragraph{Fusion.}

A related issue is fusion of transfer matrices.
\cite{Kulish:1986bb,Jimbo:1987aa}
Let us expand the identity $\qdetev_{\auxrep}\qdetev^{-1}_{\auxrep}=1$
using the relations \eqref{eq:genanti,eq:gensym}.
At second order we find a relation between the eigenvalues of transfer matrices 
in different representations
\[
T_{\srep{0,0}}(\xpm{1})\,T_{\srep{0,0}}(\xpm{2})=T_{\srep{1,0}}(\xp{1},\xm{2})+T_{\srep{0,1}}(\xp{1},\xm{2})
\qquad
\mbox{for }\xm{1}=\xp{2}.
\]
This equation is related to the tensor product 
\eqref{eq:TensorSquare} and multiplet splitting 
\eqref{eq:Splitting}
\[
\srep{0,0}\otimes\srep{0,0}=\lrep{0,0}=\srep{1,0}\oplus\srep{0,1}
\qquad
\mbox{for }\xm{1}=\xp{2}.
\]
Note that when we set $\xm{1}=\xp{2}$ in 
$T_1(\xpm{1})\,T_1(\xpm{2})$
all the terms involving $\xm{1}$ and $\xp{2}$ 
can be reexpressed using $\xp{1}$ and $\xm{2}$
(ignoring those from the undetermined factors $S^0_{12}$).

\subsection{Analytic Structure}

Let us investigate the analytic structure of the transfer matrix eigenvalue
as a function of the spectral parameter $\xpm{\auxrep}$.

\paragraph{Redefinition.}

The main complication is that $T(\xpm{\auxrep})$ depends
on the phase factor $S^0_{\auxrep j}$ on which we would like to make
no assumptions in this paper. 
Therefore we shall multiply $T(\xpm{\auxrep})$ by some function
of the external parameters $\xpm{j}$ which removes the phase factor as well as a couple
of poles. A useful redefinition is the following
\[
t(\xpm{\auxrep})=
T(\xpm{\auxrep})
\prod_{j=1}^{K}S^0_{j\auxrep}(\xpm{j},\xpm{\auxrep})
\lrbrk{1-1/\xp{\auxrep}\xp{j}}\frac{\xp{\auxrep}-\xm{j}}{\xm{\auxrep}-\xm{j}}\,.
\]
The redefined transfer matrix is the following rational function
\<\label{eq:transredef}
t(\xpm{\auxrep})\eq
+
\prod_{j=1}^{K}\lrbrk{1-1/\xp{\auxrep}\xp{j}}
\frac{\xm{\auxrep}-\xp{j}}{\xm{\auxrep}-\xm{j}}
\prod_{j=1}^{N}\frac{1}{\xi_{\auxrep}}\,\frac{\xp{\auxrep}-y_j}{\xm{\auxrep}-y_j}
\prod_{j=1}^{M}\frac{1}{\rho_{\auxrep}}
\nln\earel{}
-
\prod_{j=1}^{K}\xi_j\lrbrk{1-1/\xp{\auxrep}\xp{j}}
\prod_{j=1}^{N}\frac{1}{\xi_{\auxrep}}\,\frac{\xp{\auxrep}-y_j}{\xm{\auxrep}-y_j}
\prod_{j=1}^{M}\frac{1}{\tau\rho_{\auxrep}}\,\frac{\xm{\auxrep}+1/\xm{\auxrep}-w_j-\ihalf g^{-1}}{\xm{\auxrep}+1/\xm{\auxrep}-w_j+\ihalf g^{-1}}
\nln\earel{}
-
\prod_{j=1}^{K}\rho_j\xi_j\lrbrk{1-1/\xp{\auxrep}\xp{j}}
\prod_{j=1}^{N}\frac{\tau}{\xi_{\auxrep}}\,\frac{1-1/\xm{\auxrep}y_j}{1-1/\xp{\auxrep}y_j}
\prod_{j=1}^{M}\frac{1}{\tau\rho_{\auxrep}}\,\frac{\xp{\auxrep}+1/\xp{\auxrep}-w_j+\ihalf g^{-1}}{\xp{\auxrep}+1/\xp{\auxrep}-w_j-\ihalf g^{-1}}
\nln\earel{}
+
\prod_{j=1}^{K}\rho_j\xi_j^2\,
\frac{\xm{j}}{\xp{j}}\lrbrk{1-1/\xp{\auxrep}\xm{j}}
\prod_{j=1}^{N}\frac{\tau}{\xi_{\auxrep}}\,\frac{1-1/\xm{\auxrep}y_j}{1-1/\xp{\auxrep}y_j}
\prod_{j=1}^{M}\frac{1}{\tau^2\rho_{\auxrep}}
\,,
\>
and we can now investigate its singularities.
The new transfer matrix has a $K$-fold pole at $(0,0)$,
a $(K-N)$-fold pole at $(0,\infty)$ and
a $N$-fold pole at $(\infty,0)$.
In addition, there are $K$ poles at the positions $(\xp{j},\xm{j})$ 
which originate in the first term only.
By construction the remaining poles at $\xm{\auxrep}=1/y_j$, $\xp{\auxrep}=y_j$
and $\xm{\auxrep}+1/\xm{\auxrep}+\ihalf g^{-1}=\xp{\auxrep}+1/\xp{\auxrep}-\ihalf g^{-1}=w_j$
cancel out for periodic eigenstates by means of the Bethe equations, see \secref{sec:transeigen}.
As a rational function $t(\xpm{\auxrep})$ has the same number
of poles and zeros, namely $3K$, but their positions
are not immediately related to the $\xpm{j}$.

\paragraph{Reverse Transfer Matrix.}

We can also redefine the reverse transfer matrix 
\[
\tilde t(\xpm{\auxrep})=
\tilde T(\xpm{\auxrep})
\prod_{j=1}^{K}S^0_{j\auxrep}(\xpm{j},\xpm{\auxrep})
\lrbrk{1-1/\xm{\auxrep}\xm{j}}\frac{\xm{\auxrep}-\xp{j}}{\xp{\auxrep}-\xp{j}}\,.
\]
It has a very similar structure of poles as the forward transfer matrix:
There is a $K$-fold pole at $(0,0)$,
a $N$-fold pole at $(0,\infty)$ and
a $(K-N)$-fold pole at $(\infty,0)$.
In addition, there are $K$ poles at the positions $(\xp{j},\xm{j})$ 
which originate in the first term only.
In fact, the similarity of the analytic structures is related to the identity \eqref{eq:revrel}
which now reads
\[
\tilde t(1/\xpm{\auxrep})
=
t(\xpm{\auxrep})\,\xi_{\auxrep}^{2N}\rho_{\auxrep}^{2M}\tau^{2M-N}
\prod_{j=1}^{K}
\frac{1}{\rho_j\xi_j^2}\,
\frac{\xp{j}}{\xm{j}}\,
\frac{\xm{\auxrep}-\xm{j}}{1/\xp{\auxrep}-\xp{j}}\,.
\]
%

\paragraph{Symmetry Charges.}
The eigenvalues of the Cartan generators
can usually be read off from the eigenvalue of the transfer matrix.
In order to obtain the maximum $\alg{su}(2|2)$ symmetry for the 
periodic eigenstates, let us set $\rho=\xi=\tau=1$
as well as $\prod_j \xp{j}/\xm{j}$.
We shall furthermore split up the transfer matrix 
into four components as in \eqref{eq:transcomp}.
The expansion of these $T_k$ around the point $(\infty,\infty)$ is given by
(we assume that there are no contributions from the phase factor $S^0_{12}$
at this order)
\[
T_{1,2,3,4}(\xpm{\auxrep})=
1+
\frac{i}{g\xpm{\auxrep}}
\left\{
\begin{array}{l}
(-C-\sfrac{3}{2}K+N)\\
(-K+N-M)\\
(-K+M)\\
(-C+\sfrac{1}{2}K)
\end{array}
\right\}+\ldots,
\]
where $C$ is the overall central charge
\[
C=
\sum_{j=1}^{K}
C_k=
\sum_{j=1}^{K}
\lrbrk{\frac{1}{2}+\frac{ig}{\xp{j}}-\frac{ig}{\xm{j}}}\,.
\]
The Cartan charge eigenvalues of the two $\alg{su}(2)$ subalgebras
read $K-N$ and $N-2M$. 
The expansion of the transfer matrix itself yields
\[
T(\xpm{\auxrep})=
-\frac{i}{g\xpm{\auxrep}}(2C-K)+\ldots\,.
\]
Note that the combination $2C-K$ equals the $\superN=4$ gauge theory
anomalous dimension.

\section{Models with $\alg{psu}(2,2|4)$ Symmetry}
\label{sec:psu224}

Let us now proceed to the planar AdS/CFT correspondence
of $\superN=4$ gauge theory and string theory on $AdS_5\times S^5$
\cite{Maldacena:1998re,Gubser:1998bc,Witten:1998qj}.
The symmetry of these systems is $\alg{psu}(2,2|4)$. 
In this section we shall discuss transfer matrices and the 
(analytic) Bethe ansatz for models with $\alg{psu}(2,2|4)$ symmetry.
The discussion can merely be considered a sketch of the
integrable structures while
further work is clearly required for a full understanding.

\subsection{Particles and Residual Symmetry}

Both models, gauge and string theory, may be considered,
at least in perturbation theory, 
as particle models with
$8|8$ flavours on a circle \cite{Berenstein:2002jq}.
Out of the $30|32$ generators of $\alg{psu}(2,2|4)$,
there are $8|8$ generators which create particles
with zero momentum from the vacuum and $8|8$ generators
which annihilate these. 
The remaining $14|16$ generators form the algebra 
$\Reals\ltimes\bigbrk{\alg{psu}(2|2)\times\alg{psu}(2|2)'}\ltimes \Reals$.
Here the outer automorphism is the generator $\gen{B}$ 
discussed in \secref{sec:Alg.Ext} and the central charge is $\gen{C}$.
The central charge $\gen{C}$ is what we shall consider to be the Hamiltonian
of the system. The determination of its spectrum is our goal.
Note that in the full $\alg{psu}(2,2|4)$ algebra, 
the Hamiltonian is one of the Cartan generators obeying
non-trivial commutation relations. This is quite different
from many other particle or spin chain models where 
the Hamiltonian commutes with the symmetry algebra.

In addition to these generators, 
we shall introduce two more central charges $\gen{P},\gen{K}$ 
whose action must vanish on physical states. 
These four generators are shared between the $\alg{psu}(2|2)$ factors
so that each factor extends to the algebra $\gen{h}_+$
discussed in \secref{sec:Alg.Ext}.
The residual algebra which leaves the number of particles invariant is thus
\[
\Reals\ltimes\bigbrk{\alg{psu}(2|2)\times\alg{psu}(2|2)'}\ltimes \Reals^3.
\]
The prime shall be used in this section 
to distinguish the two factors of $\alg{psu}(2|2)$.
The particles transform as fundamentals 
under each copy of $\alg{h}$
\[\label{eq:partmult}
\srep{C,P,K}\otimes\srep{C,P,K}'.
\]
The central charge eigenvalues $C,P,K$ are the same for both algebra factors.
For simplicity, we shall assume that the maximum symmetry is
preserved, i.e.~we set $\xi_k=\rho_k=\tau=1$ throughout this section.

\subsection{Bethe Equations}
	
Our goal in this context 
is to find equations which describe
the exact and complete spectrum of planar $\superN=4$ SYM and/or
planar strings on $AdS_5\times S^5$ 
at \emph{finite coupling $g$ and finite length $L$}.
Whether or not this is too much to ask for and
whether or not the resulting equations would
resemble Bethe/Baxter equations is currently unclear.
Nevertheless it does not hurt to be optimistic:
Bethe equations exist for the leading order
in gauge theory \cite{Beisert:2003yb}. 
A description in terms of spectral curves 
is the analog for classical strings \cite{Beisert:2005bm,Dorey:2006zj}.
Higher-order corrections of the form of Bethe equations 
\cite{Beisert:2005fw} are known to yield correct results
for both types of models.

The proposed Bethe equations for planar AdS/CFT \cite{Beisert:2005fw} 
come in five flavours.%
\footnote{Two pairs of Bethe equations in \protect\cite{Beisert:2005fw} 
are equivalent which reduces the overall number
of equations from seven to five.}
Two flavours correspond to excitations related to
one $\alg{psu}(2|2)$ factor. They take precisely
the form \eqref{eq:Bethe23}.
Two further flavours correspond to the other 
$\alg{psu}(2|2)'$ factor. These equations
are as in \eqref{eq:Bethe23} with 
$y_j,w_j$ replaced by $y'_j,w'_j$.  
The main Bethe equation reads
\[\label{eq:mainbethe}
1=
\lrbrk{\frac{\xm{\auxrep}}{\xp{\auxrep}}}^L
\mathop{\prod_{j=1}^{K}}_{j\neq k}
S^0_{j\auxrep}S'^0_{j\auxrep}
\lrbrk{\frac{\xp{\auxrep}-\xm{j}}{\xm{\auxrep}-\xp{j}}}^2
\prod_{j=1}^N\frac{\xm{\auxrep}-y_j}{\xp{\auxrep}-y_j}
\prod_{j=1}^{N'}\frac{\xm{\auxrep}-y'_j}{\xp{\auxrep}-y'_j}\,,
\]
where $L$ represents the length of the chain.
At the same time, $L$ also determines
the remaining Cartan charge 
$E=L+\half N+\half N'+C-K$ of $\alg{psu}(2,2|4)$
besides $K,N,M,N',M'$. 
The Bethe equation ties up the two 
$\alg{psu}(2|2)$ factors and 
it takes precisely the form suggested 
by the elements of the diagonalised S-matrix in 
\eqref{eq:sdiagelem}.

With suitable choices of the scattering phase factor $S^0$,
the Bethe equations are known to work at least asymptotically:
The spectrum they encode agrees 
with the first few orders in planar $\superN=4$ gauge theory
at small coupling $g\approx 0$.
Agreement may potentially break down around $\order{g^{2L}}$
where the range of the interaction grows as long as the 
spin chain state. A similar problem is known 
to exist for the Bethe equations of the Inozemtsev chain \cite{Inozemtsev:2002vb,Serban:2004jf}.
Likewise, Bethe equations describe the first few orders
in string theory on $AdS_5\times S^5$ at string coupling $g\approx \infty$. 
The range of applicability is presently not clear at strong coupling
due to potential exponentially suppressed corrections
\cite{Schafer-Nameki:2005tn,Schafer-Nameki:2005is,Schafer-Nameki:2006gk}.
As the solution is quite similar in nature 
to the solution of the $\grp{O}(N)$ model in \cite{Zamolodchikov:1978xm}
one may expect it to have the same limitations regarding states
on a circle.
A thermodynamic Bethe ansatz may 
be required to impose exact periodicity
\cite{Ambjorn:2005wa}.

\subsection{Algebraic Bethe Ansatz}

The main obstruction for a rigorous treatment 
of the underlying integrable system is that there is no obvious
R-matrix with $\alg{psu}(2,2|4)$ symmetry for the model.
The Hamiltonian is long-ranged while 
an R-matrix typically leads to nearest-neighbour 
interactions. Therefore we cannot rely on the 
algebraic Bethe ansatz which would provide us 
with all necessary tools to prove
exactness of the Bethe equations. 

A promising way to bypass this problem is 
to use a model where the long-range character is merely effective.
In the $\alg{su}(2)$ sector of $\superN=4$ gauge theory 
this can be achieved, at least to some extent, 
with the Hubbard model Hamiltonian \cite{Rej:2005qt},
see \secref{sec:RSS}.
In terms of the integrable structure
another level is added to the Bethe ansatz:
The sites of the homogeneous spin chain are
lifted to momentum-carrying excitations of a more
fundamental spin chain. An extra Bethe equations
governs the distribution of the site momenta.
A similar approach for string theory was proposed in 
\cite{Mann:2004jr,Mann:2005ab,Gromov:2006dh}.
The additional level of the Bethe ansatz leads to
a proliferation of states which needs to be compensated
by stronger physical state conditions. 
In string theory the additional Bethe roots should all
occupy mode number zero and for gauge theory the distribution
of Bethe roots also has to be of a special kind. 
One can compare this approach qualitatively to a covariant approach
as opposed to the light cone approach pursued in this paper: 
In the covariant approach, the particles transform under
the full symmetry algebra. A large amount of unphysical states
are projected out by the Virasoro constraints. 
In the light cone approach, particles transform under a residual
algebra. The physicality condition reduces to a simple level matching
constraint. 

A problem which has to be dealt with in such a covariant approach is
that the Hamiltonian is part of the symmetry algebra. 
That means that we must be able to read off the energy
of a state from its constituent particles. 
This is in fact the same situation as for the light cone approach,
but the algebra $\alg{psu}(2,2|4)$ is substantially bigger
and more constraining than $\alg{h}\approx \alg{su}(2|2)$.
While $\alg{h}$ admits a simple representation with non-trivial 
contributions to the central charge, this may not be so for 
$\alg{psu}(2,2|4)$. For instance the fundamental, adjoint and 
$\superN=4$ SYM field representations all obey some atypicality conditions. 
Therefore they have quantised contributions to the energy. 
Most likely they cannot be deformed to continuous energies without
adding infinitely many new components or violating unitarity. 
The question would therefore be, in which representation of $\alg{psu}(2,2|4)$ 
should the particles transform? It might be worth
considering representations which have neither a
highest nor a lowest-weight component \cite{Teschner:2006aa,Bytsko:2006ut}.
In the Bethe ansatz for such a representation, 
the states would necessarily have infinitely many Bethe roots. 
In that case, it is often more convenient to use 
the analytic Bethe ansatz or Baxter equations. 

Another point worth mentioning in this context is that the symmetry
algebra of $\superN=4$ \emph{gauge} theory is actually larger than $\alg{psu}(2,2|4)$;
it merely reduces to $\alg{psu}(2,2|4)$ on physical states.
Otherwise it is naturally extended by gauge transformations
which arise in the commutator of two supercharges. 
These gauge transformations actually provide 
the two additional charges $\gen{P},\gen{K}$ of $\gen{h}$.
However, for $\alg{psu}(2,2|4)$ the extension is
enormous, it comprises gauge transformations with 
arbitrary transformation parameters. The latter could
be single fields, open chains of fields, but also 
variations of fields and combinations of these elements. 
In the planar case, gauge transformations
have been described in section 2.6.2 of \cite{Beisert:2004ry}.
Perhaps it is possible to construct an R-matrix 
with $\alg{psu}(2,2|4)$ symmetry extended by these gauge transformations.
This might serve as a starting point for the 
algebraic Bethe ansatz for planar AdS/CFT.

The algebraic Bethe ansatz typically starts from an
R-matrix which is used to construct a monodromy matrix
and a transfer matrix. Even though we do not know if an R-matrix 
exists, it might be possible to construct monodromy and transfer matrix operators. 
The algebraic Bethe ansatz would then provide us with 
the eigenvalue of a transfer matrix from which we could
derive the analytic Bethe ansatz. 
We shall proceed with the assumption that transfer matrix eigenvalues exist
and will attempt to reverse-engineer them in the following section.

\subsection{Transfer Matrices}

Transfer matrices are very useful objects for integrable spin chains. 
They encode the full set of mutually commuting conserved quantities.
Typically, the Hamiltonian is one of these. 
What is more, the knowledge of a functional form for their spectra
can be used to formulate an analytic Bethe ansatz or Baxter equations.
In other words, the formal expression does not only produce the
correct eigenvalue of the transfer matrix for any given state, 
but also determines which states are admissible, i.e.~consistent with 
the periodicity conditions.

Our optimistic hope is that a valid expression for the transfer matrix eigenvalues
will serve the same purpose in this model, namely `solve' the latter.%
\footnote{Or, if this is not possible, we might at least
find a reason for this unfortunate fact.}
Before we start, we have to specify which representation of
the symmetry group shall be used for the transfer matrix. 
In principle this choice should not matter. Nevertheless, it
is a priori not clear which representations are admissible.
It is however clear that among the admissible choices
some are more convenient than others. 
E.g.~for standard integrable spin chains
one almost invariably considers transfer matrices in 
fundamental representations.

Essentially, we will be trying to find the eigenvalue 
of an operator whose precise form we do not know.
What helps in this seemingly futile quest 
is the fact that transfer matrices are closely
tied to representation theory. 
Transfer matrices are the traces of monodromy matrices
in some representation of the symmetry algebra. 
Thus, we should expect the transfer matrix
(or its eigenvalues)
to be a sum over one term per component 
of the representation.
Furthermore, when decomposing the symmetry algebra 
(e.g.~to the residual algebra),
the transfer matrix 
will decompose into transfer matrices 
of the subalgebra.
In the case of $\alg{psu}(2,2|4)$ and its residual subalgebra
we can make use of the knowledge we obtained in the previous 
\secref{sec:transfer}.

\paragraph{Fundamental Representation.}

Let us start by considering the $4|4$-dimensional fundamental representation 
of $\alg{su}(2,2|4)$.
The benefit is that it is probably the simplest possible choice. 
A potential drawback is that it is not a proper
representation of $\alg{psu}(2,2|4)$ because it has
non-zero central charge.

The representation splits up into two representations of the residual
algebra
\[
\srep{\mathrm{fund}}_{\alg{su}(2,2|4)}
\to
\bigbrk{\srep{0,0;+\vec{C}}\otimes\srep{\cdot}'}\oplus
\bigbrk{\srep{\cdot}\otimes\srep{0,0;-\vec{C}}'},
\]
where $\srep{\cdot}$ is the trivial representation. 
The opposite signs for the central charge vector $\vec{C}=(\half,0,0)$
imply that one representation is the fundamental 
of one $\alg{su}(2|2)$ factor while the other is the 
conjugate fundamental of the other $\alg{su}(2|2)$.
Due to this decomposition we may expect the eigenvalue of
the fundamental transfer matrix to take the form
\[\label{eq:psufund}
T\indup{fund}(\xpm{\auxrep})=
f(\xpm{\auxrep})\,T(\xpm{\auxrep})
+
f'(\xpm{\auxrep})\,T'(1/\xpm{\auxrep}).
\]
Here $T$ and $T'$ are the transfer matrices in \eqref{eq:trans}
which depend implicitly on the Bethe roots
$\xpm{k},y_k,w_k$ and $\xpm{k},y'_k,w'_k$.
Conversely, the prefactors $f$ and $f'$ may depend implicitly 
only on the main Bethe roots $\xpm{k}$.
Note that this expression agrees with the
transfer matrix eigenvalue for the $\alg{psu}(1,1|2)$ sector conjectured in 
\cite{Beisert:2005fw}. To compare we would have to discard
two components from each $T,T'$ and multiply by suitable factors $f,f'$.

The aim would be to adjust the prefactors such that 
all singularities whose position depends on the $\xpm{k}$ 
cancel out when suitable Bethe equations are met. 
The Bethe equations for the auxiliary roots
already guarantee that the singularities
related to $y_k,w_k$ and $y'_k,w'_k$ are absent.
It may then appear favourable 
to remove as many poles depending on $\xpm{k}$ as possible
using the functions $f,f'$.
We may for instance set 
\[
f(\xpm{\auxrep})=
\prod_{j=1}^{K}S^0_{j\auxrep}\,\frac{\xp{\auxrep}-\xm{j}}{\xm{\auxrep}-\xm{j}}\,,
\qquad
f'(\xpm{\auxrep})=
\prod_{j=1}^{K}S^{\prime0}_{j\bar\auxrep}\,\frac{1/\xp{\auxrep}-\xm{j}}{1/\xm{\auxrep}-\xm{j}}\,.
\]
Then the two terms $fT_{2,3}$ in \eqref{eq:transcomp}
will not depend on $\xpm{\auxrep}$ at all.
The remaining singular terms read
\<
T\indup{fund}(\xpm{\auxrep})\eq
+
\prod_{j=1}^{K}\frac{\xm{\auxrep}-\xp{j}}{\xm{\auxrep}-\xm{j}}
\prod_{j=1}^{N}\frac{\xp{\auxrep}-y_j}{\xm{\auxrep}-y_j}
+
\prod_{j=1}^{K}\frac{1/\xp{\auxrep}-\xm{j}}{1/\xp{\auxrep}-\xp{j}}
\prod_{j=1}^{N}\frac{1/\xm{\auxrep}-y_j}{1/\xp{\auxrep}-y_j}
\nlnum\nonumber
+
\prod_{j=1}^{K}\frac{\xp{\auxrep}-\xm{j}}{\xp{\auxrep}-\xp{j}}
\prod_{j=1}^{N'}\frac{\xm{\auxrep}-y'_j}{\xp{\auxrep}-y'_j}
+
\prod_{j=1}^{K}\frac{1/\xm{\auxrep}-\xp{j}}{1/\xm{\auxrep}-\xm{j}}
\prod_{j=1}^{N'}\frac{1/\xp{\auxrep}-y'_j}{1/\xm{\auxrep}-y'_j}
+\ldots\,.
\>
This expression has various poles.
The pole at $(\xp{\auxrep},\xm{\auxrep})=(\xp{k},\xm{k})$
cancels under the condition
\[\label{eq:strangebethe}
1=
\frac{1-1/\xp{k}\xp{k}}{1-1/\xm{k}\xm{k}}
\mathop{\prod_{j=1}^{K}}_{j\neq k}
\frac{\xp{k}-\xm{j}}{\xm{k}-\xp{j}}\,
\frac{1-1/\xp{k}\xp{j}}{1-1/\xm{k}\xm{j}}
\prod_{j=1}^{N}\frac{\xm{k}-y_j}{\xp{k}-y_j}
\prod_{j=1}^{N'}\frac{\xm{k}-y'_j}{\xp{k}-y'_j}\,.
\]
The same condition also ensures cancellation of the pole
at $(\xp{\auxrep},\xm{\auxrep})=(1/\xp{k},1/\xm{k})$.
However, there still remain poles at 
$(\xp{\auxrep},\xm{\auxrep})=(1/\xp{k},\xm{k})$\
and $(\xp{\auxrep},\xm{\auxrep})=(\xp{k},1/\xm{k})$.
Their cancellation would require, in addition to \eqref{eq:strangebethe}, the conditions
\[
\prod_{j=1}^{N}\frac{1/\xm{k}-y_j}{1/\xp{k}-y_j}
=
\prod_{j=1}^{N'}\frac{\xm{k}-y'_j}{\xp{k}-y'_j}\quad
\mbox{and}\quad
\prod_{j=1}^{N}\frac{\xm{k}-y_j}{\xp{k}-y_j}
=
\prod_{j=1}^{N'}\frac{1/\xm{k}-y'_j}{1/\xp{k}-y'_j}\,.
\]
While \eqref{eq:strangebethe} reminds of the Bethe equation 
\eqref{eq:mainbethe} for $\xpm{k}$, these 
additional conditions will overconstrain the system.
Furthermore, it is not an option to consider only 
the poles at
$(\xp{k},\xm{k})$ and
$(1/\xp{k},1/\xm{k})$, but not
$(1/\xp{k},\xm{k})$ and
$(\xp{k},1/\xm{k})$
as there is no well-defined distinction between them: It is only consistent
to demand cancellation of \emph{all} dynamical poles, i.e.~those
whose position depends on the $\xpm{j}$.
The only scenario in which cancellation of some poles would
be acceptable is a perturbative one:
Assume in some limit (in particular the weak/strong coupling regime with $g\to0,\infty$) 
we can clearly distinguish the $\xpm{k}$ from the $1/\xpm{k}$.
Then we could demand cancellation of poles at, e.g., $(\xp{k},\xm{k})$ only.
In this scenario, the resulting Bethe equations would be valid asymptotically,
but at finite $g$ the analytic Bethe ansatz would fail.

In fact, we can manifestly evade an overconstrained system by setting for example
\<\label{eq:fBDS}
f(\xpm{\auxrep})\eq
\prod_{j=1}^{K}S^0_{j\auxrep}\,\frac{\xp{\auxrep}-\xm{j}}{\xm{\auxrep}-\xm{j}}
\lrbrk{1-1/\xp{\auxrep}\xp{j}}
\lrbrk{1-1/\xm{\auxrep}\xp{j}},
\nln
f'(\xpm{\auxrep})\eq
\lrbrk{\frac{\xm{\auxrep}}{\xp{\auxrep}}}^{L}\prod_{j=1}^{K}S^{\prime0}_{j\bar\auxrep}\,
\lrbrk{1-1/\xp{\auxrep}\xm{j}}^2.
\>
The resulting transfer matrix eigenvalue has 
the following singularities
\<
T\indup{fund}(\xpm{\auxrep})\eq
+
\prod_{j=1}^{K}\frac{1-1/\xp{\auxrep}\xp{j}}{\xm{\auxrep}-\xm{j}}
\lrbrk{u_{\auxrep}-u_j-ig^{-1}}
\prod_{j=1}^{N}\frac{\xp{\auxrep}-y_j}{\xm{\auxrep}-y_j}
\nlnum\nonumber
+
\lrbrk{\frac{\xm{\auxrep}}{\xp{\auxrep}}}^{L}
\prod_{j=1}^{K}
\frac{1-1/\xp{\auxrep}\xp{j}}{\xm{\auxrep}-\xm{j}}
\lrbrk{u_{\auxrep}-u_{j}+ig^{-1}}
\prod_{j=1}^{N'}\frac{\xm{\auxrep}-y'_j}{\xp{\auxrep}-y'_j}
+\ldots\,.
\>
The only potential dynamical poles are at
$(\xp{k},\xm{k})$ and they cancel under the condition
\<\label{eq:BDS}
1=
\mathop{\prod_{j=1}^{K}}_{j\neq k}
\frac{\xp{k}-\xm{j}}{\xm{k}-\xp{j}}\,
\frac{1-1/\xp{k}\xm{j}}{1-1/\xm{k}\xp{j}}
\prod_{j=1}^{N}\frac{\xm{k}-y_j}{\xp{k}-y_j}
\prod_{j=1}^{N'}\frac{\xm{k}-y'_j}{\xp{k}-y'_j}\,.
\>
This equation is the main Bethe equation for 
perturbative $\superN=4$ gauge theory at the first few loop orders 
which was proposed in \cite{Beisert:2005fw}.
However, it is clearly not the right choice for 
perturbative string theory \cite{Arutyunov:2004vx}
because of the missing phase factor.
One may try to adjust the functions $f,f'$
in order to implement it, but at
the same time no new singularities may be introduced. 
In other words, the phase factor would have to be decomposed
into poles and zeros and then distributed 
properly between $f$ and $f'$.
Further care has to be taken regarding 
various Riemann sheets that seem to
exist in the phase factor \cite{Beisert:2006ib}.
This would involve specifying 
the precise definition of the 
various $\xpm{\auxrep}$ and $1/\xpm{\auxrep}$
that appear as arguments of the phase factor.
This is beyond the scope of the present work.
Another interesting point to be understood is 
if and how the factors $f,f'$ in \eqref{eq:fBDS} could 
arise from some transfer matrix operator.

\paragraph{Field Representation.}

The spins in the $\superN=4$ SYM spin chain
belong to a non-compact multiplet,
let us denote it by $\srep{\mathrm{field}}_{\alg{psu}(2,2|4)}$.
It decomposes as follows in the residual 
algebra
\[
\srep{\mathrm{field}}_{\alg{psu}(2,2|4)}\to
\bigbrk{\srep{\cdot}\otimes\srep{\cdot}'}\oplus\bigoplus_{n=1}^\infty
\bigbrk{\srep{0,n-1;n\vec{C}}\otimes\srep{0,n-1;n\vec{C}}'},
\]
with $\vec{C}=(\half,0,0)$.
This is in fact just the decomposition
used at the first level of the coordinate Bethe ansatz:
The trivial representation corresponds to the vacuum
and the $n$-th representation in the sum 
corresponds to an $n$-fold excitation
of the vacuum at a single site. 
In particular, for $n=1$ we obtain 
$\srep{0,0}\otimes\srep{0,0}'$ which is the 
particle multiplet \eqref{eq:partmult} of the model.

The corresponding formula for 
the transfer matrix eigenvalue is 
\[\label{eq:fieldtrans}
T\indup{field}(x_{\auxrep})=
f_0(\xexp{0}{\auxrep})+
\sum_{n=1}^\infty
\ushift_{\auxrep}^{-n}
f_n(\xexp{\pm n}{\auxrep})\,
T_{\srep{0,n-1}}(\xexp{\pm n}{\auxrep})\,
T'_{\srep{0,n-1}}(\xexp{\pm n}{\auxrep})\,
\ushift^{+n}_{\auxrep}
\]
with $\ushift_{\auxrep}$ the
shift operator given in 
\eqref{eq:shiftdef}.
The transfer matrix $T_{\srep{0,n-1}}$ for the
$n$-fold excitation is
given explicitly in \eqref{eq:transanti}.
Note that this transfer matrix eigenvalue depends
explicitly on all the parameters $\xexp{-2n}{\auxrep}$
and is consequently defined on an infinite-genus 
Riemann surface \cite{Beisert:2006ib}.
This is what may make the investigation 
of this particular transfer matrix quite subtle.

\begin{figure}\centering
\includegraphics{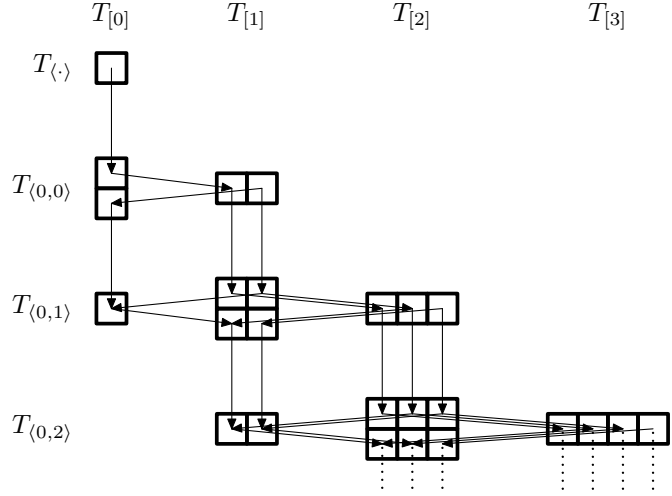}
\caption{Structure of the $\alg{psu}(2,2|4)$ field multiplet:
Boxes represent multiplet components and
clusters represent $\alg{su}(2)\times\alg{su}(2)\subset\alg{su}(2,2|4)$.
Arrows represent simple roots of $\alg{psu}(2,2|4)$.
Not shown are the multiplets of the other $\alg{su}(2|2)'$.}
\label{fig:fieldrep}
\end{figure}

In order to understand the transfer matrix eigenvalue
in \eqref{eq:fieldtrans}
it is useful to consider the structure of 
the field multiplet in \figref{fig:fieldrep}.
A box corresponds to a component of the multiplet and
thus to one term in the transfer matrix.
Horizontal stacks of $m+1$ boxes are spin-$(\half m)$
multiplets of $\alg{su}(2)$ corresponding to the
transfer matrix $T_{[m]}$ in \eqref{eq:transsu2}. 
The boxes in one line of the diagram correspond
to the $\srep{0,n-1}$ representation of $\gen{h}$,
the four horizontal stacks correspond to 
the four terms of $T_{\srep{0,n-1}}$ in \eqref{eq:transanti} 
(from left to right and from top to bottom).
Altogether the figure shows $T\indup{field}$, 
but with the components $T_{\srep{0,n-1}}'$ of the other algebra
$\alg{h}'$ hidden.

Within the analytic Bethe ansatz, dynamic poles in the transfer matrix eigenvalue 
cancel between two terms.
Cancellation should take place if the two components are related by a
simple root of the algebra. In the figure, vertical arrows correspond
to $\xpm{}$ roots and their corresponding Bethe equations. 
Likewise, right and left arrows represent $y_j$ 
and $1/y_j$ roots, respectively.
The $\alg{su}(2)$ roots $w_j$
are not depicted, they connect 
the boxes within horizontal stacks.
All the dynamical poles for $y_j$ and $w_j$ 
cancel if the $\alg{h}$ Bethe equations hold
because $T\indup{field}$ is composed from 
proper $\alg{h}$ transfer matrices.
The dynamical poles for $\xpm{j}$ remain to be investigated. 
They should be cancelled by Bethe equations corresponding to vertical arrows.

An instructive choice for the functions $f_n$ reads
\[
f_n(\xpm{\auxrep})=
\prod_{j=1}^{K}\prod_{k=1}^m S^0(\xpm{j},\xexp{+m-2k+2,+m-2k}{\auxrep})
\prod_{j=1}^{K}\frac{\xexp{+n}{\auxrep}-\xm{j}}{\xexp{-n}{\auxrep}-\xp{j}}\,
\frac{\xexp{+n}{\auxrep}-\xm{j}}{\xexp{-n}{\auxrep}-\xm{j}}\,
\frac{1/\xexp{+n}{\auxrep}-\xp{j}}{1/\xexp{+n}{\auxrep}-\xm{j}}\,.
\]
It leads to a cancellation of poles at
$(\xexp{-2n}{\auxrep},\xexp{-2n-2}{\auxrep})=(\xp{k},\xm{k})$
provided that the equation \eqref{eq:strangebethe} holds.
The cancellation takes place between 
two consecutive terms in \eqref{eq:fieldtrans}, more
precisely between the first or third term and
the second or fourth term in \eqref{eq:transanti},
see \figref{fig:fieldrep}.
Unfortunately, as above,
there are additional poles at 
$(\xexp{-2n}{\auxrep},\xexp{-2n-2}{\auxrep})=(\xp{k},1/\xm{k})$
and 
$(\xexp{-2n}{\auxrep},\xexp{-2n-2}{\auxrep})=(1/\xp{k},\xm{k})$
whose cancellation would lead to constraints which overdetermine the system.
We have not yet succeeded in finding suitable functions $f_n$ 
to reproduce a single equation such as \eqref{eq:BDS} or \eqref{eq:mainbethe}
from cancellation of all dynamical poles in the transfer matrix.

There seems to be an alternative expression for a transfer matrix eigenvalue
in the field representation for which the cancellations of poles 
follow a similar pattern.
It takes the form
\[
\tilde T\indup{field}(x_{\auxrep})=
\tilde f_0(\xexp{0}{\auxrep})+
\sum_{n=1}^\infty
\ushift^{+n}_{\auxrep}\tilde f_n(\xexp{\pm n}{\auxrep})\,
T_{\srep{0,n-1}}(\xexp{\pm n}{\auxrep})\,
T'_{\srep{0,n-1}}(\xexp{\pm n}{\auxrep})\,
\ushift^{+n}_{\auxrep}\,.
\]
It is conceivable that it corresponds to the reverse transfer matrix 
as defined in \secref{sec:reverse}. 
Furthermore, one might contemplate taking the combination 
$T\indup{field}+\tilde T\indup{field}$. 
The structure of this object is reminiscent of a 
representation without highest and lowest weights.
Such representations may play a role for the
formulation of the model on a finite chain \cite{Teschner:2006aa,Bytsko:2006ut}
and it would be worth investigating their representation theory
for the superalgebra $\alg{psu}(2,2|4)$.

It could also turn out that the analytic Bethe ansatz does not strictly
apply at finite coupling $g$ and that the dynamical poles will not cancel 
for any choice of functions. 
In this case one may hope for a covariant approach as proposed in 
\cite{Mann:2004jr,Mann:2005ab,Gromov:2006dh,Rej:2005qt}
to describe the exact spectrum. 
Then the main Bethe equation would not be rigid as in 
\eqref{eq:mainbethe}, but rather
take a dynamical form depending on further auxiliary spectral
parameters. These would have to obey further Bethe equations
which should yield the desired phase factor of the S-matrix,
at least approximately. Let us nevertheless start a final attempt
to construct the analytic Bethe ansatz using the quantum characteristic
function of \secref{sec:qchar}.

\paragraph{Quantum Characteristic Function.}

The quantum characteristic function,
which was introduced in~\secref{sec:qchar}, 
may also exist for the model with $\alg{psu}(2,2|4)$ symmetry.
It could be used to generate transfer matrices
in various symmetric representations
including those discussed above
and thus treat them in a unified fashion.
By qualitative comparison we are led
to the generic form
\[
\qdetev\supup{\alg{psu}(2,2|4)}_{\auxrep}=
A(\xexp{0}{\auxrep})\,
\qdetev_{\bar\auxrep}\,
B(\xexp{0}{\auxrep})\,
\qdetev^{\prime}_{\auxrep}\,
C(\xexp{0}{\auxrep}).
\]
Here $\qdetev_{\bar\auxrep}$ represents 
the quantum characteristic function of
one $\alg{h}$ subalgebra with
argument $1/\xpm{\auxrep}$.
The other factor 
$\qdetev'_{\auxrep}$ 
is the quantum characteristic function 
of the other $\alg{h}$ subalgebra.
Finally, $A,B,C$ are some undefined functions.

There are two useful ways to expand the characteristic function: 
First we shall expand both factors of $\qdetev$ according to 
\eqref{eq:genconj}. 
This expansion yields antisymmetric products of the fundamental
representation
\[
\qdetev\supup{\alg{psu}(2,2|4)}_{\auxrep}=
A(\xexp{0}{\auxrep})\,
B(\xexp{0}{\auxrep})\,
C(\xexp{0}{\auxrep})
-
\ushift_{\auxrep}^{-1}\,
T\indup{fund}(\xpm{\auxrep})\,
\ushift_{\auxrep}^{-1}
+\ldots\,.
\]
The first term represents some singlet transfer matrix
and the second term the fundamental transfer matrix in 
\eqref{eq:psufund}
with the coefficient functions
\<
f(\xpm{\auxrep})\eq
A(\xp{\auxrep})\,
B(\xm{\auxrep})\,
C(\xm{\auxrep})\,
\prod_{j=1}^{K}
S^0_{j\bar\auxrep}S^0_{j\auxrep}\,
\frac{\xp{\auxrep}-\xm{j}}{\xm{\auxrep}-\xm{j}}\,
\frac{1/\xp{\auxrep}-\xp{j}}{1/\xm{\auxrep}-\xp{j}}\,,
\nln
f'(\xpm{\auxrep})\eq
A(\xp{\auxrep})\,
B(\xp{\auxrep})\,
C(\xm{\auxrep})\,
\prod_{j=1}^{K}
S'^0_{j\bar\auxrep}S'^0_{j\auxrep}\,
\frac{\xp{\auxrep}-\xm{j}}{\xm{\auxrep}-\xm{j}}
\frac{1/\xp{\auxrep}-\xp{j}}{1/\xm{\auxrep}-\xp{j}}\,.
\>

The second mode of expansion is to use 
\eqref{eq:genconj} for the first factor $\qdetev_{\bar\auxrep}$, 
but \eqref{eq:genanti} for $\qdetev'_{\auxrep}$.
It yields a Laurent expansion of the type
\[
\qdetev\supup{\alg{psu}(2,2|4)}_{\auxrep}\simeq
\sum_{n,m=0}^\infty
\ushift_{\auxrep}^{-m}\,\tilde T_{\srep{0,m-1}}\,\ushift_{\auxrep}^{-m}
\ushift_{\auxrep}^{+n}\,T'_{\srep{0,n-1}}\,\ushift_{\auxrep}^{+n}
=
\sum_{n=-\infty}^\infty
\ushift_{\auxrep}^{n}\,T_{n}\,\ushift_{\auxrep}^{n}.
\]
The coefficients $T_n$ correspond to 
non-compact representations of $\alg{su}(2,2|4)$
with central charge proportional to $n$.
They have infinitely many terms of the form
\[
T_n\simeq
\sum_{k=\min(0,-n)}^\infty
\ushift_{\auxrep}^{-n-k}\,\tilde T_{\srep{0,k-1}}\,
\ushift_{\auxrep}^{+n}
\,T'_{\srep{0,n+k-1}}\,
\ushift_{\auxrep}^{+k}.
\]
The representation corresponding to $n=0$ is the field representation 
of $\alg{psu}(2,2|4)$ and
consequently $T_0=T\indup{field}$ in \eqref{eq:fieldtrans}.
The precise form for the coefficient functions $f_k$ 
reads
\<
f_0(\xexp{0}{\auxrep})\eq
A(\xexp{0}{\auxrep})\,
B(\xexp{0}{\auxrep})\,
C(\xexp{0}{\auxrep})
\prod_{j=1}^{K}
\frac{1/\xexp{0}{\auxrep}-\xp{j}}{1/\xexp{0}{\auxrep}-\xm{j}}\,
\frac{\xexp{0}{\auxrep}-\xm{j}}{\xexp{0}{\auxrep}-\xp{j}}\,,
\nln
f_1(\xpm{\auxrep})\eq
A(\xp{\auxrep})\,
B(\xm{\auxrep})\,
C(\xp{\auxrep})
\prod_{j=1}^{K}
S^0_{j\bar\auxrep}S^0_{j\auxrep}\,
\frac{1/\xp{\auxrep}-\xp{j}}{1/\xm{\auxrep}-\xm{j}}
\frac{\xp{\auxrep}-\xm{j}}{\xm{\auxrep}-\xp{j}}\,,\quad\ldots\,.
\>

The transfer matrix eigenvalues derived from
the quantum characteristic function have various poles.
Some of them cancel when the following condition holds
\[\label{eq:BetheChar}
1=-
\frac{B(\xp{\auxrep})}{B(\xm{\auxrep})}
\prod_{j=1}^{K}
S'^0_{j\auxrep}S^0_{\bar\auxrep j}\,
\frac{\xp{\auxrep}-\xm{j}}{\xp{\auxrep}-\xp{j}}\,
\frac{\xp{\auxrep}-\xm{j}}{\xm{\auxrep}-\xp{j}}\,
\frac{1/\xm{\auxrep}-\xm{j}}{1/\xp{\auxrep}-\xm{j}}
\prod_{j=1}^{N}\frac{\xm{\auxrep}-y_j}{\xp{\auxrep}-y_j}
\prod_{j=1}^{N'}\frac{\xm{\auxrep}-y'_j}{\xp{\auxrep}-y'_j}
\]
for all $\xpm{\auxrep}=\xpm{k}$.
This equation resembles the main Bethe equation \eqref{eq:mainbethe}
and depends only on one of the three undetermined functions, $B$.
For example, when we set $S_{kj}=1$ and
\[
B(\xexp{0}{\auxrep})=
\prod_{j=1}^{K}
\frac{1/\xexp{0}{\auxrep}-\xm{j}}{\xexp{0}{\auxrep}-\xm{j}}\,,
\]
we recover the above Bethe equation \eqref{eq:strangebethe}.
Note, however, that several dynamic poles will remain.
Some of them can perhaps be absorbed by a suitable choice of 
functions $A,C$.

We shall close this section with the curious observation that
the general equation \eqref{eq:BetheChar} can be rewritten 
using \eqref{eq:cross} as
\[
1=
-
\frac{B(\xp{\auxrep})}{B(\xm{\auxrep})}
\prod_{j=1}^{K}
\frac{S^0_{j\auxrep}S'^0_{j\auxrep}}{X_{j\bar\auxrep}}
\lrbrk{\frac{\xp{\auxrep}-\xm{j}}{\xm{\auxrep}-\xp{j}}}^2
\prod_{j=1}^N\frac{\xm{\auxrep}-y_j}{\xp{\auxrep}-y_j}
\prod_{j=1}^{N'}\frac{\xm{\auxrep}-y'_j}{\xp{\auxrep}-y'_j}\,.
\]
Under the assumption of crossing
symmetry, $X_{kj}=1$, and when setting
\[B(\xexp{0}{\auxrep})=(\xexp{0}{\auxrep})^{-L}\,,
\]
this is precisely the Bethe equation
for a spin chain of length $L$.
It remains to be seen whether this observation
can be extended to a full-fledged analytic Bethe ansatz
for $\alg{psu}(2,2|4)$ integrable models.

\section{Conclusions and Outlook}

In this paper we have investigated the integrable
structure of a spin chain model
with centrally extended $\alg{psu}(2|2)$ symmetry
which arises in the context of the planar AdS/CFT correspondence. 
Our focus was on the S/R-matrix and transfer matrices
of this model. 
Perhaps the most important new results of this work
are the following: 
Firstly, we have obtained a simple proof that the
S-matrix satisfies the Yang-Baxter relation.
In general this is a cumbersome task, but in the present
case the proof consists mainly of representation theory
which was outlined in \secref{sec:Alg}. 
Secondly, we have shown that the one-dimensional Hubbard model
is based on the very same integrable structure that we have
been investigating. It consequently possesses a hidden supersymmetry.
The S/R-matrix can be used to derive a host of generalisations
of the Hubbard model including some of the known ones. 
Thirdly, we have derived the spectra of some transfer matrices
for the spin chain. Curiously, when Janik{}'s crossing relation for
the S/R-matrix holds, various expressions simplify to some extent. 
Furthermore, the transfer matrix eigenvalues lead to the proper Bethe
equations via an analytic Bethe ansatz. 
Finally, we have made some attempts to generalise 
transfer matrix eigenvalues to the complete planar AdS/CFT model with
$\alg{psu}(2,2|4)$ symmetry. \Secref{sec:psu224} contains
some hopefully inspiring notes for future work on 
some exact Bethe ansatz for AdS/CFT.
Potentially some other representations of $\alg{psu}(2,2|4)$ 
(or an extension by $\superN=4$ SYM gauge transformations)
play a role. For instance, it might be worth considering
representations without highest and lowest weights. 
In the case of $\alg{sl}(2)$ these are called representations of the
principle continuous series, but here there will be a richer
structure due to the higher rank 
and partial non-compactness of $\alg{psu}(2,2|4)$.

Apart from this, there are further potentially fruitful directions of investigation:
Recently, a Hopf algebra structure for the present integrable
model has been outlined. More work is
needed to cast the integrable structure into the framework
of Hopf algebras. It is likely that some of the curious
observations made here will come out more naturally 
in that framework. 
It would also be interesting to embed
the new twists $\rho,\xi,\tau$ into this framework. 

A curious fact is that the particle momentum
or spectral parameter is already an intrinsic parameter
of the representations of centrally extended $\alg{psu}(2|2)$.
For most other spin chains, the spectral parameter
is unrelated to the classical symmetry algebra of the model. 
Are there other models with the same property?
Are there more S-matrices like the present one which are
not of a difference form?
Is there a general classification for such models?
We have for instance seen that representation theory admits two choices 
for the S/R-matrix. One choice merely leads to a trivial permutation operator
while the other one yields the discussed integrable structure. 
The existence of the second solution is due to a quadratic constraint
among the central charges which is very reminiscent of a mass shell condition.
Perhaps it is possible to construct similarly interesting models
based on Poincar\'e (super)symmetry where the site momenta 
obey some quadratic relation.

\subsection*{Acknowledgements}

I am grateful to 
V.~Dobrev, 
F.~G\"ohmann,
C.~Gomez,
X.-W.~Guan, 
V.~Kazakov, 
J.~Maldacena, 
P.~Orland, 
J.~Plefka,
N.~Reshetikhin,
B.~S.~Shastry, 
P.~Sorba,
J.~Teschner,
S.~Zieme,
B.~Zwiebel,
the referees
and, in particular, 
F.~Spill and M.~Staudacher
for interesting discussions and helpful comments
regarding this work.
The work of N.~B.~is supported in part by the U.S.~National Science
Foundation Grant No.~PHY02-43680. Any opinions, findings and conclusions or
recommendations expressed in this material are those of the authors and do not
necessarily reflect the views of the National Science Foundation.

\appendix

\section{Rapidities}
\label{sec:torus}

The parameters $\xpm{}$ subject to the constraint \eqref{eq:xpmconstr}
naturally define a Riemann surface of genus one, a torus \cite{Janik:2006dc}: 
Here we shall summarise some properties of this surface
and present a useful parametrisation.

\subsection{Rapidity Plane}

We can introduce a single complex coordinate $z$ on this torus
using elliptic functions with modulus $k$.
A particularly simple choice is
\[\label{eq:rapp}
p=2\ellAM(z,k),\qquad k=4ig=\frac{i\sqrt{\lambda}}{\pi}\,,
\]
where $\ellAM$ is Jacobi's elliptic amplitude function
and $k$ is the elliptic modulus.%
\footnote{We use the convention that $k$ appears
in squared form, $k^2=m$, in the elliptic integrals.}
The other parameters are given by
\[\label{eq:rappar}
\xpm{}=i\,\bigbrk{\ellCN z\pm i\ellSN z}\frac{1+\ellDN z}{k\ellSN z}\,,
\qquad
u=\frac{2i\ellCN z\ellDN z}{k\ellSN z}\,,\qquad
C=\half \ellDN z.
\]
Note that in this rapidity plane,
the usual relation between momentum and energy holds
\[
\frac{dp}{dz}=4C.
\]

\subsection{Periods}

The half-periods of the elliptic functions are given by 
\[\label{eq:tper}
\omega_1=2\ellK(k),\qquad
\omega_2=2i\ellK(\sqrt{1-k^2})-2\ellK(k).
\]
In other words, the kinematic parameters $\xpm{},u,C,e^{ip}$
are invariant under the shifts $z\mapsto z+2\omega_1$ and 
$z\mapsto z+2\omega_2$. 
For real coupling $g$, 
the period $2\omega_1$ is purely real
and the period $2\omega_2$ is purely imaginary.

\paragraph{Real Period.}

In a lattice model, such as a spin chain, 
the momentum is defined only modulo $2\pi$
because structures below the lattice spacing cannot be resolved.
Here this periodicity is reflected by the real period 
of the torus. 

In fact, already a half-period $\omega_1$ shifts $p$ by $2\pi$ 
and thus leaves all the other variables invariant. 
It means that the choice \eqref{eq:rapp}
with elliptic modulus $k=4ig$ is in fact a double covering of 
the actual kinematic space 

The double covering is not a necessity, but there are at least three
reasons for using it: Firstly, we are dealing with a system
of bosons and fermions. For fermions with half-integer statistics
the period is doubled in some cases, perhaps there is a similar use here as well.
Secondly, the comparison to the R-matrix of the Hubbard model leads
to the following identifications
\[U=\frac{1}{4ik}\,,\qquad
\exp(2h)=\frac{1+\ellDN z}{4k^2\ellSN z}\,,\qquad
\frac{b}{a}=-i\bigbrk{\ellCN z+ i\ellSN z}\,,
\]
which are not periodic under $z\mapsto z+\omega_1$. 
In particular, the parameter $\xi=b/a$ \eqref{eq:hubaux} is anti-periodic.
Thirdly, the expressions in \eqref{eq:rappar} are reasonably convenient
in comparison to the expressions for a
single cover of the kinematic space, which is also a torus.

\paragraph{Imaginary Period.}

The dispersion relation \eqref{eq:latticeshell}
is (almost) a relativistic mass shell condition. 
When we Wick rotate the momentum variable, 
the mass shell becomes a circle which has a certain periodicity. 
In other words, 
when we set $x=2C$ and $y=4ig\sin(\half p)$ then
\eqref{eq:latticeshell} describes a unit circle in the $x$-$y$ plane.
One full rotation corresponds to a shift by the imaginary period $2\omega_2$.

\paragraph{Complex Structure.}

The complex structure of the torus is defined as
\[\label{eq:cstruct}
\tau=\frac{\omega_2}{\omega_1}\,.
\]
For a real coupling $g$, it is purely imaginary and thus 
the torus is rectangular.

Let us investigate the weak and strong coupling limits.
At weak coupling, the complex structure asymptotes as
\[
\tau\approx \frac{\log(g^2)}{i\pi}\,\to \infty.
\]
In other words, the imaginary periodicity disappears 
and we are left only with the periodicity 
corresponding to the discreteness of the spin chain.
This is what we expect for a perturbative gauge theory.
Conversely, at strong coupling, the complex structure
asymptotes as
\[
\tau\approx \frac{i\pi}{\log(256/g^2)}\,\to 0.
\]
Here the periodicity of the lattice disappears and 
we are left with periodicity of the Wick rotated mass shell.
This is in agreement with a smooth relativistic world sheet
and thus with classical string theory.

In fact, the two limits are not unrelated. 
Considering $\tau(g)$ as a function of the coupling
constant, it obeys the self-duality relation
\[
\tau(1/16g)=-\frac{1}{\tau(g)}\,.
\]
The corresponding map for the 't~Hooft coupling
is $\lambda\mapsto \pi^4/\lambda$. At the fixed point
$g=\sfrac{1}{4}$ or $\lambda=\pi^2$ of the map $g\mapsto 1/16g$, 
the complex structure is $\tau=i$, i.e.~the fundamental domain of the
torus is a square.

One may wonder if there is any meaning to this map
which reminds of a strong/weak duality transformation.
Similarly, it would be interesting to understand if
there exists a physical model with a general complex $\tau$, 
i.e.~with a non-rectangular torus.
This looks reminiscent of a gauge theory 
with topological angle $\theta$
and of the $\grp{SL}(2,\Integers)$ modular group.
However, in that case the complex structure of the torus 
equals $\tau=\theta/2\pi+i N\indup{c}/4\pi g^2$
and not \eqref{eq:tper,eq:cstruct}.

\subsection{Discrete Transformations and Special Points}

\begin{table}\centering
$\begin{array}{|l|cc|cc|}\hline
z\mapsto&\xp{}\mapsto&\xm{}\mapsto&C\mapsto&p\mapsto\\\hline\hline
z+\omega_1                        &\xp{}   &\xm{}   &+C   &p+2\pi\\[1pt]\hline
z+\omega_2                        &1/\xp{} &1/\xm{} &-C   &-p\\[1pt]\hline
+z+\sfrac{0}{2}(\omega_1+\omega_2)&\xp{}   &\xm{}   &+C   &+p\\[1pt]
+z+\sfrac{1}{2}(\omega_1+\omega_2)&-1/\xm{}&-\xp{}  &\cdot&\cdot\\[1pt]
+z+\sfrac{2}{2}(\omega_1+\omega_2)&1/\xp{} &1/\xm{} &-C   &-p\\[1pt]
+z+\sfrac{3}{2}(\omega_1+\omega_2)&-\xm{}  &-1/\xp{}&\cdot&\cdot\\[1pt]\hline
-z+\sfrac{0}{2}(\omega_1+\omega_2)&-\xm{}  &-\xp{}  &+C   &-p\\[1pt]
-z+\sfrac{1}{2}(\omega_1+\omega_2)&1/\xp{} &\xm{}   &\cdot&\cdot\\[1pt]
-z+\sfrac{2}{2}(\omega_1+\omega_2)&-1/\xm{}&-1/\xp{}&-C   &+p\\[1pt]
-z+\sfrac{3}{2}(\omega_1+\omega_2)&\xp{}   &1/\xm{} &\cdot&\cdot\\[1pt]\hline
\end{array}
$

\caption{Simple discrete transformations of the rapidity plane.
Dots indicate longer expressions.}
\label{tab:rapsym}
\end{table}

\begin{figure}\centering
\includegraphics[scale=0.8]{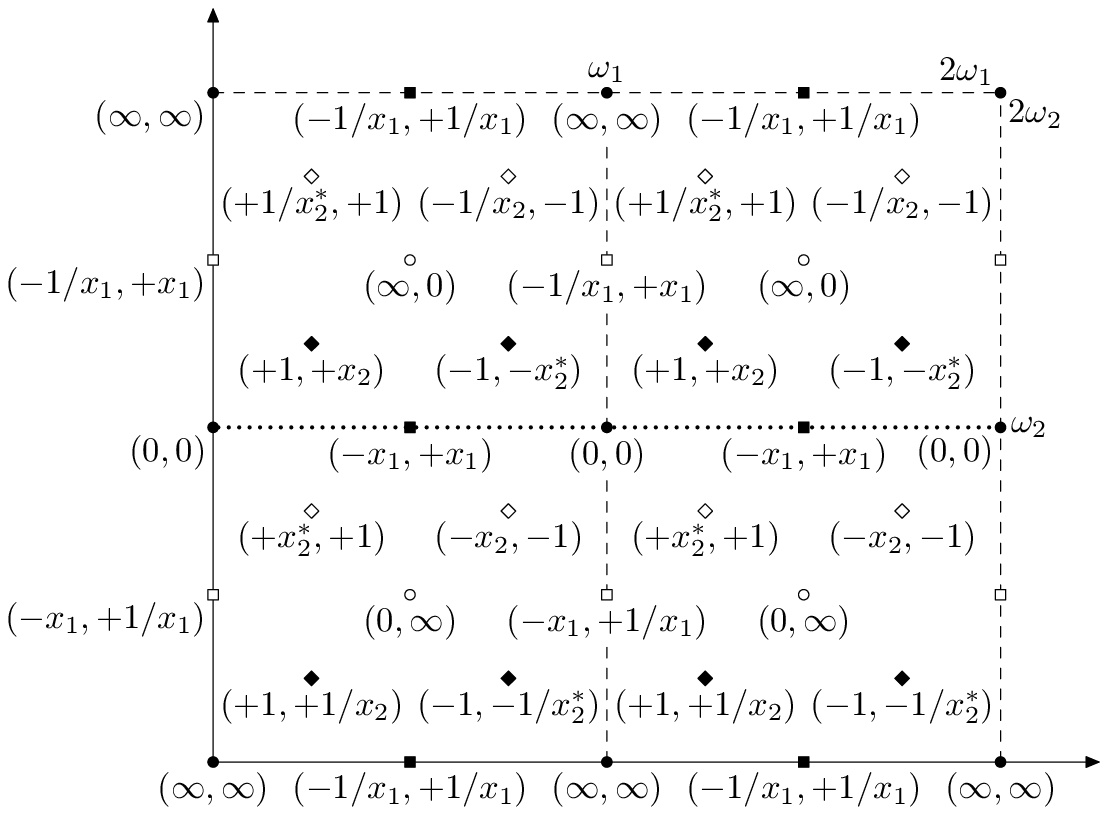}
\[\nn
x_1=i\sqrt{\frac{\sqrt{1+16g^2}-1}{\sqrt{1+16g^2}+1}}\,,\qquad
x_2=\frac{\sqrt{1+4ig}-1}{\sqrt{1+4ig}+1}\,
\]
\caption{Some special points $(\xp{},\xm{})$ in the rapidity plane.}
\label{fig:rappoints}
\end{figure}

We have already discussed shifts by the real half-period
which act trivially.
A shift by an imaginary half-period 
acts as
\[
\xpm{}(z+\omega_2)=1/\xpm{}(z)\,,
\qquad
u(z+\omega_2)=u(z)\,,\qquad
C(z+\omega_2)=-C(z).
\]
This map essentially interchanges the representation with its conjugate.
It is called the antipode map and it plays a central 
role for crossing symmetry, see \secref{sec:crossing}.
In addition to these, there are further interesting 
discrete transformations of the rapidity plane.
An extensive list is given in \tabref{tab:rapsym}.
The transformations of the maps can be verified using 
the addition formulas of elliptic functions $\ellSN,\ellCN,\ellDN$.
Of particular interest may be the map $z\mapsto -z$.
This map inverts the momentum but leaves the energy
invariant. It thus represents parity inversion. 

Points invariant under some of the transformations 
are shown in \figref{fig:rappoints}. 
They are all of the form $z=\quarter m_1\omega_1+\quarter m_2\omega_2$
with $m_1+m_2$ even. 
At these points the parameters $\xp{},\xm{}$ take special values: 
The values $0,\infty$ are invariant under taking the negative
and $\pm 1$ are invariant under inversion. The remaining points
are such that the product or quotient of $\xpm{}$ is $-1$.

\bibliography{n4trans}
\bibliographystyle{nb}

\end{document}